\documentclass{article}
\usepackage[utf8]{inputenc}
\usepackage{macros}

\usepackage{booktabs}

\usepackage{thm-restate}

\usepackage{todonotes}

\usepackage{pifont}
\newcommand{\cmark}{\ding{51}}%

\newcommand{\OL}{\mathsf{OL}}
\DeclareMathOperator{\conv}{conv}
\DeclareMathOperator{\round}{\mathsf{round}}

\DeclareMathOperator{\Vol}{Vol}
\DeclareMathOperator{\pdet}{pdet}

\title{High-Dimensional Geometric Streaming in Polynomial Space}
\author{
David P. Woodruff \\ Carnegie Mellon University \\ \texttt{dwoodruf@cs.cmu.edu} \and Taisuke Yasuda \\ Carnegie Mellon University \\ \texttt{taisukey@cs.cmu.edu}
}

\begin{document}

\maketitle

\thispagestyle{empty}
\begin{abstract}
    Many existing algorithms for streaming geometric data analysis have been plagued by exponential dependencies in the space complexity, which are undesirable for processing high-dimensional data sets. For instance, the best known algorithms for maintaining convex hulls and L\"owner--John ellipsoids use $\eps^{-\Theta(d)}$ bits of space; maintaining $\ell_p$ subspace embeddings use $d^{p/2+1}$ bits of space; and selecting $k$ out of $n$ given vectors in $\mathbb{R}^d$ maximizing the $k$-dimensional volume uses $2^k d$ bits of space. These are all intractable for large $d$, $p$, or $k$. In particular, once $d \geq \log n$, there are no known non-trivial streaming algorithms for problems such as maintaining convex hulls and L\"owner--John ellipsoids of $n$ points, despite a long line of work in high-dimensional streaming computational geometry since  \cite{AHV2004} (J.~ACM 2004). 
    
    We simultaneously improve all of these results to $\poly(d,\log n)$ bits of space by trading off with a $\poly(d,\log n)$ factor distortion. We achieve these results in a unified manner, by designing the first streaming algorithm for maintaining a coreset for $\ell_\infty$ subspace embeddings with $\poly(d,\log n)$ space and $\poly(d,\log n)$ distortion. Our algorithm also gives similar guarantees in the \emph{online coreset} model. Along the way, we sharpen known results for online numerical linear algebra by replacing a log condition number dependence with a $\log n$ dependence, answering an open question of \cite{BDMMUWZ2020} (FOCS 2020). Our techniques provide a novel connection between leverage scores, a fundamental object in numerical linear algebra, and computational geometry.
    
    For $\ell_p$ subspace embeddings, our improvements in online numerical linear algebra yield nearly optimal trade-offs between space and distortion for one-pass streaming algorithms. For instance, we obtain a deterministic coreset using $O(d^2\log n)$ space and $O((d\log n)^{\frac12-\frac1p})$ distortion for $p>2$, whereas previous deterministic algorithms incurred a $\poly(n)$ factor in the space or the  distortion \cite{CDW2018} (ICML 2018).

    Our techniques have implications also in the offline setting, where we give optimal trade-offs between the space complexity and distortion of a subspace sketch data structure, which preprocesses an $n\times d$ matrix $\bfA$ and outputs $\norm*{\bfA\bfx}_p$ up to a $\poly(d)$ factor distortion for any $\bfx$. To do this we give an elementary proof of a ``change of density'' theorem of \cite{LT1980}  (J.~Functional Analysis 1980) and make it algorithmic.
\end{abstract}


\clearpage
\setcounter{page}{1}

\newpage

\section{Introduction}

Data science has permeated modern computer science in the last few decades, leading to a surge in demand for geometric data processing algorithms on large data sets. Two decades ago, the data sets studied in practice, represented by an $n\times d$ matrix $\bfA$, had many rows (large $n$) and small dimension ($d = O(1)$). Driven by such applications, many \emph{streaming algorithms} were developed, which only require one or a few passes through a stream which allows access to the rows $\bfa_1, \bfa_2, \dots, \bfa_n\in\mathbb R^d$ one at a time. In this setting, \emph{$\eps$-kernels} were introduced by \cite{AHV2004, AHV2005}, which gave a unified approach towards obtaining $(1+\eps)$-factor approximations using $\eps^{-\Theta(d)}$ space for an enormous number of geometric problems, including width, convex hull, and minimum enclosing spherical shell, to name just a few of the applications of $\eps$-kernels.

Since then, the dimensionality of data sets encountered in practice has increased dramatically, and space complexities that scale exponentially in $d$, or even a large polynomial (say $d^{4}$), can no longer be considered practical. Some geometric problems have adapted to this high-dimensional setting, including minimum enclosing cylinder \cite[Theorem 3.1]{Cha2006}, minimum enclosing ball (MEB) \cite{ZC2006, CP2014, AS2015}, and diameter \cite[Theorem 3.2]{AS2015}, by tolerating a larger $O(1)$-factor distortion. \cite{AS2015} also give lower bounds for the MEB and diameter problems, showing that any one-pass streaming algorithm with less than an $\alpha$-factor distortion must use $\exp(\poly(d))$ bits of space, where $\alpha = \frac{1+\sqrt 2}{2}$ for MEB and $\alpha = \sqrt 2$ for diameter. Furthermore, \cite{AS2015} show that the width problem requires $\exp(\poly(d))$ bits of space for any algorithm achieving distortion smaller than $d^{1/3}/8$. Thus, distortions of at least $\poly(d)$ are necessary for some of these problems in order to achieve $\poly(d)$ bits of space. However, many problems still do not have polynomial space algorithms, even with $\poly(d)$ distortions, such as computing width, L\"owner--John ellipsoids \cite{MSS2010, AS2015}, $\ell_p$ subspace embeddings for large $p$ \cite{CDW2018}, and convex hulls \cite{BBKLY2018}.

\subsection{Our Contributions}
In this work, we address the lack of streaming algorithms for geometric problems in the high-dimensional setting by providing a unified approach towards achieving $\poly(d,\log n)$ space and distortion. As argued before, a dependence of $\poly(d)$ in the distortion is necessary for polynomial space algorithms, and is arguably natural since many geometric summarization problems inherently incur such losses in the distortion, for example for L\"owner--John ellipsoids. Distortions of $\poly\log n$ in geometric extent measures have also been previously studied in interesting cases \cite{VVYZ2007}.

To obtain our results for streaming geometry, we design the first one-pass streaming algorithm for the \emph{$\ell_\infty$ subspace sketch problem}. That is, given a row arrival stream for $\bfA\in\mathbb Z^{n\times d}$ with entries bounded by $\poly(n)$, we show how to maintain a coreset $S\subseteq[n]$ of size at most $\abs*{S} \leq O(d\log n)$ such that
\[
    \mbox{for all $\bfx\in\mathbb R^d$, }\qquad\norm*{\bfA\vert_S\bfx}_\infty \leq \norm*{\bfA\bfx}_\infty \leq O(\sqrt{d\log n})\norm*{\bfA\vert_S\bfx}_\infty.
\]
Our algorithm is deterministic and uses only $O(d^2\log^2 n)$ bits of space, which is an optimal trade-off between the space complexity and distortion, up to $\poly\log n$ factors. In fact, our algorithm has the property that each $i\in S$ is selected \emph{irrevocably}, i.e., we immediately decide whether to permanently keep or discard the row $\bfa_i$. Such algorithms can be considered under the \emph{online coreset} model, in which the input matrix $\bfA\in\mathbb R^{n\times d}$ is now allowed to take real values, and the algorithm's complexity is measured by the number of rows it stores. Under this model, our algorithm stores $O(d\log(n\kappa^\OL))$ rows and achieves a distortion of $O(\sqrt{d\log(n\kappa^\OL)})$, where $\kappa^\OL = \kappa^\OL(\bfA) \coloneqq \norm*{\bfA}_2\max_{i=1}^n\norm*{\bfA_i^-}_2$\footnote{Here, $\bfA_i$ denotes the $i\times d$ matrix formed by the first $i$ rows of $\bfA$.} is the \emph{online condition number of $\bfA$} \cite{BDMMUWZ2020}. Various linear algebraic and geometric problems have recently been considered in the online model, including spectral approximation \cite{CMP2020}, low rank approximation \cite{BLVZ2019, BDMMUWZ2020}, and $\ell_1$ subspace embeddings \cite{BDMMUWZ2020}. 

Note that the $\ell_\infty$ subspace sketch problem is of central importance in computational geometry: it is closely related to directional width estimation \cite{AHV2004, AHV2005} as well as the polytope membership problem \cite{AFM2017}. It can also be used to approximate maximum inner product search, for which sampling-based algorithms have recently received much attention in the large-scale machine learning literature \cite{BKPS2015,DYH2019,LP2021}. Even beyond these applications, we will show that the $\ell_\infty$ subspace sketch primitive in fact leads to the first $\poly(d,\log n)$ space, $\poly(d,\log n)$ distortion algorithm for a much wider variety of geometric problems, $k$-robust directional width, including $\ell_p$ subspace sketch for $p<\infty$, convex hull, L\"owner--John ellipsoids, volume maximization, minimum-width spherical shell, and solving linear programs. Our results can thus be seen as a high-dimensional and high-distortion analogue of the fact that $\eps$-kernels solve many streaming problems in the $(1+\eps)$-distortion setting \cite{AHV2004, AHV2005}.

Next, we focus on the streaming $\ell_p$ subspace sketch problem. By a known reduction to $\ell_\infty$ subspace embeddings using exponential random variables \cite{And2017, WZ2013}, one can obtain algorithms for $\ell_p$ subspace sketch from our $\ell_\infty$ result. However, this does not achieve optimal trade-offs between space complexity and distortion, and we give a different algorithm achieving optimal trade-offs, up to $\poly\log n$ factors. In particular, we obtain a deterministic algorithm achieving $O(d^2\log n)$ bits of space and $O((d\log n)^{\frac12-\frac1p})$ distortion, significantly improving upon the earlier deterministic one-pass algorithms of \cite{CDW2018}, which incurred a $\poly(n)$ factor in either the space complexity or distortion. This nearly matches the guarantee obtained by using Lewis weights in the offline setting \cite{Lew1978, SZ2001, CP2015}, and in fact achieves optimal trade-offs, up to $\poly\log n$ factors. See Table \ref{tab:subspace-sketch-stream} for a summary of our results.

Furthermore, many of our algorithms yield \emph{mergeable summaries}, i.e.,  the data structures for two matrices $\bfA_1$ and $\bfA_2$ can be merged with little overhead to form a data structure for the concatenation $[\bfA_1^\top\ \bfA_2^\top]^\top$. This allows our algorithms to apply to \emph{distributed} settings, where the rows of the input are partitioned among many servers \cite{CDW2018}.

\begin{table}
    \centering
    \begin{tabular}{ c c c c c c c c c }
    \toprule
    & Distortion & Space & & Deterministic & Online & Optimal \\
    \hline
    $p = 2$ & 1 & $d^2$ & Folklore & \cmark & & \cmark \\
    $0 < p < \infty$ & $1+\eps$ & $d^{2\lor(\frac{p}{2}+1)}$ & Thm.~\ref{thm:1+eps-merge-reduce-subspace-sketch} & & & \cmark \\
    $1 \leq p < \infty$ & $d^{O(1/\gamma)}$ & $n^\gamma d$ & Thm 4.2 \cite{CDW2018} & \cmark \\
    \rowcolor{blue!15}$p = \infty$ & $\sqrt{d\log n}$ & $d^2$ & Thm.~\ref{thm:subspace-sketch-stream-l-inf} & \cmark & \cmark & \cmark \\
    \rowcolor{blue!15}$2<p<\infty$ & $(d\log n)^{\frac12 - \frac1p}$ & $d^2$ & Cor.~\ref{cor:subspace-sketch-stream-l2-online} & \cmark & \cmark & \cmark \\
    \rowcolor{blue!15}$2<p<\infty$ & $d^{\frac12\parens*{1-\frac{q}{p}}}\log n$ & $d^{\frac{q}{2} + 1}$ & Thm.~\ref{thm:subspace-sketch-stream-lq} & & & \cmark \\
    \rowcolor{blue!15}$p=2$ & $1+\eps$ & $d^2$ & Thm.~\ref{thm:online-spectral-approx-bss} & \cmark & \cmark & \cmark \\
    \bottomrule
    \end{tabular}
    \caption{Results for one-pass streaming $\ell_p$ subspace sketch in the ``for all'' model. New results highlighted in blue. We suppress $O(1)$ factors in the distortion and $\poly(\eps^{-1},\log n)$ factors in space. ``Optimal'' means that the trade-off between the space complexity and distortion is optimal, up to $\log n$ factors (see Table \ref{tab:subspace-sketch}).}
    \label{tab:subspace-sketch-stream}
\end{table}

Although our streaming $\ell_p$ subspace sketch achieves nearly optimal trade-offs up to polylogarithmic factors, it is still possible to ask for improvements in these bounds, as well as faster algorithms, in the offline setting where we have unlimited access to $\bfA$. In a third contribution, in the offline setting, we construct $\ell_p$ subspace embeddings with nearly optimal trade-offs between space complexity and distortion, which shave all $\poly\log n$ factors off of the distortion. As a crucial step, we give a new elementary proof of a ``change of density'' theorem in geometric functional analysis due to Lewis and Tomczak-Jaegermann \cite{LT1980}, by using Lewis weights \cite{Lew1978, SZ2001, CP2015}. This allows us to make the construction algorithmic, and in fact, nearly input sparsity time. Our space complexity upper bound matches a subspace sketch lower bound due to \cite{LWW2021}. These subspace sketch lower bounds also witness the near tightness of our streaming $\ell_p$ subspace sketch algorithms. See Table \ref{tab:subspace-sketch} for our results on the offline subspace sketch problem.

\begin{table}
\centering
\begin{tabular}{ c c c c c c c }
\toprule
& & Distortion & Space \\
\hline
$0 < p \leq 2$ & Upper Bound & $1$ & $d^2$ & & \cite{Ind2006} \\
$2 < p < \infty$ & Upper Bound & $d^{\frac12-\frac1p}$ & $d^{2}$ & & \cite{Lew1978} \\
$2 < p < \infty$ & Upper Bound & $1$ & $d^{p/2+1}$ & & \cite{BLM1989, CP2015} \\
$2 < p < \infty$ & Lower Bound & $d^{\frac12\parens*{1-\frac{q}{p}}}$ & $d^{q/2 + 1}$ & $2 \leq q \leq p$ & \cite{LWW2021} \\
\rowcolor{blue!15}$2 < p < \infty$ & Upper Bound & $d^{\frac12\parens*{1-\frac{q}{p}}}$ & $d^{q/2 + 1}$ & $2 < q < p$ & Theorem \ref{thm:subspace-sketch-lp-lq} \\
\rowcolor{blue!15}$0<p\leq\infty$ & Lower Bound & $<\infty$ & $d^2$ & & Theorem \ref{thm:finite-distortion-subspace-sketch-for-all} \\
\bottomrule
\end{tabular}
\caption{Our results for subspace sketch in the ``for all'' model. Results for the ``for each'' model remove a factor of $d$ from the space bounds, for the same distortion. New results highlighted in blue. We suppress $\poly\log n$ factors in the space complexity.}
\label{tab:subspace-sketch}
\end{table}

Furthermore, our fast algorithms for computing these $\ell_p$ subspace embeddings give the fastest known running times for $\ell_p$ regression and $\ell_p$ column subset selection, when we allow for distortions which scale as $\poly(d)$ (see Table \ref{tab:lp-linear-algebra}). Note that algorithms for $\ell_p$ column subset selection already incur distortions on the order $\poly(d)$ \cite{CGKLPW2017, DWZZR2019} (as they must due to known lower bounds).

\begin{table}
    \centering
    \begin{tabular}{ c c c c c c c c c }
    \toprule
    & Distortion & Time & \\
    \hline
    $\ell_p$ Linear Regression & $1+\eps$ & $n^\omega$ & \cite{AKPS2019} \\
    & $1+\eps$ & $\nnz(\bfA) + d^{\frac{p}{2}\omega}$ & \cite{AKPS2019} + \cite{CP2015} \\
    \rowcolor{blue!15}& $d^{\frac12\parens*{1-\frac{q}{p}}}$ & $\nnz(\bfA) + d^{\frac{q}{2}\omega}$ & Theorem \ref{thm:lp-regression-large-distortion} \\
    \hline
    $\ell_p$ $k$-Column Subset Selection & $k^{1-\frac1p}$ & $n^\omega d$ & \cite{DWZZR2019} + \cite{AKPS2019} \\
    & $k^{1-\frac1p}$ & $\nnz(\bfA)d + k^{\frac{p}{2}\omega} d$ & \cite{DWZZR2019} + \cite{AKPS2019} + \cite{CP2015} \\
    \rowcolor{blue!15}& $k^{1-\frac1p+\frac12\parens*{1-\frac{q}{p}}}$ & $\nnz(\bfA)d + k^{\frac{q}{2}\omega} d$ & Theorem \ref{thm:lp-css} \\
    \bottomrule
    \end{tabular}
    \caption{Results for fast numerical linear algebra in $\ell_p$ for $p>2$, with the current matrix multiplication time $\omega \approx 2.37286$ \cite{AW2021}. Here, $q$ is any number between $2$ and $p$. We suppress $\poly(\log n,\eps^{-1})$ factors in the running time and constant factors in the distortion.}
    \label{tab:lp-linear-algebra}
\end{table}

Finally, we return to the study of streaming algorithms for $\ell_p$ subspace embeddings, and implement the above offline trade-off by showing how to compute Lewis weights using a small number of passes. This requires generalizations of Lewis weight computation and sampling results \cite{CP2015} as well as our ``change of density'' theorem that work with only $O(\log n)$ bit complexity, which may be of independent interest.


\subsection{Streaming Algorithms for Geometric Problems}

We first introduce two models of streaming algorithms which we study: the \emph{row arrival streaming model} and the \emph{online coreset model}. In these models, we have an $n\times d$ input matrix $\bfA$ with rows $\bfa_1, \bfa_2, \dots, \bfa_n$, where $n$ is so large that we cannot observe the entire matrix at once, and we can only observe one row at a time. 

In the row arrival streaming model, we assume that $\bfA\in\mathbb Z^{n\times d}$ is an integer matrix with entries bounded by $\poly(n)$. Then, the rows $\bfa_1, \bfa_2, \dots, \bfa_n$ are presented in a stream one at a time in that order, and we must minimize the number of bits that we store while making only one pass\footnote{We also consider algorithms which make multiple passes through the stream, but we will restrict to one pass for now.} through the stream of rows of $\bfA$. 

On the other hand, in the online coreset model, the input matrix $\bfA$ takes real values $\mathbb R^{n\times d}$. Again, the rows $\bfa_1, \bfa_2, \dots, \bfa_n$ are presented in one pass over a stream, one at a time, in that order. However, in this model, for each $i\in[n]$, we must irrevocably choose whether to store $\bfa_i$ or not. That is, if we choose to store $\bfa_i$, then we may not discard it at a later time. For each stored row, we allow for the row $\bfa_i$ to be scaled by some weight $\bfw_i\in\mathbb R$. The goal is to minimize the number of rows of $\bfA$ that are stored. We assume that we may perform exact arithmetic and linear algebra on the stored rows.

\subsubsection{Online Coresets for \texorpdfstring{$\ell_\infty$}{l inf} Subspace Sketch}

We first discuss our results for the $\ell_\infty$ subspace sketch problem, in both the row arrival streaming and online coreset models, which is the basis for all of our algorithms for geometric problems.

\begin{Definition}[Streaming/Online $\ell_\infty$ Subspace Sketch]\label{def:streaming-subspace-sketch-l-inf}
The \emph{streaming $\ell_\infty$ subspace sketch problem} is defined as follows\footnote{Although one may define randomized versions of this problem \cite{LWW2021}, as we consider later, we restrict ourselves to deterministic algorithms in this section.}. We are given an $n\times d$ matrix $\bfA$ over one pass through a row arrival stream. Then:
\begin{itemize}
    \item In the \emph{row arrival streaming model}, $\bfA\in\mathbb Z^{n\times d}$ with entries bounded by $\poly(n)$, and we must maintain a data structure $Q:\mathbb R^d\to\mathbb R$ such that, at the end the stream, we have for some $\Delta\geq1$ that
    \[
        \mbox{for all $\bfx\in\mathbb R^d$, } \qquad \norm*{\bfA\bfx}_\infty \leq Q(\bfx) \leq \Delta\norm*{\bfA\bfx}_\infty
    \]
    \item In the \emph{online coreset model}, $\bfA\in\mathbb R^{n\times d}$ is a real matrix, and we must irrevocably choose a subset of entries $S\subseteq[n]$ and weights $\bfw \in \mathbb R^S$ as well as a function $Q:\mathbb R^d\to\mathbb R$ depending only on $\diag(\bfw)\bfA\vert_S$ such that, at the end of the stream, we have for some $\Delta\geq1$ that
    \[
        \mbox{for all $\bfx\in\mathbb R^d$, } \qquad \norm*{\bfA\bfx}_\infty \leq Q(\bfx) \leq \Delta\norm*{\bfA\bfx}_\infty
    \]
\end{itemize}
\end{Definition}

To motivate and discuss the streaming $\ell_\infty$ subspace sketch problem, we first illustrate some connections with computational geometry. It is not hard to see that L\"owner--John ellipsoids can be used to achieve roughly $\sqrt d$ distortion and $d^2$ words of space for $\ell_\infty$ subspace sketch in the offline setting, which is a nearly optimal trade-off. Thus, one may wonder whether there are algorithms for maintaining L\"owner--John ellipsoids in a row arrival stream. This is, however, a fundamental unresolved problem in streaming computational geometry that has been suggested several times \cite{MSS2010, AS2015}. In fact, we show the negative result that L\"owner--John ellipsoids require $\Omega(n)$ bits of space to maintain up to a distortion of less than $\Theta(\sqrt{d/\log n})$:

\begin{Theorem}[Informal Restatement of Theorem \ref{thm:ellipsoid-lb}]\label{thm:ellipsoid-lb-informal}
    Suppose an algorithm $\mathcal A$ maintains the L\"owner--John ellipsoid of the convex hull of $\bfa_1,\bfa_2,\dots,\bfa_n\in\mathbb R^d$, up to a factor of $\sqrt{d/\log n}$, in one pass over a row arrival stream, and $\mathcal A$ succeeds with probability at least $2/3$, over its coin tosses. Then, $\mathcal A$ uses $\Omega(n)$ bits space.
\end{Theorem}

This is perhaps surprising, given that for the syntactically similar MEB problem, constant factor approximations are possible using $\poly(d,\log n)$ bits of space \cite{CP2014, AS2015}. More relevantly to us, Theorem \ref{thm:ellipsoid-lb-informal} suggests that a drastically new approach is needed to tackle the streaming subspace sketch problem. Despite the above negative result, we nonetheless obtain a deterministic streaming algorithm for maintaining an $\ell_\infty$ subspace embedding, and in fact, a deterministic online coreset algorithm \cite{CMP2020,BDMMUWZ2020}:

\begin{Theorem}[Informal Restatement of Theorem \ref{thm:subspace-sketch-stream-l-inf}]
    Let $\bfA$ be an $n\times d$ matrix presented in one pass over a row arrival stream. There is an algorithm $\mathcal A$ which maintains a coreset $S\subseteq[n]$ such that
    \[
        \mbox{for all $\bfx\in\mathbb R^d$, }\qquad\norm*{\bfA\vert_S\bfx}_\infty \leq \norm*{\bfA\bfx}_\infty \leq \Delta\norm*{\bfA\vert_S\bfx}_\infty
    \]
    where
    \begin{itemize}
        \setlength\itemsep{-0.1em}
        \item in the streaming model, $\Delta = O(\sqrt{d\log n})$, $\abs*{S} = O(d\log n)$, and $\mathcal A$ uses $O(d^2\log^2 n)$ bits of space.
        \item in the online coreset model, $\Delta = O(\sqrt{d\log(n\kappa^\OL)})$ and $\abs*{S} = O(d\log(n\kappa^\OL))$.
    \end{itemize}
\end{Theorem}

As we show in Theorem \ref{thm:restricted-subspace-sketch-lb}, any data structure $Q$ which satisfies
\[
    \Pr\braces*{Q(\bfx) \leq \norm{\bfA\bfx}_\infty \leq \Delta \cdot Q(\bfx)} \geq \frac23
\]
for each $\bfx\in\mathbb R^d$ must either have $\Delta = \Omega(\sqrt{d/\log n})$ or use $\Omega(n)$ bits of space, even if $Q$ is constructed with unlimited computation and access to $\bfA$. Furthermore, as we show in Theorem \ref{thm:finite-distortion-subspace-sketch-for-all}, if
\[
    \Pr\braces*{\mbox{for all $\bfx\in\mathbb R^d$, } Q(\bfx) \leq \norm{\bfA\bfx}_\infty \leq \Delta \cdot Q(\bfx)} \geq \frac23
\]
for any $\Delta < \infty$, then $Q$ must use $\Omega(d^2)$ bits of space. Thus, our deterministic streaming algorithm achieves the best distortion and space that is possible for any randomized offline algorithm, up to $\poly\log n$ factors.

\begin{Remark}
Theorem \ref{thm:subspace-sketch-stream-l-inf} implies that $\ell_\infty$ regression can be solved up to a factor of $O(\sqrt{d\log n})$ using $O(d^2\log^2 n)$ bits of space in a row arrival stream. Note that this is not possible in other models of streaming, such as the \emph{turnstile streaming model}, in which the matrix $\bfA$ and target vector $\bfb$ are revealed by a stream of coordinate-wise additive updates $\bfA_{i,j}\gets\bfA_{i,j} + \Delta_{i,j}$ and $\bfb_i\gets\bfb_i + \Delta_i$ for $\Delta_{i,j},\Delta_i\in\mathbb Z$. Indeed, under this model, Theorem 8.1 of \cite{BJKS2004} shows that for vectors $\bfx,\bfy\in([m]\cup\{0\})^n$, deciding whether $\norm*{\bfx - \bfy}_\infty$ is $\leq 1$ or $\geq m$ requires at least $\Omega(n/m^2)$ bits of space. If we set $\bfA\gets [\bfx^\top\ R]^\top\in\mathbb Z^{(n+1)\times 1}$ and $\bfb\gets [\bfy^\top\ R]^\top\in\mathbb Z^{n+1}$ for some large $R$, then estimating the optimal $\ell_\infty$ regression cost implies estimating $\norm*{\bfx-\bfy}_\infty$.
\end{Remark}

Our algorithm is based on recent developments on numerical linear algebra in the online setting \cite{CMP2020,BDMMUWZ2020}. In the streaming model, the trade-off between the space complexity and distortion is nearly optimal, as we will discuss later in the introduction. In the online coreset model, the number of rows stored is nearly optimal by a lower bound that we show in Theorem \ref{thm:online-coreset-lb-l-inf}.

\subsubsection{Techniques for Online \texorpdfstring{$\ell_\infty$}{l inf} Subspace Sketch}
\paragraph{Strawman Solutions.}
We first discuss certain natural coreset approaches to the streaming $\ell_\infty$ subspace sketch problem and why they do not work, in order to illustrate the difficulty of the problem. We assume for simplicity for now that all input vectors have norm $\Theta(1)$.

Intuitively, we want a small number of input rows that are well spread apart, so that we have a small number of rows that approximate the entire data set $\bfA$ in all directions. One way to do this is to add a new row to our coreset if and only if it has a small inner product, say at most some threshold $\tau = 1/\poly(d)$, with each of the stored rows. Certainly, such a row must be included in the coreset, otherwise that row itself as a query would fail to achieve a $1/\tau$-approximation. This can  also be shown to yield a small coreset of size at most $\poly(d)$. However, such an algorithm could fail to store a row which is very well-aligned with an earlier row, but also has a tiny component pointing outside of the span of every other row, which means the coreset would fail to have any multiplicative error. One could try to fix this by adding the condition that we add a row if it increases the rank of the coreset; this also does not work, since there could be future rows which significantly increase the maximum component in this direction, but also have large inner product with the stored rows. 

Another approach, which attempts to address the problem of having rows which increase the maximum component in a given direction, is to maintain the maximum component for $\poly(d)$ random directions. That is, one can first choose a set of $\poly(d)$ random directions $S$, and for each $\bfv\in S$, store the input row which has the maximum inner product with $\bfv$. However, it can be shown that $\poly(d)$ directions is in fact not enough to ``catch'' hidden growing components. Indeed, suppose that that input rows consist of the standard basis vectors $\pm\bfe_1$. These vectors will be stored. Then, suppose that the algorithm receives the vector $\bfa \coloneqq (1-1/n)\bfe_1 + (1/n^{10}) \bfe_2$. In order for this vector to be stored by a random vector $\bfv$, we must have that
\[
    \angle*{\bfa,\bfv} = \angle*{(1-1/n)\bfe_1 + (1/n^{10}) \bfe_2, \bfv} > \abs*{\angle*{\bfe_1, \bfv}},
\]
or $\bfv_2 \geq n^9 \abs*{\bfv_1}$ by rearranging. The probability that this occurs for a random vector $\bfv$ is at most $O(1/n^9)$, and thus by a union bound over the $\poly(d)$ many random vectors, no direction stores $\bfa$. However, $\bfa$ has a component outside of the span of the previous rows, so even for vectors whose norms are within $1\pm 1/\poly(n)$ factors of each other, this algorithm fails. It is easy to see that even if we store rows that increase the rank of the coreset, it would still fail to store rows which increase the component along $\bfe_2$ by $\poly(n)$ factors. 

\paragraph{Our Approach.}
We now give a high-level proof of our online $\ell_\infty$ subspace sketch result. We seek a subset $S\subseteq[n]$ such that
\[
    \norm*{\bfA\bfx}_\infty \leq \Delta \norm*{\bfA\vert_S\bfx}_\infty
\]
for some $\Delta\geq 1$, so suppose we have maintained such an $S$, and let $\bfa\in\mathbb R^d$ be a new row to process. As hinted previously, we encounter a problem if there exists any direction $\bfx\in\mathbb R^d$ along which the new row updates the maximum component by more than a $\poly(d)$ factor. That is, if there exists $\bfx\in\mathbb R^d$ such that $\abs*{\angle*{\bfa,\bfx}} \gg \norm*{\bfA\vert_S\bfx}_\infty$, then we must include $\bfa$ in our new coreset. However, we are unable to analyze such an algorithm, due to the lack of structure of the $\ell_\infty$ norm. Now note that if $\abs*{S} = \poly(d, \log n)$, then $\norm*{\bfA\vert_S\bfx}_2 = \poly(d,\log n)\norm*{\bfA\vert_S\bfx}_\infty$, so using $\norm*{\bfA\vert_S\bfx}_2$ is just as good of a condition for adding $\bfa$. The advantage is that the $\ell_2$ norm has significantly more structure than the $\ell_\infty$ norm, which we can take advantage of in order to bound the size of the coreset. 

Suppose now that we add $\bfa$ to our coreset whenever there exists some $\bfx\in\mathbb R^d$ such that $\abs*{\angle*{\bfa,\bfx}}^2 \geq \norm*{\bfA\vert_S\bfx}_2^2$. In the language of numerical linear algebra, this corresponds to the condition that \emph{the leverage score of $\bfa$ with respect to $\bfA\vert_S$ is at least $1$}. Once we have made the connection to leverage scores, we are now in the position to bound the size of $S$. Note that in the final coreset $\bfA\vert_S$, we have by construction that every row $\bfa_i$ has leverage score at least $1$ with respect to the previous rows. This can be phrased as the fact that all of the \emph{online leverage scores} $\bftau_i^{\OL}$ (Definition \ref{def:online-leverage-scores}) of $\bfA\vert_S$ are at least $1$. Now, it can be shown that the $i$th online leverage score bounds the incremental difference between the log-volume spanned by columns of the first $i$ rows $\bfA_i$ of $\bfA$ and $\bfA_{i+1}$, which gives a bound of $O(d\log\kappa^\OL)$ on the sum of online leverage scores, where $\kappa^\OL = \norm*{\bfA}_2\max_{i=1}^n \norm*{\bfA^-}_2$ is the online pseudo condition number of $\bfA$ \cite{CMP2020, BDMMUWZ2020}. This means that $S$ must have at most $O(d\log\kappa^\OL)$ rows. In turn, this allows us to bound the distortion of the subspace sketch data structure as
\[
    \norm*{\bfA\bfx}_\infty = \max_{i=1}^n \abs*{\angle*{\bfa_i,\bfx}} \leq \norm*{\bfA\vert_S\bfx}_2 \leq \sqrt{\abs*{S}}\norm*{\bfA\vert_S\bfx}_\infty \leq O(\sqrt{d\log\kappa^\OL})\norm*{\bfA\vert_S\bfx}_\infty.
\]
Although the $\kappa^\OL$ here is for the submatrix $\bfA\vert_S$, it can be shown that this is only a $\poly(n)$ factor away from $\kappa^\OL$ of $\bfA$. While this discussion contains a number of ideas for our online coreset algorithm for the $\ell_\infty$ subspace sketch problem, we still need to improve our result from $O(\sqrt{d\log\kappa^\OL})$ to $O(\sqrt{d\log n})$ distortion for integer matrices with entries bounded by $\poly(n)$ for the row arrival streaming model. For this, we will improve the bound on the sum of online leverage scores for such matrices. We discuss this result in the next section.

\subsubsection{Techniques for Sharper Online Numerical Linear Algebra}

We now discuss our techniques for improving the sum of online leverage scores for integer matrices with entries bounded by $\poly(n)$. Na\"ively, the earlier condition number bound gives a bound of $O(d^2\log n)$ by using that for such matrices, $\kappa \leq \poly(n)^d$ (see, e.g., \cite[Lemma 4.1]{CW2009}). Note that $\kappa$ can indeed be as large as $\exp(\poly(d))$, even for sign matrices \cite{AV1997}. We improve this to the following:

\begin{restatable}[Sum of Online Leverage Scores]{Theorem}{SumOnlineLeverageScore}\label{thm:sharp-online-leverage-scores}
    Let $\bfA\in\mathbb Z^{n\times d}$ have entries bounded by $\poly(n)$. Then,
    \[
        \sum_{i=1}^n \bftau_i^{\mathsf{OL}}(\bfA) = O(d\log n).
    \]
\end{restatable}

We start with the proof of \cite{CMP2020}, which gives a bound of $O(d\log\kappa^\OL)$. This is done by analyzing the quantity $\det(\bfA^\top\bfA+\lambda\bfI_d)$, for $\lambda = (\max_{i=1}^n\norm*{\bfA_i^-}_2)^{-1}$. This quantity is at most $O(\norm*{\bfA}_2)^{d}$, and can be shown to be lower bounded by $\exp(\frac12\sum_i \bftau_i^{\mathsf{OL}}(\bfA))\cdot \det(\lambda\bfI_d)^d$ by the matrix determinant lemma, which gives
\[
    \det(\bfA_{i+1}^\top\bfA_{i+1}+\lambda\bfI_d) \geq \det(\bfA_{i}^\top\bfA_{i}+\lambda\bfI_d)\exp(\bftau_i^{\mathsf{OL}}(\bfA)/2)
\]
where $\bfA_j$ is the first $j$ rows of $\bfA$. Taking logarithms on both sides and rearranging yields that
\[
    \sum_{i=1}^n \bftau_i^{\mathsf{OL}}(\bfA) \leq O\parens*{d\log\frac{\norm*{\bfA}_2}{\lambda}} = O(d\log\kappa^\OL).
\]
Now, one may question whether regularizing by $\lambda$ is necessary, as it leads to the undesirable $\log \frac1\lambda$ factor. Indeed, we set $\lambda = 0$ and instead analyze the \emph{pseudodeterminant} $\pdet(\bfA^\top\bfA)$, which is the product of the nonzero eigenvalues. With this change, we have almost the same result, except that we must treat rows $i$ which do not lie in $\rowspan(\bfA_{i-1})$ differently. In this case, we have
\[
    \pdet(\bfA_i^\top\bfA_i) \geq \pdet(\bfA_{i-1}^\top\bfA_{i-1}) \norm*{\bfa_i^{\bot}}_2^2
\]
where $\bfa_i^{\bot}$ is the component of $\bfa_i$ orthogonal to $\rowspan(\bfA_{i-1})$. Now observe that the product of $\norm*{\bfa_i^{\bot}}_2^2$ for all rows $i$ which do not lie in $\rowspan(\bfA_{i-1})$ is exactly the volume spanned by these vectors, which is a positive integer, and thus $\geq 1$. We thus avoid the $\log\frac1\lambda$ factor and instead get the upper bound of $O(d\log n)$. 

As a result of Theorem \ref{thm:sharp-online-leverage-scores}, we immediately remove condition number dependencies from a variety of results in online numerical linear algebra which rely on Theorem \ref{thm:sharp-online-leverage-scores}, and answer an open question of \cite{BDMMUWZ2020} on removing the condition number dependence from the online spectral approximation problem, under bit complexity assumptions.

\begin{Theorem}[Online Coreset for Spectral Approximation]\label{thm:online-spectral-approx-bss}
    Let $\bfA\in\mathbb Z^{n\times d}$ have entries bounded by $\poly(n)$. There is a deterministic online coreset algorithm which outputs $\tilde\bfA$ such that
    \[
        (1-\eps)\bfA^\top\bfA \preceq \tilde\bfA^\top\tilde\bfA \preceq (1+\eps)\bfA^\top\bfA
    \]
    and the number of rows in $\tilde\bfA$ is $O(d(\log n)^2/\eps^2)$. 
\end{Theorem}
\begin{proof}
This is obtained simply by plugging in the improved guarantee of Theorem \ref{thm:sharp-online-leverage-scores} into the derandomized \textsc{OnlineBSS} algorithm described in Appendix B.2 of \cite{BDMMUWZ2020}.
\end{proof}

We also implement the simpler sampling algorithm with a similar randomized guarantee. The proof is not quite a direct application of Theorem \ref{thm:sharp-online-leverage-scores}, so we provide a full proof in Appendix \ref{sec:online-spectral-approx} for completeness.

\begin{restatable}[Online Coreset for Spectral Approximation via Leverage Score Sampling]{Theorem}{OnlineSpectralApproximation}\label{thm:online-spectral-approx}
    Let $\bfA\in\mathbb Z^{n\times d}$ have entries bounded by $\poly(n)$. There is an online coreset algorithm which outputs $\tilde\bfA$ such that
    \[
        \Pr\braces*{(1-\eps)\bfA^\top\bfA \preceq \tilde\bfA^\top\tilde\bfA \preceq (1+\eps)\bfA^\top\bfA} \geq \frac23
    \]
    and the number of rows in $\tilde\bfA$ is $O(d(\log d)(\log n)/\eps^2)$. 
\end{restatable}

\subsubsection{High-Dimensional Computational Geometry in Polynomial Space}\label{sec:high-dim-cg}

We now show that our $\ell_\infty$ subspace sketch algorithm gives the first polynomial space algorithms for many important problems in streaming computational geometry, including the basic problems of symmetric width, convex hull, and L\"owner--John ellipsoids. Previous algorithms for these problems had an exponential dependence on $d$, due to reliance on $\eps$-kernels \cite{AHV2004, AHV2005}. In particular, in the high-dimensional regime of $d \geq C\log n$ for a large enough constant $C$, the memory bound for known results becomes larger than $\tilde\Theta(nd)$, and thus \emph{there were no previously known nontrivial algorithms in this regime}, despite the fact that algorithms that work in the high-dimensional regime have been sought after for over a decade since they were suggested by \cite{AHV2004, AHV2005, Cha2006, ZC2006} and others.

In the following discussion, we generally assume a centrally symmetric input instance, that is, if $\bfa\in\mathbb R^d$ is a point in the input point set, then so is $-\bfa$. Note that for most geometric problems falling under the class of \emph{extent measure} problems \cite{AHV2004, AS2015}, considering only centrally symmetric instances is without loss of generality, up to constant factor losses in the distortion. For illustration, consider the directional width problem, in which we wish to estimate $\max_{i=1}^n \angle*{\bfa_i,\bfx} - \min_{j=1}^n \angle*{\bfa_j,\bfx}$ for any query direction $\bfx\in\mathbb R^d$. One can translate the entire point set by one of the input points, say $\bfa_1$, so that $0\in\mathbb R^d$ is one of the elements of the point set. This preserves the directional width. Note then that $\max_{i=1}^n \angle*{\bfa_i-\bfa_1,\bfx} \geq \angle*{0,\bfx} = 0$ and $\min_{j=1}^n \angle*{\bfa_j-\bfa_1,\bfx} \leq \angle*{0,\bfx} = 0$, so
\begin{align*}
    \max_{i=1}^n \angle*{\bfa_i-\bfa_1,\bfx} - \min_{j=1}^n \angle*{\bfa_j-\bfa_1,\bfx} &= \abs*{\max_{i=1}^n \angle*{\bfa_i-\bfa_1,\bfx}} + \abs*{\min_{j=1}^n \angle*{\bfa_j-\bfa_1,\bfx}} = \max_{i=1}^n \abs*{\angle*{\bfa_i-\bfa_1,\bfx}}.
\end{align*}
Then for each translated point $\bfa_i - \bfa_1$, we add its negation $-(\bfa_i - \bfa_1)$, which preserves the latter value. Similar arguments apply to other problems, such as convex hull, L\"owner--John ellipsoids, etc.

Because our $\ell_\infty$ subspace sketch algorithm is online, many of our algorithms for streaming geometry are online as well, and we present results in both the row arrival streaming and online coreset models. Full proofs are included in Section \ref{sec:streaming-comp-geo}.

\paragraph{$k$-Robust Directional Width.}

Perhaps the most straightforward of our applications is directional width \cite{AHV2004, AHV2005} for centrally symmetric instances. Indeed, this is exactly equivalent to the $\ell_\infty$ subspace sketch problem. Using the ``peeling'' technique of \cite{AHY2008}, we also obtain algorithms for \emph{$k$-robust directional width} $\mathcal E_k(\bfx, \bfA)$, which measures the width of a point set in a given direction $\bfx\in\mathbb R^d$, after removing the $k$ most extreme points in that direction (see Definition \ref{def:k-robust-width} for a formal definition).

\begin{restatable}[$k$-Robust Directional Width in Polynomial Space]{Theorem}{kRobustWidth}\label{thm:k-robust-width}
    Let $\bfA$ be an $n\times d$ matrix presented in one pass over a row arrival stream. There is an algorithm $\mathcal A$ which maintains a coreset $S\subseteq[n]$ such that
    \[
        \frac1\Delta\mathcal E_k(\bfx,\bfA) \leq \mathcal E_k(\bfx,\bfA\vert_S) \leq \mathcal E_k(\bfx,\bfA)
    \]
    where
    \begin{itemize}
        \setlength\itemsep{-0.1em}
        \item in the streaming model, $\Delta = O(\sqrt{d\log n})$, $\abs*{S} = O(kd\log n)$, and $\mathcal A$ uses $O(kd^2\log^2 n)$ bits of space.
        \item in the online coreset model, $\Delta = O(\sqrt{d\log(n\kappa^\OL)})$ and $\abs*{S} = O(kd\log(n\kappa^\OL))$.
    \end{itemize}
\end{restatable}

\paragraph{Convex Hull.}

A fundamental problem in computational geometry is the approximation of the convex hull of $n$ points $\bfa_1,\bfa_2,\dots,\bfa_n\in\mathbb R^d$. For $(1+\eps)$-approximation, $\eps$-kernels \cite{AHV2004, AHV2005} give coresets of near-optimal size of $\eps^{-\Theta(d)}$, even in the streaming model \cite{Cha2006,Cha2016}. More recently, \cite{BBKLY2018} removed the exponential dependence on $d$ for certain beyond-worst-case instances. However, a general streaming algorithm for convex hull in $\poly(d,\log n)$ bits of space, even with $\poly(d,\log n)$ distortion, remained elusive. In the offline setting, this is possible via coresets for L\"owner--John ellipsoids (see Section 3.6 of \cite{Tod2016}).

By using our coreset for $\ell_\infty$ subspace sketch, we obtain coresets for approximating symmetric convex hulls, with $\poly(d,\log n)$ bits of space and distortion. This is done by noticing that our $\ell_\infty$ subspace sketch result yields an online coreset for approximating a polytope defined by the intersection of the linear inequalities specified by each of the rows, and then using the fact that this linear inequality polytope is the \emph{polar body} of the symmetric convex hulls of the corresponding rows \cite{HW2020}.

\begin{restatable}[Streaming Convex Hulls in Polynomial Space]{Theorem}{ConvexHullStream}\label{thm:convex-hull-stream}
    Let $\bfA$ be an $n\times d$ matrix presented in one pass over a row arrival stream. There is an algorithm $\mathcal A$ which maintains a coreset $S\subseteq[n]$ such that
    \[
        \conv(\{\pm\bfa_i\}_{i\in S}) \subseteq \conv(\{\pm\bfa_i\}_{i=1}^n) \subseteq \Delta\conv(\{\pm\bfa_i\}_{i\in S}).
    \]
    where
    \begin{itemize}
        \setlength\itemsep{-0.1em}
        \item in the streaming model, $\Delta = O(\sqrt{d\log n})$, $\abs*{S} = O(d\log n)$, and $\mathcal A$ uses $O(d^2\log^2 n)$ bits of space.
        \item in the online coreset model, $\Delta = O(\sqrt{d\log(n\kappa^\OL)})$ and $\abs*{S} = O(d\log(n\kappa^\OL))$.
    \end{itemize}
\end{restatable}

Note that this also gives us a $O(\sqrt{d\log n})^d$-factor approximation to the volume of convex hull.

\paragraph{L\"owner--John Ellipsoids.}

As previously discussed, streaming L\"owner--John ellipsoids in the high-dimensional setting has been open \cite{MSS2010, AS2015}: \cite{MSS2010} proposed a simple algorithm of iteratively adding points to a L\"owner--John ellipsoid which does not yield $\poly(d,\log n)$ distortion, while \cite{AS2015} gave an $O(1)$-approximation for MEB in $\poly(d)$ space, and asked whether their ideas applied to L\"owner--John ellipsoids.

We first note that our streaming $\ell_\infty$ subspace sketch result immediately gives a result for L\"owner--John ellipsoids for linear inequality polytopes (see Remark \ref{rem:ellipsoid} following Theorem \ref{thm:subspace-sketch-stream-l-inf}).

\begin{restatable}[L\"owner--John Ellipsoids in Polynomial Space]{Theorem}{EllipsoidStream}\label{thm:john-ellipsoid-streaming-ub}
    Let $\bfA$ be an $n\times d$ matrix presented in one pass over a row arrival stream. Define the polytope $K = \braces*{\bfx\in\mathbb R^d : \norm*{\bfA\bfx}_\infty \leq 1}$. There is an algorithm $\mathcal A$ which maintains a coreset $S\subseteq[n]$ from which we can compute an ellipsoid $E'$ such that
    \[
        E' \subseteq K \subseteq \Delta E'
    \]
    where
    \begin{itemize}
        \setlength\itemsep{-0.1em}
        \item in the streaming model, $\Delta = O(\sqrt{d\log n})$, $\abs*{S} = O(d\log n)$, and $\mathcal A$ uses $O(d^2\log^2 n)$ bits of space.
        \item in the online coreset model, $\Delta = O(\sqrt{d\log(n\kappa^\OL)})$ and $\abs*{S} = O(d\log(n\kappa^\OL))$.
    \end{itemize}
    Since $K\subseteq E\subseteq \sqrt d K$, $E'$ is an $O(\Delta\sqrt d)$-approximate L\"owner--John ellipsoid.
\end{restatable}

We then show in Corollary \ref{cor:ellipsoids-convex-hull} that the earlier polar body trick yields L\"owner--John ellipsoids for symmetric convex hulls as well. Note that for streaming, a $\sqrt d$ distortion is necessary by Theorem \ref{thm:ellipsoid-lb-informal}.

\paragraph{Volume Maximization.}

We next consider the problem of selecting $k$ rows that approximately maximizes the volume of the parallelepiped spanned by the rows, known as \emph{volume maximization}, or maximum a posteriori (MAP) inference of determinantal point processes (DPPs) \cite{BKLZ2020}. Relative error guarantees for this problem have been studied by \cite{IMGR2019, IMGR2020,MRWZ2020}, culminating in the following:

\begin{Theorem}[Streaming Volume Maximization, Theorem 1.9 of \cite{MRWZ2020}]
    Let $\bfA\in\mathbb Z^{n\times d}$ have entries bounded by $\poly(n)$ and $k\geq 1$. Let $C\in [1,(\log n)/k]$. There is a one-pass streaming algorithm that computes a subset $S\subseteq[n]$ of $k$ points such that
    \[
        \Pr\braces*{O(Ck)^{k/2}\Vol(\bfA\vert_S) \geq \Vol(\bfA\vert_{S_*})} \geq \frac23
    \]
    where $\Vol(\bfA\vert_S)$ is the volume of the parallelepiped spanned by the rows $\bfA\vert_S$ indexed by $S$ and $\bfA\vert_{S_*}$ is a set of $k$ rows that maximizes the volume. The algorithm uses $O(n^{O(1/C)}d)$ bits of space.
\end{Theorem}

This result is obtained by combining coresets for volume maximization \cite{IMGR2019} with streaming $\eps$-kernels for directional width \cite{Cha2006}. Note that even when $C = (\log n)/k$, the space complexity is $\exp(O(k))d$ and thus still exponential in $k$. By replacing $\eps$-kernels for directional width with our $\ell_\infty$ subspace sketch result, we obtain the first relative error polynomial space algorithms for volume maximization\footnote{The algorithm of \cite{BKLZ2020} has polynomial space as well, but has an additive error guarantee}.

\begin{restatable}[Polynomial Space Streaming Volume Maximization]{Theorem}{VolMaxStream}
    Let $\bfA\in\mathbb Z^{n\times d}$ with entries bounded by $\poly(n)$ and $k\geq 1$. Let $1 < C < (\log n)/k$ and $r = (\log n)/C$. There is a one-pass streaming algorithm that computes a subset $S\subseteq[n]$ of $k$ points such that
    \[
        \Pr\braces*{O(r^2 Ck\log^2 n)^{k/2}\Vol(\bfA\vert_S) \geq \Vol(\bfA\vert_{S_*})} \geq \frac23
    \]
    where $\Vol(\bfA\vert_S)$ is the volume of the parallelepiped spanned by the rows $\bfA\vert_S$ indexed by $S$ and $\bfA\vert_{S_*}$ is a set of $k$ rows that maximizes the volume. The algorithm uses $O(rd\log^2 n)$ bits of space.
    
    If only the indices (rather than the $d$-dimensional rows) are required, there is an algorithm using $O(k^2\log^3 n)$ bits of space with $O(k\log n)^k$ distortion.
\end{restatable}

See Section \ref{sec:vol-max} for details. With $C = \Theta((\log n)/k)$, we obtain an $O(k\log^{3/2} n)^{k}$-approximation using $O(kd\log^2 n)$ bits of space. Note that we do not get an online coreset for this algorithm, since the $k$ points are selected as a subset of a larger online coreset from our $\ell_\infty$ coreset algorithm.

\paragraph{Minimum-Width Spherical Shell.}

Our next application is the problem of approximating the \emph{spherical shell} of minimum width which encloses a set of points. Formally, a spherical shell centered at $\bfc\in\mathbb R^d$ with inner radius $r$ and outer radius $R$ is $\sigma(\bfc,r,R)\coloneqq \braces*{\bfx\in\mathbb R^d : r \leq \norm*{\bfx-\bfc}_2 \leq R}$, and we seek relative error approximations to $R - r$. This problem has received much attention in the computational geometry literature \cite{AS1998, AAHS1999, Cha2002, Cha2006}. We show the following result in Section \ref{sec:spherical-shell}:
\begin{restatable}[Minimum Width Spherical Shell in Polynomial Space]{Theorem}{SphericalShell}\label{thm:spherical-shell}
    Let $\bfA$ be an $n\times d$ matrix presented in one pass over a row arrival stream.  There is an algorithm $\mathcal A$ which maintains a coreset $S\subseteq[n]$ from which we can compute find a center $\hat\bfc$, inner radius $\hat r$ and outer radius $\hat R$ such that $\sigma(\hat\bfc,\hat r,\hat R)\supseteq \braces*{\bfa_i}_{i=1}^n$ and
    \[
        \hat R - \hat r \leq \Delta^{3/2}\min_{\sigma(\bfc,r,R)\supseteq \braces*{\bfa_i}_{i=1}^n} R - r
    \]
    where
    \begin{itemize}
        \setlength\itemsep{-0.1em}
        \item in the streaming model, $\Delta = O(\sqrt{d\log n})$, $\abs*{S} = O(d\log n)$, and $\mathcal A$ uses $O(d^2\log^2 n)$ bits of space.
        \item in the online coreset model, $\Delta = O(\sqrt{d\log(n\kappa^\OL)})$ and $\abs*{S} = O(d\log(n\kappa^\OL))$.
    \end{itemize}
\end{restatable}
For this result, our earlier reduction to centrally symmetric instances requires more care, and is worked out in our proof.

\paragraph{Linear Programming.} Finally, we consider linear programming for instances with a centrally symmetric constraint polytope $\braces*{\bfx\in\mathbb R^d:\norm*{\bfA\bfx}_\infty\leq 1}$. More formally, we seek to approximate the optimal value of the following optimization problem
\begin{equation*}\label{eq:lp-obj}
\begin{aligned}
    &\mbox{maximize } &&\angle*{\bfc,\bfx} \\
    &\mbox{subject to } &&\bfx\in\mathbb R^d, \norm*{\bfA\bfx}_\infty \leq 1
\end{aligned}
\end{equation*}
where the rows of $\bfA$ arrive in a row arrival stream. The problem of solving classes of linear programs in the streaming model has been studied intensely over the years \cite{CC2007, AG2013, IMRUVY2017, AKZ2019, AKL2021}, as linear programs can solve many important problems in machine learning and big data analysis including $\ell_1$ and $\ell_\infty$ regression, support vector machines, and maximum matching, and set cover. However, solving linear programs in a stream is difficult in a small number of passes, and as far as we are aware, known results all require a large number of passes or at least exponential in $d$ space. By maintaining a multiplicative approximation to the constraint set via our Theorem \ref{thm:subspace-sketch-stream-l-inf}, we obtain the following:
\begin{Theorem}[Streaming Linear Programming]
Let $\bfA$ be an $n\times d$ matrix presented in one pass over a row arrival stream. Define the polytope $K = \braces*{\bfx\in\mathbb R^d : \norm*{\bfA\bfx}_\infty \leq 1}$. There is an algorithm $\mathcal A$ which maintains a coreset $S\subseteq[n]$ such that for any $\bfc\in\mathbb R^d$, one can compute from $\bfA\vert_S$ a vector $\hat\bfx\in K$ such that
    \[
        \max_{\bfx\in K}\angle*{\bfc,\bfx} \leq \Delta \cdot \angle*{\bfc,\hat\bfx}
    \]
    where
    \begin{itemize}
        \setlength\itemsep{-0.1em}
        \item in the streaming model, $\Delta = O(\sqrt{d\log n})$, $\abs*{S} = O(d\log n)$, and $\mathcal A$ uses $O(d^2\log^2 n)$ bits of space.
        \item in the online coreset model, $\Delta = O(\sqrt{d\log(n\kappa^\OL)})$ and $\abs*{S} = O(d\log(n\kappa^\OL))$.
    \end{itemize}
\end{Theorem}

\subsection{Streaming and Online \texorpdfstring{$\ell_p$}{lp} Subspace Sketch}


\subsubsection{The Subspace Sketch Problem}

We now consider the $\ell_p$ subspace sketch problem, which is defined analogously to $\ell_\infty$ in Definition \ref{def:streaming-subspace-sketch-l-inf}. This problem in the offline setting, as well as its randomized variants, was introduced by \cite{LWW2021}. When defining the randomized version of this guarantee, \cite{LWW2021} define two versions, known as the ``for each'' guarantee and the ``for all'' guarantee. For our streaming algorithms, we focus on the stronger ``for all'' guarantee.
\begin{Definition}
Let $\bfA\in\mathbb R^{n\times d}$ and $\Delta\geq1$. Then:
\begin{itemize}
    \item \textbf{For each guarantee}: $Q_p$ satisfies the \emph{``for each'' guarantee} if for each $\bfx\in\mathbb R^d$,
    \[
        \Pr\braces*{\norm*{\bfA\bfx}_p \leq Q_p(\bfx) \leq \Delta \norm*{\bfA\bfx}_p} \geq \frac23
    \]
    \item\textbf{For all guarantee}: $Q_p$ satisfies the \emph{``for all'' guarantee} if
    \[
        \Pr\braces*{\forall \bfx\in\mathbb R^d, \norm*{\bfA\bfx}_p \leq Q_p(\bfx) \leq \Delta \norm*{\bfA\bfx}_p} \geq \frac23,
    \]
\end{itemize}
\end{Definition}

\subsubsection{Prior Work on Streaming Subspace Sketch}

The subspace sketch problem is a vast generalization of the more well-known \emph{subspace embedding} problem, in which $Q_p$ specifically takes the form $\norm*{\bfS\bfA\bfx}$, for some norm $\norm*{\cdot}$ and a linear map $\bfS\in\mathbb R^{s\times n}$. 
Many, but not all, of our upper bounds on the subspace sketch problem will actually be subspace embeddings.

In the regime of $\Delta = (1+\eps)$ for $\eps\to0$, near-optimal streaming algorithms can be obtained quite straightforwardly by leveraging $\ell_p$ subspace embeddings algorithms due to \cite{CP2015}. These subspace embedding results achieve near-optimal space complexity by sampling methods. One can then use the \emph{merge-and-reduce} framework, in which one repeatedly finds subsets of rows that provide a $(1+\eps)$ approximation for blocks of rows, and then combines them in a binary tree fashion (see \cite{CDW2018, BDMMUWZ2020}), to get streaming subspace embedding algorithms of approximately the same quality. Since the approximation is composed with a depth of $\log n$, our distortion is $(1+\eps)^{\log n}$; by replacing $\eps$ by $\frac\eps{\log n}$, we recover the same trade-off as the offline subspace sketch problem, up to $\poly\log n$ factors. The space complexity is roughly $d^{2\lor(p/2+1)}$ words of space. However, the exponential dependence on $p$ makes this intractable when $p$ is large.

The previous work of \cite{CDW2018} studied the problem of maintaining a subspace sketch data structure \emph{deterministically} using a similar merge-and-reduce strategy, but their results unfortunately incur an  $n^{\Omega(1)}$ factor either in the distortion or the space complexity. Similar composable coreset approaches have been explored by other works, e.g., \cite{IMGR2019}.

\subsubsection{Streaming Algorithms for \texorpdfstring{$\ell_p$}{lp} Subspace Sketch}

We now discuss our results for the $\ell_p$ streaming subspace sketch problem. We first develop the following deterministic streaming algorithm, which runs in polynomial time and achieves an optimal trade-off between distortion and space. In particular, this result greatly improves upon the trade-offs achieved by \cite{CDW2018}. 

\begin{Theorem}[Informal restatement of Theorem \ref{thm:subspace-sketch-stream-l2}]
    Let $\bfA\in\mathbb Z^{n\times d}$ have entries bounded by $\poly(n)$. Let $2 < p < \infty$. There is a one-pass streaming algorithm maintaining a data structure $Q$ using $O(d^2\log n)$ bits of space such that
    \[
        \mbox{for all $\bfx\in\mathbb R^d$, }\qquad\norm*{\bfA\bfx}_p \leq Q(\bfx) \leq O((d\log n)^{\frac12-\frac1p})\norm*{\bfA\bfx}_p.
    \]
\end{Theorem}

Our result proceeds by defining an online set of weights that behave similarly to Lewis weights.

By tolerating randomization and exponential time, we also obtain a full set of near-optimal trade-offs:

\begin{Theorem}[Informal restatement of Theorem \ref{thm:subspace-sketch-stream-lq}]
    Let $\bfA\in\mathbb Z^{n\times d}$ have entries bounded by $\poly(n)$. Let $2 < q < p < \infty$. There is a one-pass streaming algorithm which maintains a data structure $Q$ using $O(d^{q/2+1}\log n)$ bits of space such that
    \[
        \Pr\braces*{\forall\bfx\in\mathbb R^d,\norm*{\bfA\bfx}_p \leq Q(\bfx) \leq O(d^{\frac12\parens*{1-\frac{q}{p}}}\log n)\norm*{\bfA\bfx}_p} \geq \frac23.
    \]
\end{Theorem}

Both Theorems \ref{thm:subspace-sketch-stream-l2} and \ref{thm:subspace-sketch-stream-lq} proceed by constructing weights $\bfw_i$ for $i\in[n]$ on the fly and then approximating $\norm*{\bfW\bfA\bfx}_q$ using an $O(1)$-approximate subspace sketch data structure for $q = 2$ for Theorem \ref{thm:subspace-sketch-stream-l2} or $2 < q < p$ for Theorem \ref{thm:subspace-sketch-stream-lq}. Because the $O(1)$-approximate subspace sketch results we use are mergeable, Theorems \ref{thm:subspace-sketch-stream-l2} and \ref{thm:subspace-sketch-stream-lq} are mergeable as well, allowing for use in distributed settings. Furthermore, for $q = 2$, our result can be combined with $\ell_2$ online coresets to yield online coresets for $\ell_p$:

\begin{Corollary}\label{cor:subspace-sketch-stream-l2-online}
    Let $2 < p < \infty$. Let $\bfA$ be an $n\times d$ matrix presented in one pass over a row arrival stream. There is an algorithm $\mathcal A$ which maintains a coreset $S\subseteq[n]$ and weights $\bfw\in\mathbb R^S$ such that
    \[
        \mbox{for all $\bfx\in\mathbb R^d$, }\qquad\norm*{\bfA\bfx}_p \leq \norm*{\diag(\bfw)\bfA\vert_S\bfx}_2 \leq \Delta\norm*{\bfA\bfx}_p
    \]
    where
    \begin{itemize}
        \setlength\itemsep{-0.1em}
        \item in the streaming model, $\Delta = O((d\log n)^{\frac12-\frac1p})$, $\abs*{S} = O(d\log n)$, and $\mathcal A$ uses $O(d^2\log^2 n)$ bits of space.
        \item in the online coreset model, $\Delta = O((d\log\kappa^\OL)^{\frac12-\frac1p})$ and $\abs*{S} = O(d\log\kappa^\OL)$.
    \end{itemize}
\end{Corollary}

As a corollary, we immediately obtain streaming algorithms for solving $\ell_p$ regression.

\begin{Corollary}[Online Coresets for $\ell_p$ Regression]\label{cor:streaming-lp-regression}
    Let $2 < p \leq \infty$. Let $\bfA$ be an $n\times d$ matrix and let $\bfb$ be a vector and suppose that the $n\times(d+1)$ matrix $\bfA' \coloneqq [\bfA\ \bfb]$ is presented in a row arrival stream. There is an algorithm $\mathcal A$ which maintains a coreset $S\subseteq[n]$ and weights $\bfw\in\mathbb R^S$ from which we can compute $\hat\bfx\in\mathbb R^d$ such that $\norm*{\bfA\hat\bfx-\bfb}_p\leq \Delta\min_\bfx\norm*{\bfA\bfx-\bfb}_p$, where
    \begin{itemize}
        \setlength\itemsep{-0.1em}
        \item in the streaming model, $\Delta = O((d\log n)^{\frac12-\frac1p})$, $\abs*{S} = O(d\log n)$, and $\mathcal A$ uses $O(d^2\log^2 n)$ bits of space.
        \item in the online coreset model, $\Delta = O((d\log\kappa^\OL)^{\frac12-\frac1p})$ and $\abs*{S} = O(d\log\kappa^\OL)$.
    \end{itemize}
\end{Corollary}

Our results are summarized in Table \ref{tab:subspace-sketch-stream}.

\subsection{Change of Density}

We now turn to the offline $\ell_p$ subspace sketch problem. We first investigate the problem of approximating $\ell_p$ norms of vectors in a $d$-dimensional subspace of $\ell_p^n$ by a reweighted $\ell_q$ norm of the same vector, formalized by the following definition from the functional analysis literature:

\begin{Definition}[Change of Density \cite{LT1980}]\label{def:lp-lq-change-of-density}
Let $0 < p, q \leq \infty$ and let $d\in\mathbb N$. Then, $c(d, p, q)$ denotes the smallest $c > 0$ such that for any $\bfA\in\mathbb R^{n\times d}$, there exists a nonnegative $\bfw\in\mathbb R^n$ such that, for $\bfW = \diag(\bfw)$,
\[
    \mbox{for all $\bfx\in\mathbb R^d$, }\qquad\norm*{\bfA\bfx}_p \leq \norm*{\bfW^{\frac1q-\frac1p}\bfA\bfx}_q \leq c\norm*{\bfA\bfx}_p.
\] 
\end{Definition}

Here, we think of $\bfw$ as weights (or a \emph{measure}) on the rows of $\bfA$ when evaluating $\ell_q$ norms, i.e., $\norm*{\bfy}_{q,\bfw} = \parens*{\sum_{i=1}^n \bfw_i \cdot \abs*{\bfy_i}^q}^{1/q}$. Note then that $\norm*{\bfW^{-1/p}\bfA\bfx}_{p,\bfw} = \norm*{\bfA\bfx}_p$ so the map $\bfA\bfx\mapsto\bfW^{-1/p}\bfA\bfx$ equipped with the appropriate norm is an \emph{isometry}. On the other hand, the weighted $\ell_q$ norm is $\norm*{\bfW^{-1/p}\bfA\bfx}_{q,\bfw} = \norm*{\bfW^{1/q-1/p}\bfA\bfx}_q$, which is comparable to $\norm*{\bfA\bfx}_p$ if $\bfw$ satisfies the guarantee of Definition \ref{def:lp-lq-change-of-density}. See Section 1.2 in \cite{JS2001} for an exposition on the role of change of densities in the theory of subspaces of $L_p$.

\paragraph{Lewis weights.}
The following seminal result is known about the parameter $c(d,p,q)$ for $q = 2$:

\begin{Theorem}[\cite{Lew1978, Joh1948, SZ2001}]\label{thm:lewis-weights}
Let $d\in\mathbb N$. For all $0 < p \leq \infty$,
\[
    c(d, p, 2) = c(d, 2, p) = d^{\abs*{1/2 - 1/p}}.
\]
\end{Theorem}

Theorem \ref{thm:lewis-weights} is due to Lewis \cite{Lew1978} in the regime of $1 \leq p < \infty$, and the weights $\bfw$ achieving the guarantee of Definition \ref{def:lp-lq-change-of-density} are known as \emph{Lewis weights}, in honor of \cite{Lew1978}. For the remaining parameter regimes, the case of $p = \infty$ follows from L\"owner--John ellipsoids \cite{Joh1948}, while the case of $0 < p < 1$ was proven in \cite{SZ2001}. Although the original proof by Lewis in \cite{Lew1978} uses involved theorems from Banach space theory, particularly the theory of $p$-summing operators \cite{DJT1995}, the proofs of \cite{SZ2001,CP2015} notably provide elementary proofs based only on analyzing the Lagrange multipliers for a convex program.

The use of Lewis weights was introduced to the theoretical computer science community by \cite{CP2015}, whose work made Lewis weights algorithmic by giving input sparsity time algorithms for approximating Lewis weights, and used them to obtain fast algorithms for solving $\ell_p$ regression. Subsequently, Lewis weights have become widely used in algorithms research, playing important roles in recent developments in optimization \cite{DLS2018, LS2019, JLS2021}, convex geometry \cite{LLV2020}, randomized numerical linear algebra \cite{CP2015, CWW2019, LWYZ2020, CCDS2020, MW2021}, and machine learning \cite{MRM2021, CD2021, PPP2021, MMWY2021}. Algorithms for computing Lewis weights themselves have also been refined over the years, both for $0<p<\infty$ \cite{Lee2016, LS2019, FLPS2021} as well as $p=\infty$, corresponding to L\"owner--John ellipsoids \cite{Tod2016, CCLY2019}. 

\paragraph{Change of density to $\ell_q$, $q\neq 2$.} The following result is lesser-known to the theoretical computer science community, and gives an optimal bound on $c(d,p,q)$ for $q\neq 2$.

\begin{Theorem}[Theorem 1.2 of \cite{LT1980}]\label{thm:lp-lq-change-of-density}
Let $d\in\mathbb N$ and let $1\leq p, q < \infty$. The following holds:
\begin{itemize}
    \item If $\min(p,q) \leq 2$, then $c(d,p,q) \leq d^{\abs*{\frac1q-\frac1p}}$.
    \item If $\min(p,q) \geq 2$, then $c(d,p,q) \leq d^{\frac12\parens*{1-\frac{p\land q}{p\lor q}}}$.
\end{itemize}
\end{Theorem}

As \cite{LT1980} show, the quantity $c(d,p,q)$ is intimately related to various other quantities, including $p$-summing norms and $p$-integral norms of operators, and is of independent interest in the functional analysis literature. For instance, an important corollary of this result is the best known upper bound on the \emph{Banach--Mazur distance} \cite{Tom1989} between a subspace of $\ell_p^n$ and any subspace of $\ell_q^n$, which formalizes the notion of distance between $\ell_p$ and $\ell_q$ for subspaces. As the authors note in Corollary 1.9 \cite{LT1980}, this result is optimal for $1\leq p < q < 2$. In fact, we will show that the proof of a result of \cite{LWW2021} implies that this is optimal in the regime of $\min(p,q) \geq 2$ as well, when $n$ is large enough. Thus, Theorem \ref{thm:lp-lq-change-of-density} obtains a tight characterization of the distance between subspaces of $\ell_p$ and $\ell_q$, in the sense of Banach--Mazur distance.


For $\min(p,q) \leq 2$, Theorem \ref{thm:lp-lq-change-of-density} follows from properties of Lewis weights, enjoying simple proofs and fast algorithms due to our refined understanding of Lewis weights. However, for $\min(p,q)\geq 2$, the proof is much more complicated. The authors first relate the problem of bounding $c(d,p,q)$ to bounding the smallest constant $\alpha>0$ such that $\pi_q(u) \leq \alpha\pi_p(u)$ for all linear maps $u$ (Definition 1.3 of \cite{LT1980})\footnote{In fact, they show that these two parameters are \emph{equal}.}, where $\pi_p(u)$ is the \emph{$p$-summing norm} of $u$ \cite{DJT1995}. To prove that $\alpha$ bounds $c(d,p,q)$, the authors invoke a factorization theorem of Maurey \cite{Mau1974}\footnote{See also Proposition 10 in Chapter III.H of \cite{Woj1991} for a proof and exposition in English of a similar theorem from \cite{Mau1974}, which gives the ``transposed'' result.}, which replaces Lewis's theorem and gives weights $\bfw$ for the change of density. Finally, the bound on $\alpha$ follows from a result of \cite{Car1980}, which uses results from the theory of operator ideals \cite{Pie1980}.

Our main result of this section is an elementary proof of Theorem \ref{thm:lp-lq-change-of-density} using only Lewis weights. Due to the simplicity of our proof, we obtain a robust version of the theorem, which still yields bounds when we replace Lewis weights with coarse overestimates of Lewis weights (see Definition \ref{def:approximate-lewis-overestimate}). More specifically, we show the following version of Theorem \ref{thm:lp-lq-change-of-density}, which is more refined than \cite{LT1980} in the sense that:
\begin{enumerate}
    \item The change of density is specifically the $\ell_p$ Lewis weights, rather than a tailor-made construction.
    \item The error guarantees degrade gracefully when the change of density is replaced by an approximation.
\end{enumerate}

\begin{restatable}[Change of density via approximate Lewis weights]{Theorem}{ChangeOfDensity}
\label{thm:lp-lq-change-of-density-lewis}
Let $\bfA\in\mathbb R^{n\times d}$ and $0<p,q<\infty$. Let $\bfw\in\mathbb R^n$ be $\alpha$-approximate $\ell_p$ Lewis weight overestimates (Definition \ref{def:approximate-lewis-overestimate}) and $\bfW = \diag(\bfw)$. For $p\geq q$, let
\begin{align*}
    \kappa_{d,p,q} &\coloneqq (\alpha d)^{1/q - 1/p} \\
    \lambda_{d,p,q} &\coloneqq (\alpha d)^{[0\lor(1/2-1/q)](p-q)/p}
\end{align*}
and let $\kappa_{d,p,q}\coloneqq \kappa_{d,q,p}^{-1}$, $\lambda_{d,p,q}\coloneqq \lambda_{d,q,p}^{-1}$ if $q\leq p$. Then for all $\bfx\in\mathbb R^d$ we have the following:
\begin{align*}
    \norm*{\bfA\bfx}_p &\leq \norm*{\lambda_{d,p,q}\cdot \bfW^{1/q-1/p}\bfA\bfx}_q \leq \kappa_{d,p,q}\lambda_{d,p,q}\norm*{\bfA\bfx}_p \qquad \mbox{ if $p\geq q$} \\
    \norm*{\bfA\bfx}_p &\leq \norm*{\kappa_{d,p,q}\cdot \bfW^{1/q-1/p}\bfA\bfx}_q \leq \kappa_{d,p,q}\lambda_{d,p,q}\norm*{\bfA\bfx}_p \qquad \mbox{ if $q\geq p$}
\end{align*}
Note that
\[
    \kappa_{d,p,q}\lambda_{d,p,q} = \begin{cases}
        (\alpha d)^{\abs*{\frac1q-\frac1p}} & \text{if $\min(p,q)\leq 2$} \\
        (\alpha d)^{\frac12\parens*{1-\frac{p\land q}{p\lor q}}} & \text{if $\min(p,q)\geq 2$}
    \end{cases}
\]
By instantiating with $\alpha = 1$, i.e.,\ exact Lewis weights, we exactly recover the guarantees of Theorem \ref{thm:lp-lq-change-of-density}.
\end{restatable}


Our main technique is a new simple identity (Lemma \ref{lem:lewis-weight-switching}) for Lewis weights which may be of independent interest, which shows that if we reweight the rows of $\bfA$ with $\ell_p$ Lewis weights, then the $\ell_q$ Lewis weights of the resulting matrix coincide with the $\ell_p$ Lewis weights of $\bfA$. Given this identity, the proof follows from just a few lines of estimates, which substantially simplifies the original proof of \cite{LT1980}. Furthermore, because our change of density uses Lewis weights, we inherit fast algorithms for computing these weights. Note that although polynomial time algorithms are known for many factorization theorems \cite{Tro2009}, known algorithms require solving constrained eigenvalue minimization problems, and are not known to have fast input sparsity time algorithms as Lewis weights do. Our result shows the following surprising message:
\begin{center}
\emph{$\ell_p$ Lewis weights provide optimal approximations of $\ell_p$ by $\ell_q$, for all $q$}.
\end{center}
We hope that our techniques will find further applications in functional analysis and theoretical computer science. In Section \ref{sec:LTJ80}, we give our proof for the ``change of density'' theorem, and in Section \ref{sec:app-lin-alg}, we give several immediate applications of our results to numerical linear algebra. In particular, we give the fastest known algorithms for $\ell_p$ linear regression with $\poly(d)$-factor relative error distortion and $\ell_p$ column subset selection. Both $\ell_p$ regression \cite{Cla2005, DDHKM2009, BCLL2018, APS2019, AKPS2019, AS2020, ABKS2021, JLS2021} and $\ell_p$ column subset selection \cite{SWZ2017, CGKLPW2017, BKW2017, BBBKLW2019, DWZZR2019, MW2021, JLLMW2021} are extremely well-studied, and obtaining fast algorithms for these problems is important. Our results are summarized in Table \ref{tab:lp-linear-algebra}.

\subsection{The Subspace Sketch Problem with Large Approximation}

As an application of Theorem \ref{thm:lp-lq-change-of-density-lewis}, we obtain new tight bounds on the offline $\ell_p$ subspace sketch problem. 

The offline subspace sketch problem captures the fundamental limits of dimension reduction in $\ell_p$: with unbounded computation and access to $\bfA$, how much can $\bfA$ be compressed, as a function of the distortion $\Delta$? The work of \cite{LWW2021} studied this problem in the regime of $\Delta = (1+\eps)$ for $\eps\to 0$. Here, \cite{LWW2021} found surprising separations between $p\in2\mathbb Z$ and all other $p$, showing a lower bound of $\tilde\Omega(d/\eps^2)$ bits of space required to store $Q_p$ for $p\in[1,\infty)\setminus 2\mathbb Z$ for the ``for each'' guarantee, which separates these $p$ from $p\in 2\mathbb Z$ due to an upper bound of $\tilde O(d^{p})$ due to \cite{Sch2011}. For $\eps = \Theta(1)$, they showed a lower bound of $\tilde\Omega(d^{p/2})$ for the ``for each'' guarantee and $\tilde\Omega(d^{p/2+1})$ for the ``for all'' guarantee, matching known upper bounds.

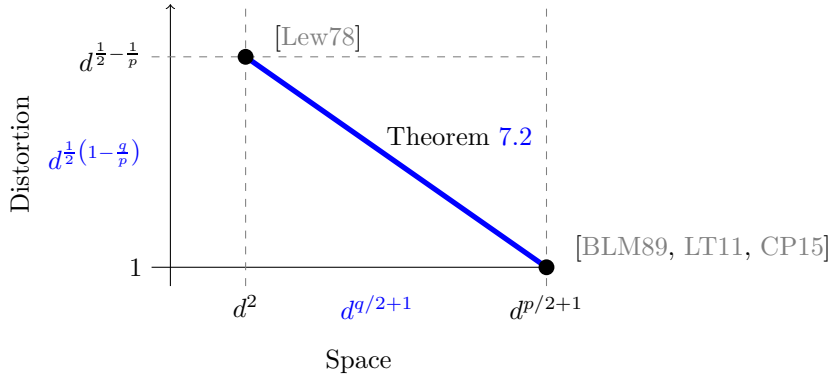
\begin{figure}[ht]
    \centering
    \begin{tikzpicture}[scale=0.5]
    
    \draw[->] (-0.5,0) -- (10,0);
    \draw[->] (0,-0.5) -- (0,7);
    \draw[blue, domain=2:10,smooth,variable=\q,thick,line width=0.7mm,samples=100] plot ({\q},{7 * (1 - \q/10)});
    
    \draw[gray, dashed] (2,-0.5)--(2,7);
    \node[below] (a) at (2,-0.5) {$d^2$};
    \draw[gray, dashed] (10,-0.5)--(10,7);
    \node[below] (a) at (10,-0.5) {$d^{p/2+1}$};
    \node[below] (a) at (5.5,-0.5) {$\color{blue}d^{q/2+1}$};
    
    \draw[gray, dashed] (-0.5,7 * 0.8)--(10,7 * 0.8);
    \node[left] (a) at (-0.5,7 * 0.8) {$d^{\frac12-\frac1p}$};
    \node[left] (a) at (-0.5,3) {$\color{blue}d^{\frac12\parens*{1-\frac{q}{p}}}$};
    \node[left] (a) at (-0.5,0) {$1$};
    
    \node[circle,fill=black,inner sep=0pt,minimum size=6pt] (a) at (2,7*0.8) {};
    \node[right] (a) at (2.5,7*0.8+0.5) {\cite{Lew1978}};
    \node[circle,fill=black,inner sep=0pt,minimum size=6pt] (a) at (10,0) {};
    \node[right] (a) at (10.5,0.5) {\cite{BLM1989, LT2011, CP2015}};
    
    \node[right] (a) at (5.5,3.5) {Theorem \ref{thm:subspace-sketch-lp-lq}};
    
    \node[] (a) at (5,-2.5) {Space};
    \node[rotate = 90] (a) at (-4,3) {Distortion};
    
    \end{tikzpicture}
    \caption{Space vs distortion for the $\ell_p$ subspace sketch problem.}
	\label{fig:trade-off}
\end{figure}

Although space bounds of the form $\tilde O(d^{p/2}/\eps^2)$ are possible for achieving $(1+\eps)$ distortion \cite{LT2011}, for large constant $p$, this space usage may already be problematic, especially if one is willing to tolerate a larger approximation factor. One could first observe that even for $p=\infty$, if one is willing to tolerate a distortion of $\sqrt d$, then it is possible to do better by using L\"owner--John ellipsoids, since it only takes $O(d^2)$ words of space (or $O(d^2\log n)$ bits) to store the quadratic form for the L\"owner--John ellipsoid for the convex set $\braces{\bfx\in \mathbb R^d: \norm*{\bfA\bfx}_p\leq 1}$. Taking this idea a step further, one could also store the quadratic form for the Lewis ellipsoid for $\bfA$ using $O(d^2)$ words to achieve a distortion of $O(d^{\frac12-\frac1p})$. However, these two upper bounds jump from $d^{p/2+1}$ space to $d^2$ space, which raises the question of whether it is possible to obtain a smooth trade-off. As another contribution, we answer this question in the affirmative, by applying our Theorem \ref{thm:lp-lq-change-of-density-lewis}. Our trade-offs are summarized in Table \ref{tab:subspace-sketch} and visualized in Figure \ref{fig:trade-off}. The lower bounds of \cite{LWW2021} extend to the parameter regime we consider, and shows that our upper bounds are nearly optimal, up to logarithmic factors. Our algorithmic technique is to first approximate the $\ell_p$ norm by the $\ell_q$ norm using Theorem \ref{thm:lp-lq-change-of-density-lewis} with some $q < p$, and then to use a constant factor approximation to the $\ell_q$ norm using $O(d^{q/2})$ words of space for the ``for each'' guarantee or $\tilde O(d^{q/2+1})$ for the ``for all'' guarantee. This shows the following:
\begin{center}
    \emph{Optimal dimension reduction trade-offs for $\ell_p$ are achieved through approximations by $\ell_q$.}
\end{center}
We discuss these results in Section \ref{sec:subspace-sketch}.


Finally, we also implement the algorithms above using multiple passes through the row arrival stream. The main technique is the implementation of Lewis weight approximation with $O(d^2\log n)$ bits of space, which requires suitable generalizations of Lewis weights that can be maintained with $O(\log n)$ bit complexity, as well as corresponding modifications to the Lewis weight sampling \cite{CP2015} and our change of density theorems. Note that although the pass complexity suffers compared to our earlier one-pass streaming algorithms, our results relying on Lewis weight computation over multiple passes allows for a new set of trade-offs in polynomial time. We summarize these results in Table \ref{tab:subspace-sketch-stream-lewis}.

\begin{table}
    \centering
    \begin{tabular}{ c c c c c c }
    \toprule
    & Distortion & Space & Passes \\
    \hline
    $2 < p < 4$ & $d^{\frac12\parens*{1-\frac{q}{p}}}$ & $d^{q/2+1}$ & $\log\log n$ & Theorem \ref{thm:subspace-sketch-stream-2<p<4} \\
    $4 \leq p < \infty$ & $d^{\frac12\parens*{1-\frac{q}{p}}}$ & $d^{q/2 + 1} + d^{\floor{p/4} + 2}$ & $\log\log n$ & Theorem \ref{thm:subspace-sketch-stream-p>4} \\
    $4 \leq p < \infty$ & $d^{\frac12\parens*{1-\frac{q}{p}}}$ & $d^{q/2+1}$ & $\log n$ & Theorem \ref{thm:subspace-sketch-stream-p>4-log-n-pass} \\
    \bottomrule
    \end{tabular}
    \caption{Results for streaming subspace sketch in the ``for all'' model, via Lewis weights. We suppress $O(1)$ factors in the distortion and $\poly\log n$ factors in the space complexity.}
    \label{tab:subspace-sketch-stream-lewis}
\end{table}

\subsection{Independent and Concurrent Work}

In \cite{MMO2022} which appeared in COLT 2022, the authors give a one-pass streaming algorithm for approximating the L\"owner--John ellipsoid of a convex hull which stores $O(d^2)$ floating point numbers and achieves a distortion of $O(\sqrt{d\log(R/r + 1)})$, where $R$ is the radius of the smallest ball containing the input points and $r$ is the radius of the largest ball contained in the input points. The algorithm in \cite{MMO2022} is quite different from ours, analyzing an algorithm similar to that of \cite{MSS2010}. For real-valued inputs, their distortion is independent of $n$ while our Theorem \ref{thm:john-ellipsoid-streaming-ub} incurs a dependence on $\log n$. However, our algorithm offers a couple of other advantages over \cite{MMO2022}: (1) for integer matrices with polynomially bounded entries, our result improves upon \cite{MMO2022} by providing a $O(\sqrt{d\log n})$ distortion without further assumptions, whereas the aspect ratio $R/r$ could be exponential in $d$; (2) our algorithm is a coreset algorithm, i.e., it only relies on storing a subset of the input points. We also note that they do not solve related $\ell_p$ subspace sketch algorithms, as we do.

\section{Preliminaries}

\subsection{Notation}

We often denote $\bfA\in\mathbb R^{n\times d}$ as an input matrix, and $\bfa_i\in\mathbb R^d$ by its rows as column vectors for $i\in[n]$. We use both $\bfv_i$ and $\bfv(i)$ to denote the $i$th entry of a vector $\bfv\in\mathbb R^n$. For a matrix $\bfA\in\mathbb R^{n\times d}$ and a set $S\subseteq[n]$, we write $\bfA\vert_S$ to mean the $\abs*{S}\times d$ matrix obtained by the rows of $\bfA$ indexed by $S$. For a matrix $\bfA\in\mathbb R^{n\times d}$, we denote the Moore--Penrose pseudoinverse of $\bfA$ as $\bfA^-$. 

\subsection{Lewis Weights}

We recall the definition of $\ell_p$ Lewis weights and related notions. We start with the definition of leverage scores, which is used to define Lewis weights.

\begin{Definition}[Leverage scores \cite{CLMMPS2015}]
Let $\bfA\in\mathbb R^{n\times d}$. The \emph{leverage scores of $\bfA$} are defined as
\[
    \bftau_i(\bfA) \coloneqq \bfa_i^\top(\bfA^\top\bfA)^{-}\bfa_i
\]
for each $i\in[n]$.
\end{Definition}

Lewis weights are then given by the following:

\begin{Definition}[$\ell_p$ Lewis weights and bases \cite{CP2015}]\label{def:lewis-weights}
Let $\bfA\in\mathbb R^{n\times d}$ and $0 < p < \infty$. The \emph{$\ell_p$ Lewis weights of $\bfA$} are defined as the unique weights $\bfw_i^p(\bfA)$ satisfying
\[
    \bfw_i^p(\bfA) = \bftau_i(\bfW^p(\bfA)^{1/2-1/p}\bfA),
\]
where $\bfW^p(\bfA) \coloneqq \diag(\bfw^p(\bfA))$. Let $\bfR\in\mathbb R^{d\times d}$ be a change of basis matrix such that $\bfW^p(\bfA)^{1/2-1/p}\bfA\bfR$ is orthonormal. Then, we call $\bfA\bfR$ the \emph{Lewis basis}.
\end{Definition}

Although the definition in Definition \ref{def:lewis-weights} is circular, it was shown in \cite{Lew1978, SZ2001, CP2015} that these indeed exist and are unique.

\subsubsection{Approximate Lewis Weights}

Many of our theorems will only rely on \emph{approximate} Lewis weights, in the following sense:

\begin{Definition}[One-sided $\ell_p$ Lewis weights and bases (Definition 2.4 of \cite{JLS2021}\footnote{Note that \cite{JLS2021} call this definition ``overestimates'', but we use a different word for this notion to reserve the word ``overestimate'' for the next definition.})]\label{def:one-sided-lewis}
Let $\bfA\in\mathbb R^{n\times d}$ and let $p>0$ be an index. We say that $\bfw\in\mathbb R^n$ are \emph{one-sided $\ell_p$ Lewis weights} if
\[
    \bfw_i \geq \bftau_i(\bfW^{1/2-1/p}\bfA),
\]
where $\bfW \coloneqq \diag(\bfw)$. Let $\bfR\in\mathbb R^{d\times d}$ be a change of basis matrix such that $\bfW^{1/2-1/p}\bfA\bfR$ is orthonormal. Then, we call $\bfA\bfR$ a \emph{one-sided Lewis basis}.
\end{Definition}

\begin{Definition}[$\alpha$-approximate $\ell_p$ Lewis overestimates]\label{def:approximate-lewis-overestimate}
Let $\bfA\in\mathbb R^{n\times d}$, let $\alpha\geq 1$, and let $p>0$ be an index. We say that $\bfw\in\mathbb R^n$ are \emph{$\alpha$-approximate $\ell_p$ Lewis overestimates} if the following hold:
\begin{itemize}
    \item $\bfw_i \geq \bfv_i$ for all $i\in[n]$ for one-sided $\ell_p$ Lewis weights $\bfv\in\mathbb R^n$
    \item $\norm*{\bfw}_1 = \sum_{i=1}^n \bfw_i \leq \alpha \cdot d$
\end{itemize}
\end{Definition}

\begin{Remark}
The above notion of overestimation of $\ell_p$ Lewis weights combines two notions of overestimation in the literature, one in which $\bfw_i$ is greater than $\bftau_i(\bfW^{1/2-1/p}\bfA)$ (e.g.\ Definition 5.3.2 of \cite{Lee2016}, Definition 2.4 of \cite{JLS2021}), and another in which $\bfw_i$ is greater than the true $\ell_p$ Lewis weights (e.g.\ \cite{CP2015}, Theorem 2.11 of \cite{MMWY2021}, Equation (1.5) of \cite{FLPS2021}). This additional flexibility will be useful to us when approximating Lewis weights in a stream.
\end{Remark}

Note that the above approximation is weaker than an $\alpha$-factor relative error approximation of every weight, which is considered in, e.g., \cite{FLPS2021}. In particular, this allows small Lewis weights to be distorted by an arbitrarily large factor, as long as the sum is bounded by $\alpha\cdot d$. For example, we can take every weight to be at least $1/n$ by taking $\alpha \geq (1+1/d)$, which allows us to obtain good bounds on bit complexity in streaming settings, which we will consider.

\begin{Theorem}[Fast Approximation of Lewis Weights]\label{thm:cohen-peng-fast-lewis-weights}
    Let $\bfA\in\mathbb R^{n\times d}$ and let $0 < p < \infty$. There is an algorithm which computes $O(1)$-approximate $\ell_p$ Lewis weights which runs in time
    \[
        \tilde O(\nnz(\bfA) + d^\omega),
    \]
    due to Lemma 7 of \cite{CLMMPS2015} for $p = 2$, Lemma 2.4 of \cite{CP2015} for $0<p<4$, and Theorem 5.3.4 of \cite{Lee2016} for $p\geq 2$. Here, $\omega \approx 2.37286$ is the current exponent of fast matrix multiplication \cite{AW2021}.
\end{Theorem}

\subsubsection{Properties of Lewis Weights}

We start with basic facts about leverage scores inherited by Lewis weights.

\begin{Lemma}\label{lem:lewis-weights-sum-to-d}
Let $\bfA\in\mathbb R^{n\times d}$ and let $0<p<\infty$. Then, $\bfw_i^p(\bfA)\in[0,1]$ for each $i\in[n]$ and
\[
    \sum_{i=1}^n \bfw_i^p(\bfA) = d.
\]
\end{Lemma}
\begin{proof}
This follows from the fact that leverage scores sum to $d$, and that the $\ell_p$ Lewis weights are the leverage scores of a reweighted matrix.
\end{proof}

It is well-known that the $\ell_p$ Lewis weights are exactly given by the row norms of Lewis bases, raised to the $p$th power \cite{Lew1978, SZ2001}. A similar relationship holds for one-sided Lewis weights:
\begin{Lemma}\label{lem:one-sided-lewis-basis-vs-weights}
Let $\bfA\in\mathbb R^{n\times d}$ and let $0<p<\infty$. The following hold:
Let $\bfw\in\mathbb R^n$ be one-sided $\ell_p$ Lewis weights, and let $\bfR$ be the corresponding one-sided Lewis basis. Then, for each $i\in[n]$,
\[
    \bfw_i \geq \norm*{\bfe_i^\top\bfA\bfR}_2^p.
\]
\end{Lemma}
\begin{proof}
We have that
\[
    \bfw_i \geq \bftau_i(\bfW^{1/2-1/p}\bfA) = \norm*{\bfe_i^\top\bfW^{1/2-1/p}\bfA\bfR}_2^2 = \bfw_i^{1-2/p}\norm*{\bfe_i^\top\bfA\bfR}_2^2
\]
which rearranges to the desired result.
\end{proof}

Another well-known fact is that Lewis weights bound the \emph{$\ell_p$ sensitivities} of $\bfA$. The proof is elementary and is included in Lemma 2.5 of \cite{MMWY2021}.

\begin{Lemma}[Lemma 2.5 of \cite{MMWY2021}]\label{lem:lewis-weights-bound-sensitivities}
Let $\bfA\in\mathbb R^{n\times d}$ and $0<p<\infty$. Then,
\[
    \sup_{\bfx\in\rowspan(\bfA)\setminus\{0\}} \frac{\abs*{[\bfA\bfx](i)}^p}{\norm*{\bfA\bfx}_p^p} \leq d^{0\lor (p/2-1)}\bfw_i^p(\bfA).
\]
\end{Lemma}

We generalize the result to one-sided $\ell_p$ Lewis weights.

\begin{Lemma}[One-sided Lewis weights bound sensitivities]\label{lem:one-sided-lewis-weights-bound-sensitivities}
Let $\bfA\in\mathbb R^{n\times d}$ and $0<p<\infty$. Let $\bfw\in\mathbb R^n$ be one-sided $\ell_p$ Lewis weights. Then,
\[
    \sup_{\bfx\in\rowspan(\bfA)\setminus\{0\}} \frac{\abs*{[\bfA\bfx](i)}^p}{\norm*{\bfA\bfx}_p^p} \leq \norm*{\bfw}_1^{0\lor (p/2-1)}\cdot\bfw_i.
\]
\end{Lemma}
\begin{proof}
The proofs follow the proofs for the usual $\ell_p$ Lewis weights. Let $\bfR$ be the one-sided $\ell_p$ Lewis basis for $\bfw$. We first claim that
\[
    \norm*{\bfx}_2 \leq \norm*{\bfw}_1^{0\lor (1/2-1/p)}\norm*{\bfA\bfR\bfx}_p
\]
for any $\bfx\in\mathbb R^d$. Indeed, if $p\geq 2$, then
\begin{align*}
    \norm*{\bfx}_2^2 &= \norm*{\bfW^{1/2-1/p}\bfA\bfR\bfx}_2^2 = \sum_{i=1}^n \bfw_i^{1-2/p}\bracks*{\bfe_i^\top\bfA\bfR\bfx}^2 && \text{Orthonormality} \\
    &\leq \bracks*{\sum_{i=1}^n \bfw_i}^{1-2/p}\bracks*{\sum_{i=1}^n \abs*{\bfe_i^\top\bfA\bfR\bfx}^p}^{2/p} = \norm*{\bfw}_1^{1-2/p}\norm*{\bfA\bfR\bfx}_p^2 && \text{H\"older's inequality}
\end{align*}
and if $p\leq 2$, then
\begin{align*}
    \norm*{\bfx}_2^2 &= \norm*{\bfW^{1/2-1/p}\bfA\bfR\bfx}_2^2 = \sum_{i=1}^n \bfw_i^{1-2/p}\bracks*{\bfe_i^\top\bfA\bfR\bfx}^{2-p}\bracks*{\bfe_i^\top\bfA\bfR\bfx}^p && \text{Orthonormality} \\
    &\leq \sum_{i=1}^n \bfw_i^{1-2/p}\norm*{\bfe_i^\top\bfA\bfR}_2^{2-p}\norm*{\bfx}_2^{2-p}\bracks*{\bfe_i^\top\bfA\bfR\bfx}^p && \text{Cauchy--Schwarz} \\
    &\leq \sum_{i=1}^n \bfw_i^{1-2/p}\cdot\bfw_i^{2/p-1}\norm*{\bfx}_2^{2-p}\bracks*{\bfe_i^\top\bfA\bfR\bfx}^p = \norm*{\bfx}_2^{2-p}\norm*{\bfA\bfR\bfx}_p^p && \text{Lemma \ref{lem:one-sided-lewis-basis-vs-weights}}
\end{align*}
We then conclude by the following:
\begin{align*}
    \frac{\abs*{[\bfA\bfR\bfx](i)}^p}{\norm*{\bfA\bfR\bfx}_p^p} &\leq \frac{\norm*{\bfe_i^\top \bfA\bfR}_2^p \norm*{\bfx}_2^p }{\norm*{\bfA\bfR\bfx}_p^p} && \text{Cauchy--Schwarz} \\
    &\leq \bfw_i \frac{\norm*{\bfw}_1^{p/2-1}\norm*{\bfA\bfR\bfx}_p^p}{\norm*{\bfA\bfR\bfx}_p^p} = \norm*{\bfw}_1^{0\lor (p/2-1)}\cdot \bfw_i && \text{Lemma \ref{lem:one-sided-lewis-basis-vs-weights}}\qedhere
\end{align*}
\end{proof}

The fact that $\ell_p$ Lewis weights bound the $\ell_p$ sensitivities, along with several other properties, shows that sampling rows by Lewis weights is an algorithm for obtaining \emph{subspace embeddings} with near-optimal dimension reduction, by a seminal result of \cite{BLM1989}, which was introduced to the theoretical computer science literature by \cite{CP2015} in the context of fast algorithms for $\ell_p$ regression. For the purposes of our streaming algorithms that require bounded bit complexity, we need a generalization of these results that allow for the substitution of Lewis weights by one-sided $\ell_p$ Lewis weights. We show how to modify the proofs of \cite{BLM1989} to allow for this modification.

\begin{Theorem}[Subspace Embeddings via Lewis Weights]\label{thm:lewis-weight-subspace-embedding}
    Let $\bfA\in\mathbb R^{n\times d}$ and let $0 < p<\infty$. Let $\bfS\in\mathbb R^{s\times n}$ be the matrix that samples $s$ rows independently and proportionally to $O(1)$-approximate $\ell_p$ Lewis weight overestimates $\bfw$ of $\bfA$ (Definition \ref{def:approximate-lewis-overestimate}), and rescales them by $1/(s\bfw_i)^{1/p}$ whenever the $i$th row is sampled. There is a $s = d^{1\lor(p/2)}\poly(\log d,\eps^{-1},\log\delta^{-1})$\footnote{The polynomial can have degree depending on $p$ for $p\in(0,1)$, see \cite{MMWY2021}.} such that
    \[
        \Pr\braces*{\forall \bfx\in\mathbb R^d, \norm*{\bfA\bfx}_p \leq \norm*{\bfS\bfA\bfx}_q \leq (1+\eps)\norm*{\bfA\bfx}_p} \geq 1 - \delta
    \]
\end{Theorem}
\begin{proof}
Theorem \ref{thm:lewis-weight-subspace-embedding} was stated in Theorem 7.1 of \cite{CP2015} only with $\delta = \Theta(1)$, with samples proportional to constant factor approximations to the Lewis weights, and $p\geq 1$, by using \cite{BLM1989}. However, by modifying the proof of \cite{BLM1989}, extensions to $1-\delta$ probability are possible, as well as replacing Lewis weights by one-sided Lewis weights.

Note first that $p\in(0,1)$ was shown by \cite{MMWY2021}, by using Lewis bases and entropy estimates developed by \cite{SZ2001}.

The extension to $1-\delta$ probability has been previously discussed in Theorem 4.2 in Appendix B.1 of \cite{JLLMW2021}, as well as \cite{LWW2021}. This is done by replacing the use of Bernstein bounds that hold with constant probability to $1-\delta$ probability by increasing the number of samples by a factor of $O(\log \delta^{-1})$, wherever appropriate.

The extension to sampling using one-sided Lewis weights is new to the best of our knowledge, and we describe the necessary changes to \cite{BLM1989} here. The two places of \cite{BLM1989} where the properties of Lewis bases are used are in Lemma 4.5 and Proposition 4.6. In our language, Equation (4.26) of \cite{BLM1989} use the fact that the exact Lewis weights $\bfw$ and the Lewis basis $\bfA\bfR$, satisfy that
\[
    \norm*{\bfe_i^\top\bfW^{-1/p}\bfA\bfR}_{2}^2 = \bfw_i^{-2/p}\norm*{\bfe_i^\top\bfA\bfR}_2^2  = 1
\]
for every $i\in[n]$ and that $\bfW^{-1/p}\bfA\bfR$ has orthonormal columns under the measure $\bfW$, that is, that $\bfW^{1/2-1/p}\bfA\bfR$ has orthonormal columns in the usual sense. If $\bfW$ and $\bfR$ are instead the one-sided analogues, then $\bfW^{1/2-1/p}\bfA\bfR$ still has orthonormal columns by Definition \ref{def:one-sided-lewis}, and Lemma \ref{lem:one-sided-lewis-basis-vs-weights} guarantees us that
\[
    \norm*{\bfe_i^\top\bfW^{-1/p}\bfA\bfR}_{2}^2 = \bfw_i^{-2/p}\norm*{\bfe_i^\top\bfA\bfR}_2^2  \leq \bfw_i^{-2/p}\bfw_i^{2/p} = 1.
\]
This inequality goes in the right direction in Equation (4.27) of Lemma 4.5 as well as the inequality following the application of the Maurey--Khintchine inequality in Proposition 4.6, which means that the result still stays the same as well.
\end{proof}

\subsection{Streaming Algorithms}

\subsubsection{\textsf{INDEX}}

In the \textsf{INDEX} problem, Alice has a set $A\subseteq[n]$ and Bob has an index $i\in[n]$. Alice then generates a message $M$ using a possibly randomized algorithm $\mathcal A$ as a function of $A$, and then passes the message $M$ to Bob. Bob must then determine whether $i\in A$ or not. 

\begin{Theorem}[\textsf{INDEX} lower bound \cite{KNR1999}]\label{thm:index}
Suppose $\mathcal A$ solves the \textsf{INDEX} problem with probability at least $2/3$. Then, $\mathcal A$ must use at least $\Omega(n)$ bits.
\end{Theorem}

\subsubsection{Frequency Moment Sketches}

A fundamental problem in the streaming literature is to estimate the $\ell_p$ norm of a fixed vector $\bfx\in\mathbb R^n$ up to $(1+\eps)$ relative error. Following a long line of research since \cite{AMS1999} (see \cite{GW2018} for the history of this results for this problem), the following is known:

\begin{Theorem}[Frequency Moment Sketches \cite{GW2018}]\label{thm:frequency-moment-sketch}
    There is a distribution over $s\times n$ matrices $\bfS$ such that, given $\bfS\bfx$ for any $\bfx\in\mathbb R^n$, one can recover $\norm*{\bfx}_p^p$ up to a $(1+\eps)$ factor distortion with probability at least $1-\delta$, for some
    \[
        s = O\parens*{n^{1-2/p}\eps^{-2}\log(1/\delta) + n^{1-2/p}\eps^{-4/p}\log^{2/p}(1/\delta)\log n}.
    \]
\end{Theorem}
\section{Streaming Subspace Sketch}

In this section, we consider subspace sketch problem in the row arrival streaming setting. Throughout this section, we assume that $\bfA\in\mathbb Z^{n\times d}$ has integer entries bounded by $\poly(n)$, unless explicitly stated. By replacing our use of bounds on online leverage scores for such matrices of $O(d\log n)$ (which we prove in Theorem \ref{thm:sharp-online-leverage-scores}) with $O(d\log\kappa^\OL)$ for real matrices (Lemma \ref{lem:online-leverage-score-bound}), we obtain similar results in the online coreset model for real matrices $\bfA\in\mathbb R^{n\times d}$ by replacing $\log n$ dependencies with $\log\kappa^\OL$ dependencies, where $\kappa^\OL$ is the online pseudo condition number of $\bfA$.

\subsection{Sharper Online Numerical Linear Algebra}

In this section, we introduce various definitions and lemmas concerning the so-called \emph{online leverage scores} \cite{CMP2020,BDMMUWZ2020}, which are an ``online'' variant of the usual statistical leverage scores. Under standard bit complexity assumptions that the input points are integers bounded by $\poly(n)$, we will substantially sharpen known results in this area by replacing $\log\kappa^\OL$ dependencies with $\log n$ dependencies, where $\kappa$ is the condition number of $\bfA$. 

\begin{Definition}[Definition 2.1 of \cite{BDMMUWZ2020}, Theorem 2.2 of \cite{CMP2020}]\label{def:online-leverage-scores}
    Let $\bfA\in\mathbb R^{n\times d}$. Then, for each $i\in[n]$, the $i$th online leverage score is defined as
    \[
        \bftau_i^{\mathsf{OL}}(\bfA) \coloneqq \begin{cases}
            \min\braces*{\bfa_i^\top(\bfA_{i-1}^\top\bfA_{i-1})^- \bfa_i,1} & \bfa_i\in\rowspan(\bfA_{i-1}) \\
            1 & \text{otherwise}
        \end{cases}
    \]
    where $\bfA_{j}\in\mathbb R^{j\times d}$ denotes the submatrix of $\bfA$ formed by the first $j$ rows. 
\end{Definition}

We require the following standard lemma relating online leverage scores and sensitivities:

\begin{Lemma}\label{lem:generalized-sensitivity}
    Let $\bfA\in\mathbb R^{n\times d}$ and $\bfa\in\rowspan(\bfA)$. Then,
    \[
        \sup_{\bfx\in\rowspan(\bfA)\setminus\{0\}} \frac{\angle*{\bfa,\bfx}^2}{\norm*{\bfA\bfx}_2^2} = \bfa^\top(\bfA^\top\bfA)^-\bfa.
    \]
\end{Lemma}
\begin{proof}
    Let $\bfA = \bfU\bfSigma\bfV^\top$ be the truncated SVD of $\bfA$. Then, $\bfa = \bfV\bfb$ for some $\bfb\in\mathbb R^{\rank(\bfA)}$. For any $\bfx\in\rowspan(\bfA)\setminus\{0\}$, let $\bfy = \bfSigma\bfV^\top\bfx$. Then,
    \begin{align*}
        \frac{\angle*{\bfa,\bfx}^2}{\norm*{\bfA\bfx}_2^2} &= \frac{\angle*{\bfV\bfSigma\bfSigma^{-1}\bfb,\bfx}^2}{\norm*{\bfU\bfy}_2^2} && \text{$\bfa = \bfV\bfb$} \\
        &= \frac{\angle*{\bfSigma^{-1}\bfb,\bfSigma\bfV^\top\bfx}^2}{\norm*{\bfy}_2^2} && \text{$\bfU$ has orthonormal columns} \\
        &= \angle*{\bfSigma^{-1}\bfb,\bfy/\norm*{\bfy}}^2 && \text{$\bfy = \bfSigma\bfV^\top\bfx$} \\
    \end{align*}
    The supremum over all $\bfx$ is equal to
    \[
        \norm*{\bfSigma^{-1}\bfb}_2^2 = \bfb^\top\bfSigma^{-2}\bfb
    \]
    by the Cauchy--Schwarz inequality and its tightness. This is equal to
    \[
        \bfa^\top(\bfA^\top\bfA)^-\bfa = \bfb^\top \bfV^\top (\bfV\bfSigma^2\bfV^\top)^- \bfV\bfb = \bfb^\top \bfSigma^{-2}\bfb.
    \]
    as claimed.
\end{proof}

Just as the sum of leverage scores is bounded by $d$ (Lemma \ref{lem:lewis-weights-sum-to-d}), the sum of online \emph{ridge} leverage scores, which is a regularized variant of online leverage scores, is known to have a small sum, although not as small as their offline counterparts.

\begin{Lemma}[Lemma 2.2 of \cite{BDMMUWZ2020}, Theorem 2.2 of \cite{CMP2020}]\label{lem:online-leverage-score-bound}
    Let $\bfA\in\mathbb R^{n\times d}$ and $\lambda>0$. Define the \emph{ridge leverage scores} of $\bfA$ as
    \[
        \bftau_i^{\lambda,\mathsf{OL}}(\bfA) \coloneqq \bfa_i^\top(\bfA^\top\bfA+\lambda\bfI_d)^{-1}\bfa_i.
    \]
    Then,
    \[
        \sum_{i=1}^n \bftau_i^{\lambda,\mathsf{OL}}(\bfA) = O\parens*{d\log\frac{\norm*{\bfA}_2}{\lambda}}.
    \]
\end{Lemma}

By setting $\lambda$ to be smaller than the minimum singular value of $\bfA_i^\top\bfA_i$ for any $i\in[n]$, this corresponds to a bound of $O(d\log\kappa^\OL)$, where $\kappa^\OL$ is the online condition number of $\bfA$.

We now sharpen the above bound for integer matrices with bounded entries. To do so, we will need the notion of a pseudodeterminant.

\begin{Definition}[Pseudodeterminant]
    Let $\bfM\in\mathbb R^{d\times d}$ be a symmetric matrix of rank $r$. Then, the \emph{pseudodeterminant} $\pdet(\bfM)$ of $\bfM$ is the product of the nonzero eigenvalues of $\bfM$.
\end{Definition}

We need the following simple lemmas which dictate the evolution of pseudodeterminants under row additions. The first shows how to handle the additional of orthogonal rows.

\begin{Lemma}\label{lem:pdet-evolution-bot}
    Let $\bfA\in\mathbb R^{n\times d}$ and let $\bfa\in\mathbb R^d$ be a vector that is orthogonal to the row span of $\bfA$. Then,
    \[
        \pdet(\bfA^\top\bfA + \bfa\bfa^\top) = \norm*{\bfa}_2^2\cdot\pdet(\bfA^\top\bfA).
    \]
\end{Lemma}
\begin{proof}
    Let $\bfA = \bfU\bfSigma\bfV^\top$. Note that the SVD of the concatenation $\bfA'\in\mathbb R^{(n+1)\times d}$ of $\bfA$ and $\bfa$ is
    \[
        \bfA' = \begin{pmatrix}\bfA \\ \bfa\end{pmatrix} = \begin{pmatrix}\bfU & 0 \\ 0 & 1\end{pmatrix}\begin{pmatrix}\bfSigma & 0\\0 & \norm*{\bfa}_2\end{pmatrix}\begin{pmatrix}\bfV^\top \\ \bfa^\top / \norm*{\bfa}_2\end{pmatrix}.
    \]
    Thus,
    \[
        \pdet(\bfA^\top\bfA + \bfa\bfa^\top) = \pdet(\bfA'^\top\bfA') = \norm*{\bfa}_2^2\prod_{j=1}^d \sigma_j^2 = \norm*{\bfa}_2^2 \cdot \pdet(\bfA^\top\bfA),
    \]
    as claimed.
\end{proof}

Our second lemma is a generalization of the matrix determinant lemma to pseudodeterminants. 

\begin{Lemma}[Matrix Pseudodeterminant Lemma]\label{lem:pdet-evolution-par}
    Let $\bfA\in\mathbb R^{n\times d}$ and let $\bfa\in\mathbb R^d$ be a vector that is in the row span of $\bfA$. Then,
    \[
        \pdet(\bfA^\top\bfA + \bfa\bfa^\top) = \pdet(\bfA^\top\bfA)(1+\bfa(\bfA^\top\bfA)^-\bfa).
    \]
\end{Lemma}
\begin{proof}
    Let $\bfA = \bfU\bfSigma\bfV^\top$ be the truncated SVD of $\bfA$. Since $\bfa$ is in the row span of $\bfV$, we may write $\bfa = \bfV\bfb$ for some $\bfb\in\mathbb R^r$, where $r = \rank(\bfA)$. Then,
    \begin{align*}
        \pdet(\bfA^\top\bfA + \bfa\bfa^\top) &= \pdet(\bfSigma^2 + \bfb\bfb^\top) \\
        &= \det(\bfSigma^2 + \bfb\bfb^\top) \\
        &= \det(\bfSigma^2) (1+\bfb^\top\bfSigma^{-2}\bfb) && \text{matrix determinant lemma} \\
        &= \pdet(\bfV\bfSigma^2\bfV^\top)(1+\bfb^\top\bfV^\top(\bfV\bfSigma^{-2}\bfV^\top)\bfV\bfb) \\
        &= \pdet(\bfA^\top\bfA)(1+\bfa(\bfA^\top\bfA)^-\bfa)
    \end{align*}
    as desired.
\end{proof}

We also need the following identity found in \cite{GK2010}, which states that the volume of a parallelotope is given by both the determinant of the Gram matrix as well as the product of the heights of the dimensions of the parallelotope.

\begin{Lemma}[Determinant volume identity \cite{GK2010}]\label{lem:det-vol}
    Let $\bfA\in\mathbb R^{r\times d}$ have linearly independent rows. Then,
    \[
        \sqrt{\det(\bfA\bfA^\top)} = \prod_{i=1}^r \norm*{\bfa_i^\bot}_2
    \]
    where $\bfa_1^\bot = \bfa_1$ and $\bfa_i^\bot$ is the projection of $\bfa_i$ onto the orthogonal complement of the row span of $\bfA_{i-1}$ for $i\geq 2$.
\end{Lemma}

We now prove the following main theorem of this section.

\SumOnlineLeverageScore*
\begin{proof}
    Our proof is a careful improvement of the original proof by \cite{CMP2020} under our bit complexity assumption. Let $i\in[n]$. If $\bfa_{i+1}$ is in the row span of $\bfA_i$, then by Lemma \ref{lem:pdet-evolution-par}, we have that
    \begin{align*}
        \pdet(\bfA_{i+1}^\top \bfA_{i+1}) &= \pdet(\bfA_i^\top \bfA_i)(1+\bfa_{i+1}(\bfA_i^\top\bfA_i)^-\bfa_{i+1}) \\
        &\geq \pdet(\bfA_i^\top \bfA_i)(1+\bftau_{i+1}^{\mathsf{OL}}(\bfA)) \\
        &\geq \pdet(\bfA_i^\top \bfA_i)\exp(\bftau_{i+1}^{\mathsf{OL}}(\bfA)/2)
    \end{align*}
    and otherwise, let $\bfa_{i+1} = \bfa_{i+1}^{\parallel} + \bfa_{i+1}^{\bot}$, where $\bfa^{\parallel}$ is the projection of $\bfa_{i+1}$ onto the row span of $\bfA_i$ and $\bfa^{\bot}$ is the residual. We have that
    \begin{align*}
        \pdet(\bfA_{i+1}^\top \bfA_{i+1}) &= \pdet(\bfA_i^\top \bfA_i + \bfa_{i+1}\bfa_{i+1}^\top) \\
        &= \pdet(\bfA_i^\top \bfA_i + \bfa_{i+1}^{\parallel}(\bfa_{i+1}^{\parallel})^\top + \bfa_{i+1}^{\bot}(\bfa_{i+1}^{\bot})^\top) \\
        &= \norm*{\bfa_{i+1}^{\bot}}_2^2\cdot \pdet(\bfA_i^\top \bfA_i + \bfa_{i+1}^{\parallel}(\bfa_{i+1}^{\parallel})^\top) && \text{Lemma \ref{lem:pdet-evolution-bot}} \\
        &= \norm*{\bfa_{i+1}^{\bot}}_2^2\cdot \pdet(\bfA_i^\top \bfA_i)(1+\bfa_{i+1}^{\parallel}(\bfA_i^\top\bfA)^-\bfa_{i+1}^{\parallel}) && \text{Lemma \ref{lem:pdet-evolution-par}} \\
        &\geq \norm*{\bfa_{i+1}^{\bot}}_2^2\cdot \pdet(\bfA_i^\top \bfA_i)
    \end{align*}
    Now let $S\subseteq[n]$ denote the at most $d$ indices such that $\bfa_{i}$ is not in the row span of $\bfA_{i-1}$. Note that we take $1\in S$ so that $\bfa_1^\bot = \bfa_1$. We then have by induction that
    \begin{align*}
        \pdet(\bfA^\top\bfA) = \pdet(\bfA_n^\top\bfA_n) &\geq \prod_{i\in[n]\setminus S}\exp(\bftau_i^{\mathsf{OL}}(\bfA)/2)\prod_{j\in S}\norm*{\bfa_j^{\bot}}_2^2 \\
        &= \exp\parens*{\frac12\sum_{i\in[n]\setminus S} \bftau_i^{\mathsf{OL}}(\bfA)}\prod_{j\in S}\norm*{\bfa_j^{\bot}}_2^2 \\
        &= \exp\parens*{\frac12\sum_{i\in[n]\setminus S} \bftau_i^{\mathsf{OL}}(\bfA)}\det(\bfA\vert_S\bfA\vert_S^\top) && \text{Lemma \ref{lem:det-vol}}
    \end{align*}
    where $\bfA\vert_S$ is the restriction of $\bfA$ to the rows indexed by $S$. By bounding each eigenvalue by the operator norm, we have that $\pdet(\bfA^\top\bfA) \leq \norm*{\bfA^\top\bfA}_2^d \leq \poly(n)^d$. Furthermore, since $\bfA\vert_S\bfA\vert_S^\top$ is a nonsingular integer Gram matrix, it has positive integer determinant, which is in particular at least $1$. We thus have that
    \[
        \exp\parens*{\frac12\sum_{i\in[n]\setminus S} \bftau_i^{\mathsf{OL}}(\bfA)} \leq \poly(n)^d \implies \sum_{i\in[n]\setminus S}\bftau_i^{\mathsf{OL}}(\bfA) \leq O(d\log n).
    \]
    Finally, $\abs*{S} \leq d$, which implies that
    \[
        \sum_{i=1}^n \bftau_i^{\mathsf{OL}}(\bfA) = \sum_{i\in[n]\setminus S}\bftau_i^{\mathsf{OL}}(\bfA) + \sum_{i\in S}\bftau_i^{\mathsf{OL}}(\bfA) \leq O(d\log n) + d = O(d\log n),
    \]
    as claimed.
\end{proof}

We next recall the notion of \emph{online $\ell_p$ sensitivities}.

\begin{Definition}[Online $\ell_p$ Sensitivities (Definition 4.1 of \cite{BDMMUWZ2020})]
    Let $\bfA\in\mathbb R^{n\times d}$ and let $1\leq p < \infty$. Then, for each $i\in[n]$, the $i$th online $\ell_p$ sensitivity is defined as
    \[
        \bfs_i^{p,\OL}(\bfA) \coloneqq \begin{cases}
            \min\braces*{\sup_{\bfx\in\rowspan(\bfA_{i-1})\setminus\{0\}}\frac{\abs*{\angle*{\bfa_i,\bfx}}^p}{\norm*{\bfA_{i-1}\bfx}_p^p},1} & \bfa_i \in \rowspan(\bfA_{i-1}) \\
            1 & \text{otherwise}
        \end{cases}
    \]
    where $\bfA_j\in\mathbb R^{j\times d}$ denotes the submatrix of $\bfA$ formed by the first $j$ rows.
\end{Definition}

Note that the online $\ell_2$ sensitivities coincide with the online leverage scores. By a reduction in \cite{BDMMUWZ2020}, this also implies improvements in the sum of \emph{online $\ell_p$ sensitivities}. We will use this in our algorithms.

\begin{Theorem}[Sum of Online $\ell_p$ Sensitivities]\label{thm:sum-online-lp-sensitivity}
    Let $\bfA\in\mathbb Z^{n\times d}$ be an integer matix with entries bounded by $\poly(n)$. Let $2 < p < \infty$. Then,
    \[
        \sum_{i=1}^n \bfs_i^{p,\OL}(\bfA) = O(d\log n)^{p/2}.
    \]
    If $\bfA\in\mathbb R^{n\times d}$ is a real matrix with online pseudo condition number $\kappa^{\OL}$, then
    \[
        \sum_{i=1}^n \bfs_i^{p,\OL}(\bfA) = O(d\log(n\kappa^{\OL}))^{p/2}.
    \]
\end{Theorem}
\begin{proof}
The second result for real-valued matrices is Corollary 3.9 of \cite{WY2022}. For the first result, we apply Corollary 3.11 of \cite{WY2022} with a carefully rounded set of online one-sided $\ell_p$ Lewis weights $\bfw$ (Definition 3.10 of \cite{WY2022}) so that the reweighted matrix $\bfW^{1/2-1/p}\bfA$ is integer, which allows us to bound the sum $\norm*{\bfw}_1$ by $O(d\log n)$.

We slightly modify the deterministic construction of online one-sided $\ell_p$ Lewis weights in \cite{WY2022}, in which weights $\bfw$ are iteratively generated by computing
\[
    \bfw_i \gets \Theta\parens*{\min\braces*{1, \bracks*{\bfa_i^\top(\bfA_{i-1}^\top\bfW_{i-1}^{1-2/p}\bfA_{i-1})^- \bfa_i}^{p/2}}}.
\]
We instead compute
\[
    \bfw_i \gets \round\parens*{\max\braces*{\min\braces*{1, \bracks*{\bfa_i^\top(\bfA_{i-1}^\top\bfW_{i-1}^{1-2/p}\bfA_{i-1})^- \bfa_i}^{p/2}}, \frac1n}}
\]
where $\round$ rounds up to the nearest integer power of $2$. We first bound the sum. If $\bfw_i = 1/n$, then the sum of these is bounded by $1$, so consider $\bfw_i$ that is larger than $1/n$. Then,
\[
    \bfw_i \leq 2\bracks*{\bfa_i^\top(\bfA_{i-1}^\top\bfW_{i-1}^{1-2/p}\bfA_{i-1})^- \bfa_i}^{p/2} \implies \bfw_i \leq 2^{2/p} \cdot \bftau_i^\OL(\bfW^{1/2-1/p}\bfA)
\]
and since $\bfW^{1/2-1/p}\bfA$ scaled up by $\Theta(n^{1/2-1/p})$ is an integer matrix, we have that
\[
    \sum_{i=1}^n \bfw_i \leq O(d\log n)
\]
by Theorem \ref{thm:sharp-online-leverage-scores}. Since $\bfw$ also has the online one-sided $\ell_p$ Lewis weight property, Corollary 3.10 \cite{WY2022} shows the desired result.
\end{proof}

\subsection{Online Coresets for \texorpdfstring{$\ell_\infty$}{l inf} Subspace Sketch: \texorpdfstring{$\poly(d)$}{poly(d)} Distortion, \texorpdfstring{$\poly(d)$}{poly(d)} Space}

We now show our streaming subspace sketch algorithm for $\ell_\infty$.

\subsubsection{Algorithm for \texorpdfstring{$\ell_\infty$}{l inf}}

\begin{algorithm}
	\caption{Online $\ell_\infty$ Subspace Sketch Coreset}
	\textbf{input:} $\bfA\in\mathbb Z^{n\times d}$. \\
	\textbf{output:} Coreset $S\subseteq[n]$.
	\begin{algorithmic}[1] 
        \State $S\gets \varnothing$
        \For{$i\in[n]$}
            \If{$\exists\bfx\in\mathbb R^n$ : $\angle*{\bfa_i,\bfx}^2 \geq \norm*{\bfA\vert_S\bfx}_2^2$}\label{line:sensitivity-add-point}
                \State $S \gets S\cup\{i\}$
            \EndIf
        \EndFor
        \State \Return $S$
	\end{algorithmic}\label{alg:subspace-sketch-stream-l-inf}
\end{algorithm}

We show the following guarantee for Algorithm \ref{alg:subspace-sketch-stream-l-inf}:

\begin{Theorem}\label{thm:subspace-sketch-stream-l-inf}
    Let $\bfA\in\mathbb R^{n\times d}$ such that for any subset $S'\subseteq[n]$, the sum of online leverage scores is bounded by 
    \[
        \sum_{i\in S'} \bftau_i^{\mathsf{OL}}(\bfA\vert_{S'}) \leq T
    \]
    and let $S$ be the output of Algorithm \ref{alg:subspace-sketch-stream-l-inf}. Then:
    \begin{itemize}
        \item $\abs*{S} \leq O(T)$
        \item $\frac1\Delta \norm*{\bfA\bfx}_\infty \leq \norm*{\bfA\vert_S\bfx}_\infty \leq \norm*{\bfA\bfx}_\infty$ for all $\bfx\in\mathbb R^d$, for $\Delta = O(\sqrt{T})$. 
    \end{itemize}
    In particular, if $\bfA\in\mathbb Z^{n\times d}$ is an integer matrix with entries bounded by $\poly(d)$, then by storing the rows of $S$, we obtain an algorithm for the streaming $\ell_\infty$ subspace sketch problem using $O(d^2\log^2 n)$ bits of space, and if $\bfA\in\mathbb R^{n\times d}$ has online pseudo condition number $\kappa^\OL$, then we obtain an online coreset algorithm storing at most $O(d\log(n\kappa^\OL))$ rows and achieves distortion at most $O(\sqrt{d\log(n\kappa^\OL}))$. 
\end{Theorem}
\begin{proof}
We first bound $\abs*{S}$. Note that for every $i\in S$,
\[
    \bftau_i^{\mathsf{OL}}(\bfA\vert_S) = \Omega(1)
\]
since if $\bfa_i\in\rowspan((\bfA\vert_S)_{i-1})$, then by line \ref{line:sensitivity-add-point} and Lemma \ref{lem:generalized-sensitivity},
\[
    \bftau_i^{\mathsf{OL}}(\bfA\vert_S) = \sup_{\bfx\in\rowspan((\bfA\vert_S)_{i-1})\setminus\{0\}} \frac{\angle*{\bfa_i,\bfx}^2}{\norm*{(\bfA\vert_S)_{i-1}\bfx}_2^2} \geq 1
\]
while if $\bfa_i\notin\rowspan((\bfA\vert_S)_{i-1})$, then $\bftau_i^{\mathsf{OL}}(\bfA\vert_S) = 1$. Since the online leverage scores of $\bfA_S$ sum to at most $T$, it follows that $\abs*{S} \leq O(T)$.

Next, we bound the distortion $\Delta$. Note that $\norm*{\bfA\vert_S\bfx}_\infty \leq \norm*{\bfA\bfx}_\infty$ is trivial, so it suffices to show the lower bound. Let $\bfx\in\mathbb R^d$ and let $i_*\in[n]$ satisfy $\norm*{\bfA\bfx}_\infty = \abs*{\angle*{\bfa_{i_*},\bfx}}$, i.e., the row that witnesses the max. If $i_*\in S$, then we already have that
\[
    \norm*{\bfA\vert_S\bfx}_\infty \geq \norm*{\bfA\bfx}_\infty
\]
so assume that $i_*\notin S$. Then,
\begin{align*}
    \norm*{\bfA\bfx}_\infty^2 &= \angle*{\bfa_{i_*},\bfx}^2 \\
    &\leq \norm*{\bfA\vert_S\bfx}_2^2 && \text{Line \ref{line:sensitivity-add-point}} \\
    &\leq \abs*{S}\cdot \norm*{\bfA\vert_S\bfx}_\infty^2 \\
    &\leq O(T)\norm*{\bfA\vert_S\bfx}_\infty^2
\end{align*}
which yields the claimed bound on $\Delta$. The guarantee for streaming algorithms for $\bfA$ with bounded bit complexity follow from online leverage score bound from Theorem \ref{thm:sharp-online-leverage-scores}. The guarantee for online coreset algorithms follows from Lemma \ref{lem:online-leverage-score-bound} and by noting that 
\[
    \kappa^\OL(\bfA_S) \leq n\cdot\kappa^\OL(\bfA)
\]
since for any $i\in[n]$,
\begin{align*}
    \norm*{(\bfA_S)_i^-}_2^{-1} &= \min_{\norm*{\bfx}_2 = 1, \bfx\in\rowspan((\bfA\vert_S)_i)}\norm*{(\bfA\vert_S)_i\bfx}_2 \\
    &\geq \min_{\norm*{\bfx}_2 = 1, \bfx\in\rowspan((\bfA\vert_S)_i)}\norm*{(\bfA\vert_S)_i\bfx}_\infty \\
    &\geq \frac1{\Delta}\min_{\norm*{\bfx}_2 = 1, \bfx\in\rowspan((\bfA\vert_S)_i)}\norm*{\bfA_i\bfx}_\infty \\
    &\geq \frac1{\Delta\sqrt n}\min_{\norm*{\bfx}_2 = 1, \bfx\in\rowspan((\bfA\vert_S)_i)}\norm*{\bfA_i\bfx}_2 \geq \frac1{n}\norm*{\bfA_i^-}_2^{-1}.
\end{align*}
Note that we use that $S$ has the $\ell_\infty$ subspace embedding guarantee here, rather than using an upper bound on $T$.
\end{proof}

\begin{Remark}\label{rem:ellipsoid}
It is not hard to see that $\norm*{\bfA\vert_S\bfx}_2$ can be used as the subspace sketch estimator in Theorem \ref{thm:subspace-sketch-stream-l-inf} instead of $\norm*{\bfA\vert_S\bfx}_\infty$. In this case, one can obtain a space complexity of $O(d^2\log n)$ bits of space instead of $O(d^2\log^2 n)$, by storing the quadratic form $\bfA\vert_S^\top\bfA\vert_S\in\mathbb Z^{d\times d}$. In particular, we obtain an ellipsoid $\braces*{\bfx\in\mathbb R^d : \norm*{\bfA\vert_S\bfx}_2 \leq 1}$ approximating the polytope $\braces*{\bfx\in\mathbb R^d : \norm*{\bfA\bfx}_\infty \leq 1}$ up to a factor of $O(\sqrt{d\log n})$. If we instead just store the rows themselves in the online coreset model, we store $O(d\log(n\kappa^\OL))$ rows for a distortion of $O(\sqrt{d\log(n\kappa^\OL)})$ betwen the polytope and ellipsoid. See also Theorem \ref{thm:john-ellipsoid-streaming-ub}.
\end{Remark}

\subsubsection{Nearly Input Sparsity Time}\label{sec:streaming-subspace-sketch-opt-var}

Here, we show how to process the rows of $\bfA$ in batches to make Algorithm \ref{alg:subspace-sketch-stream-l-inf} run in input sparsity time. We first use the observation of \cite{CMP2020,BDMMUWZ2020} that batch processing of rows can only shrink the online leverage scores, since for a batch $B\subseteq[n]$ and a row $j\in B$,
\[
    \bftau_j^{\mathsf{OL}}(\bfA) = \sup_{\bfx}\frac{\angle*{\bfa_j,\bfx}^2}{\norm*{\bfA_{j-1}\bfx}_2^2} \geq \sup_{\bfx}\frac{\angle*{\bfa_j,\bfx}^2}{\norm*{\bfA_{B}\bfx}_2^2} = \bftau_j(\bfA_B).
\]
To determine the appropriate batch size, note that the size of the coreset $S$ is still at most $\abs*{S} \leq O(d\log n)$, so we only ever compute leverage scores of a matrix with $B + O(d\log n)$ rows. By Theorem \ref{thm:cohen-peng-fast-lewis-weights}, this requires time at most
\[
    \tilde O((\nnz(\bfA\vert_{B}) + d^2\log n) + d^\omega)
\]
Thus, if the batch size is $O(d^\omega\log n)$ rows, then the running time is at most $\tilde O(\nnz(\bfA))$.

\subsubsection{Online Coreset Lower Bounds}

We now show a lower bound for the number of rows stored by an online coreset algorithm for the $\ell_\infty$ subspace sketch problem. Our reasoning is similar to the online coreset lower bound in Section 5 of \cite{CMP2020}. 

\begin{Theorem}[$\ell_\infty$ Online Coreset Lower Bound]\label{thm:online-coreset-lb-l-inf}
    Let $\bfA\in\mathbb R^{n\times d}$. Suppose an algorithm $\mathcal A$ selects rows $S\subseteq[n]$ and corresponding weights $\bfw\in\mathbb R^S$ in an online manner such that
    \[
        \mbox{for all $\bfx\in\mathbb R^d$, }\qquad\norm*{\bfA\bfx}_\infty \leq \norm*{\diag(\bfw)\bfA\vert_S\bfx}_\infty \leq \Delta\norm*{\bfA\bfx}_\infty
    \]
    with probability at least $1/2$. Then, $\mathcal A$ must select at least $\Omega(d\log_\Delta\kappa^\OL)$ in expectation. 
\end{Theorem}
\begin{proof}
For any large $M$, we construct a distribution on inputs $\bfA$ with $\kappa^\OL(\bfA) \leq O(M)$ for which any deterministic online row selection algorithm $\mathcal A$ that succeeds with probability at least $1/2$ must output $\Omega(d\log_\Delta\kappa^\OL)$ rows in expectation. This yields the conclusion by Yao's minimax principle.

Our input distribution over $n\times(d+1)$ matrices is as follows. We select an integer $N$ uniformly at random from $[\ceil*{\log_{2\Delta} M}]$. We always stream a copy of the $(d+1)\times (d+1)$ identity matrix. We then stream the rows of $N$ copies of the $d\times d$ identity matrix concatenated with an all zeros column, where the $i$th copy of the identity matrix gets scaled by $(2\Delta)^i$. Note that for this $\bfA$, the condition number is at most $O(N) \leq O(M)$. Note that the deterministic algorithm must be correct for at least $1/2$ of the $\ceil*{\log_{2\Delta} M}$ choices of $N$.

Let $i$ be the lowest choice of $N$ for which $\mathcal A$ is correct. Note that $\mathcal A$ must output at least $d$ rows even to preserve the rank. Let $\bfA_i$ denote the concatenation of the first $i$ scaled identities, and let the weighted coreset output be $\tilde\bfA_i$. We then have that
\begin{equation}\label{eq:alg-correct-l-inf}
    \mbox{for all $\bfx\in\mathbb R^d$, }\qquad\norm*{\bfA_i\bfx}_\infty \leq \norm{\tilde\bfA_i\bfx}_\infty \leq \Delta\norm*{\bfA_i\bfx}_\infty
\end{equation}
Now let $j$ be the second lowest choice of $N$ for which $\mathcal A$ is correct. On this input, note that $\mathcal A$ has already kept $d$ rows from being correct on the first $i$ copies of the scaled identities, which satisfies Equation \eqref{eq:alg-correct-l-inf}. Now between, suppose for contradiction that $\mathcal A$ fails to store at least $d$ rows between inputs $i$ and $j$. Then, for some column, say the first column, $\mathcal A$ fails to store any row with a nonzero entry in that column. Thus,
\[
    \norm{\tilde\bfA_j\bfe_1}_\infty = \norm{\tilde\bfA_i\bfe_1}_\infty \leq \Delta \norm*{\bfA_i\bfe_1}_\infty = (2\Delta)^i \Delta < (2\Delta)^j = \norm*{\bfA_j\bfe_1}_\infty
\]
which contradicts that $\mathcal A$ is correct. Thus, $\mathcal A$ must keep at least $d$ rows between inputs $i$ and $j$. Iterating this argument over all the $\ceil*{\log_{2\Delta}M} / 2$ inputs on which the algorithm is correct, it must select a total of $\Omega(d\log_{2\Delta}M)$ rows in expectation over all inputs.
\end{proof}

\begin{Remark}
Note that scaling the instance down by $\ceil*{\log_{2\Delta}M}$ does not affect the hardness. Thus, the hardness holds even when the entries are required to be bounded by $O(1)$. Furthermore, by adding the all ones vector to every column after  scaling down by $\ceil*{\log_{2\Delta}M}$, and by appending a single column of all ones, we contain the original instance as a subspace, and we have the property that all entries are in $[1, 2]$. Thus, the hardness holds also holds even when the maximum and minimum nonzero entries have ratio bounded by $O(1)$.
\end{Remark}

\subsection{One Pass Streaming \texorpdfstring{$\ell_p$}{lp} Subspace Sketch: \texorpdfstring{$\poly(d)$}{poly(d)} Distortion, \texorpdfstring{$\poly(d)$}{poly(d)} Space}

Our main result for this section is an algorithm for computing a $\poly(d)$-distortion subspace sketch data structure for $\bfA\in\mathbb R^{n\times d}$ using $\poly(d)$ space, with one pass over a row arrival stream. Here, our space distortion bound is a polynomial with degree independent of $p$.

We recall the definition of $\ell_p$ sensitivity of a vector $\bfa$ with respect to any other matrix $\bfA$, which is similar to the generalized leverage scores defined in \cite{CLMMPS2015}:

\begin{Definition}
    Let $\bfA\in\mathbb R^{n\times d}$ and $\bfa\in\mathbb R^d$. Let $0 <  p < \infty$. Then, the $\ell_p$ sensitivity of $\bfa$ with respect to $\bfA$ is defined as 
    \[
        \bfs^{p,\bfA}(\bfa) \coloneqq \begin{cases}
            \min\braces*{\sup_{\bfx\in\rowspan(\bfA)\setminus\{0\}}\frac{\abs*{\angle*{\bfa,\bfx}}^p}{\norm*{\bfA\bfx}_p^p}, 1} & \bfa \in \rowspan(\bfA) \\
            1 & \text{otherwise}
        \end{cases}
    \]
\end{Definition}

We first show a simple version of our algorithm, and then later present a full set of trade-offs between distortion and space, which is nearly optimal. 

\begin{algorithm}
	\caption{Streaming $\ell_p$ Subspace Sketch}
	\textbf{input:} $\bfA\in\mathbb Z^{n\times d}$, $2 < p < \infty$. \\
	\textbf{output:} Quadratic form $\bfQ\in\mathbb R^{d\times d}$
	\begin{algorithmic}[1] 
        \State $\bfQ = 0$
        \For{$i\in[n]$}
            \State $s\gets\max\{\bfs^{2,\bfQ^{1/2}}(\bfa_i), 1/n\}$ \Comment{Thresholded $\ell_2$ sensitivity of $\bfa_i$ with respect to $\bfQ^{1/2}$}
            \State $w\gets s^{p/4-1/2}$ 
            \State $w'\gets 2^{\ceil{\log_2(w)}}$ \Comment{Round to a power of $2$}
            \State $\bfQ\gets \bfQ + (w'\cdot \bfa_i)(w'\cdot \bfa_i)^\top$
        \EndFor
        Return $\bfQ$
	\end{algorithmic}\label{alg:subspace-sketch-stream-l2}
\end{algorithm}

\begin{Theorem}\label{thm:subspace-sketch-stream-l2}
    Let $\bfA\in\mathbb Z^{n\times d}$ be an integer matrix with entries bounded by $\poly(n)$. Let $2 < p < \infty$. Let $\bfQ\in\mathbb R^{d\times d}$ be the output of Algorithm \ref{alg:subspace-sketch-stream-l2}. Then, for all $\bfx\in\mathbb R^d$,
    \[
        \norm*{\bfA\bfx}_p \leq \bfx^\top\bfQ\bfx \leq O((d\log n)^{1/2-1/p})\norm*{\bfA\bfx}_p.
    \]
    Furthermore, Algorithm \ref{alg:subspace-sketch-stream-l2} uses at most $O(d^2\log n)$ bits of space.
\end{Theorem}
\begin{proof}
    Let $\bfQ_i$ denote the state of $\bfQ$ after the addition of the $i$th row. We will inductively show that $\norm*{\bfQ_i^{1/2}\bfx}_2^p \geq \norm*{\bfA_i\bfx}_p^p$. For $i = 1$, $\bfQ_1^{1/2}$ is just $\bfa_1$ so the result follows by monotonicity of $\ell_p$ norms. Now note that for any $\bfx\in\mathbb R^d$,
    \begin{align*}
        \norm*{\bfQ_{i+1}^{1/2}\bfx}_2^p &= \bracks*{\norm*{\bfQ_i^{1/2}\bfx}_2^2 + \angle*{w'\bfa_{i+1},\bfx}^2}^{p/2} \\
        &= \norm*{\bfQ_i^{1/2}\bfx}_2^p\bracks*{1 + w'^2\frac{\angle*{\bfa_{i+1},\bfx}^2}{\norm*{\bfQ_i^{1/2}\bfx}_2^2}}^{p/2} \\
        &\geq \norm*{\bfQ_i^{1/2}\bfx}_2^p\bracks*{1 + \bfs^{2,\bfQ_i^{1/2}}(\bfa_{i+1})^{p/2-1}\frac{\angle*{\bfa_{i+1},\bfx}^2}{\norm*{\bfQ_i^{1/2}\bfx}_2^2}} \\
        &\geq \norm*{\bfQ_i^{1/2}\bfx}_2^p\bracks*{1 + \frac{\abs*{\angle*{\bfa_{i+1},\bfx}}^p}{\norm*{\bfQ_i^{1/2}\bfx}_2^p}} \\
        &= \norm*{\bfQ_i^{1/2}\bfx}_2^p + \abs*{\angle*{\bfa_{i+1},\bfx}}^p
    \end{align*}
    Inductively, $\norm*{\bfQ\bfx}_2^p \geq \norm*{\bfA_i\bfx}_p^p$, which establishes the lower bound on $\norm*{\bfQ\bfx}_2$.

    For the upper bound, let $\bfw$ denote the vector with the $w'$ for the $i$th row in the $i$th entry, and let $\bfW = \diag(\bfw)$. Then for any $\bfx\in\mathbb R^d$,
    \begin{equation}\label{eq:qx-upper-bound}
    \begin{aligned}
        \norm*{\bfQ_i^{1/2}\bfx}_2^2 &= \norm*{\bfW\bfA\bfx}_2^2 \\
        &= \sum_{i=1}^n \bfw_i^2 \angle*{\bfa_i,\bfx}^2 \\
        &\leq 2\sum_{i=1}^n \parens*{\max\braces*{\bfs^{2,\bfQ_{i-1}^{1/2}}(\bfa_i),\frac1n}}^{p/2-1} \angle*{\bfa_i,\bfx}^2 \\
        &\leq 2\bracks*{\sum_{i=1}^n \parens*{\max\braces*{\bfs^{2,\bfQ_{i-1}^{1/2}}(\bfa_i),\frac1n}}^{(p/2-1)/(1-2/p)}}^{1-2/p} \bracks*{\sum_{i=1}^n \abs*{\angle*{\bfa_i,\bfx}}^p}^{2/p} && \text{H\"older's inequality} \\
        &= 2\bracks*{\sum_{i=1}^n\parens*{\max\braces*{\bfs^{2,\bfQ_{i-1}^{1/2}}(\bfa_i),\frac1n}}^{p/2}}^{1-2/p} \norm*{\bfA\bfx}_p^2 \\
        &\leq 2\bracks*{1+\sum_{i=1}^n \bfs^{2,\bfQ_{i-1}^{1/2}}(\bfa_i)^{p/2}}^{1-2/p} \norm*{\bfA\bfx}_p^2
    \end{aligned}
    \end{equation}
    Now note that if $\bfa_i\in \rowspan(\bfA_{i-1})$, then
    \begin{align*}
        \bftau_i^{\OL}(\bfW\bfA) &= \sup_{\bfx\in\rowspan([\bfW\bfA]_{i-1})} \frac{\angle*{\bfw_i\bfa_i,\bfx}^2}{\norm*{[\bfW\bfA]_{i-1}\bfx}_2^2} \\
        &= \bfw_i^2 \sup_{\bfx\in\rowspan([\bfW\bfA]_{i-1})} \frac{\angle*{\bfa_i,\bfx}^2}{\norm*{[\bfW\bfA]_{i-1}\bfx}_2^2} \\
        &= \bfw_i^2 \sup_{\bfx\in\rowspan(\bfQ_{i-1}^{1/2})} \frac{\angle*{\bfa_i,\bfx}^2}{\norm*{\bfQ_{i-1}^{1/2}\bfx}_2^2} \\
        &= \bfs^{2,\bfQ_{i-1}^{1/2}}(\bfa_i)^{p/2 - 1}\cdot \bfs^{2,\bfQ_{i-1}^{1/2}}(\bfa_i) \\
        &= \bfs^{2,\bfQ_{i-1}^{1/2}}(\bfa_i)^{p/2}
    \end{align*}
    and otherwise, $\bftau_i^{\OL}(\bfW\bfA) = \bfs^{2,\bfQ_{i-1}^{1/2}}(\bfa_i)^{p/2} = 1$. Now note that $\bfW\bfA$ can be multiplied by some power of $2$ of order $\poly(n)$ to obtain an integer matrix, since the $\bfw_i$ are powers of $2$ of size at least $\Omega(1/n)$. The sum of online leverage scores for $\bfW\bfA$ coincide with this scaled up integer matrix with entries bounded by $\poly(n)$, so we have that
    \[
        \sum_{i=1}^n \bfs^{2,\bfQ_{i-1}^{1/2}}(\bfa_i)^{p/2} \leq \sum_{i=1}^n \bftau_i^{\OL}(\bfW\bfA) = O(d\log n)
    \]
    by Theorem \ref{thm:sharp-online-leverage-scores}. Combining the above with Equation \eqref{eq:qx-upper-bound} completes the proof. 
\end{proof}

By replacing the use of $\ell_2$ sensitivities with $\ell_q$ sensitivities for $2 < q < p$, we can obtain the following trade-off between the distortion and the space complexity, at the cost of making the algorithm randomized and exponential time. This trade-off is nearly optimal up to $\poly\log n$ factors, as discussed in Section \ref{sec:subspace-sketch}, even for the offline subspace sketch problem.

\begin{algorithm}
	\caption{Streaming $\ell_p$ Subspace Sketch Trade-off}
	\textbf{input:} $\bfA\in\mathbb Z^{n\times d}$, $2 < q < p < \infty$. \\
	\textbf{output:} Subspace sketch data structure $Q$.
	\begin{algorithmic}[1] 
        \State Initialize an $O(1)$-approximate streaming $\ell_q$ subspace sketch data structure $Q$ (Theorem \ref{thm:1+eps-merge-reduce-subspace-sketch})
        \For{$i\in[n]$}
            \State Approximate $\bfs^{q,[\bfW\bfA]_{i-1}}(\bfa_i)$ up to a $\Theta(1)$ factor using $Q$ \Comment{Requires exponential time}
            \State $s\gets \Theta(1)\max\{\bfs^{q,[\bfW\bfA]_{i-1}}(\bfa_i), 1/n\}$ \Comment{Thresholded $\ell_q$ sensitivity of $\bfa_i$ with respect to $Q$}
            \State $w\gets s^{(p/q-1)/q}$ 
            \State $w'\gets 2^{\ceil{\log_2(w)}}$ \Comment{Round to a power of $2$}
            \State Add row $w'\cdot \bfa_i$ to $Q$
        \EndFor
        Return $Q$
	\end{algorithmic}\label{alg:subspace-sketch-stream-lq}
\end{algorithm}

\begin{Theorem}\label{thm:subspace-sketch-stream-lq}
    Let $\bfA\in\mathbb Z^{n\times d}$ be an integer matrix with entries bounded by $\poly(n)$. Let $2 < q < p < \infty$. Let $\bfQ\in\mathbb R^{d\times d}$ be the output of Algorithm \ref{alg:subspace-sketch-stream-lq}. Then, over the randomness of $Q$,
    \[
        \Pr\braces*{\forall\bfx\in\mathbb R^d, \norm*{\bfA\bfx}_p \leq Q(\bfx) \leq O(d^{\frac12\parens*{1-\frac{q}{p}}}\log n)\norm*{\bfA\bfx}_p} \geq \frac23.
    \]
    Furthermore, Algorithm \ref{alg:subspace-sketch-stream-lq} uses at most $O(d^{q/2+1}\poly\log n)$ bits of space.
\end{Theorem}
\begin{proof}
    We describe our proof more briefly than Theorem \ref{thm:subspace-sketch-stream-l2}, since many of the calculations are similar. Let $\bfw\in\mathbb R^n$ be the vector with $w'$ after the addition of row $\bfa_i$ in its $i$th entry and let $\bfW = \diag(\bfw)$. Then,
    \[
        \norm*{[\bfW\bfA]_{i+1}\bfx}_q^p \geq \norm*{[\bfW\bfA]_{i}\bfx}_q^p + \abs*{\angle*{\bfa_{i+1},\bfx}}^p
    \]
    for all $\bfx\in\mathbb R^d$, which inductively implies that $\norm*{\bfA\bfx}_p \leq \norm*{\bfW\bfA\bfx}_q$ for all $\bfx\in\mathbb R^d$. We also have that
    \begin{align*}
        \norm*{\bfW\bfA}_q^q &\leq \Theta(1)\bracks*{\sum_{i=1}^n \parens*{\max\braces*{\bfs^{q,[\bfW\bfA]_{i-1}}(\bfa_i), \frac1n}}^{(p/q-1)/(1-q/p)}}^{1-q/p}\bracks*{\sum_{i=1}^n\abs*{\angle*{\bfa_i,\bfx}}^p}^{q/p} \\
        &= \Theta(1)\bracks*{\sum_{i=1}^n \bfs^{q,[\bfW\bfA]_{i-1}}(\bfa_i)^{p/q}}^{1-q/p}\norm*{\bfA\bfx}_p^q \\
        &= \Theta(1)\bracks*{\sum_{i=1}^n \bfs_i^{q,\OL}(\bfW\bfA)}^{1-q/p}\norm*{\bfA\bfx}_p^q.
    \end{align*}
    Then by Theorem \ref{thm:sum-online-lp-sensitivity}, we have that
    \[
        \norm*{\bfW\bfA}_q \leq O(d^{q/2}\log^{q/2} n)^{1/q-1/p}\norm*{\bfA\bfx}_q \leq O(d^{\frac12\parens*{1-\frac{q}{p}}}\log n)\norm*{\bfA\bfx}_q
    \]
    as claimed.
\end{proof}

\subsection{One Pass Streaming Subspace Sketch: \texorpdfstring{$(1+\eps)$}{(1+eps)} Distortion, \texorpdfstring{$d^{p/2+1}$}{d**(p/2+1)} Space}

We show the simple result of obtaining $(1+\eps)$ distortion in one pass via the merge and reduce framework (see, e.g., \cite{CDW2018,BDMMUWZ2020}).

\begin{Theorem}\label{thm:1+eps-merge-reduce-subspace-sketch}
    Let $\bfA\in\mathbb R^{n\times d}$, $0 < p < \infty$, and $\eps>0$. Then, there is a one pass streaming algorithm which uses $O(d^{2\lor(p/2+1)}\poly(\log n,\eps^{-1}))$ space and outputs a data structure $Q_p$ such that
    \[
        \Pr\braces*{\forall\bfx\in\mathbb R^d, \norm*{\bfA\bfx}_p \leq Q_p(\bfx) \leq (1+\eps)\norm*{\bfA\bfx}_p} \geq \frac23
    \]
\end{Theorem}
\begin{proof}
    We partition the $n$ rows of $\bfA$ into blocks of $B \coloneqq O(d^{1\lor (p/2)}\poly(\log n,\eps^{-1}))$ rows, and then merge them in a binary tree-like fashion, with depth $O(\log n)$. To merge two submatrices $\bfA_1$ and $\bfA_2$ each with $B$ rows, we apply Theorem \ref{thm:lewis-weight-subspace-embedding} with accuracy parameter set to $\eps/\log n$ and failure rate parameter set to $1/\poly(n)$ on the $2B\times d$ matrix $\bfB \coloneqq [\bfA_1^\top\ \bfA_2^\top]^\top$ to get back a $B\times d$ matrix $\bfB'$ such that
    \[
        \Pr\braces*{\forall\bfx\in\mathbb R^d, \norm*{\bfB\bfx}_p \leq \norm*{\bfB'\bfx}_p \leq \parens*{1+\frac{\eps}{\log n}}\norm*{\bfB\bfx}_p} \geq 1 - \frac1{\poly(n)}
    \]
    By a union bound over all $O(n)$ merge operations required to combine all $O(n)$ blocks, the above guarantee holds every time we merge two blocks. Furthermore, because the depth of the binary tree is at most $\floor{\log_2 n}$, the distortion is at most $1+\eps$.

    Finally, to argue the space bound, we inductively argue that the above merging scheme can be implemented by storing at most $O(\log n)$ blocks of rows at a time. Assume first that $n$ is a power of $2$. Then, for $n = 2$, we clearly store at most $\log_2 n + 1 = 2$ blocks at once. Now to merge $n$ blocks, we use at most $\log_2(n/2) + 1 = \log_2 n$ at once to get a single block merging the first $n/2$ blocks. We can then do this again to obtain a single block merging the next $n/2$ blocks. This used at most $1 + \log_2 n$ blocks at once. Padding proves the result for general $n$. Then, storing $O(\log n)$ blocks of size $B\times d$ requires $O(Bd\log n)$ space, which is the claimed space complexity.
\end{proof}

\subsection{Near-Optimal Bounds for Restricted Instances}

In this section, we study a restricted variant of the $\ell_\infty$ subspace sketch problem, and give near optimal algorithms and lower bounds, i.e., without extra $\log n$ factors. 

\begin{Definition}[Restricted $\ell_\infty$ Subspace Sketch]\label{def:restricted-subspace-sketch}
We define the restricted $\ell_\infty$ subspace sketch problem as follows. Let $\bfA\in\mathbb R^{n\times d}$ be a matrix with row norms all $\Theta(1)$. Then, we must design a data structure $Q$ that receives a row arrival stream of $\bfA$ and answers queries $\bfx\in\mathbb R^d$ at the end of the stream. Furthermore, we must output
\[
    Q(\bfx) \leq \kappa\norm*{\bfA\bfx}_\infty
\]
for all $\bfx\in\mathbb R^d$, while we must output
\[
    Q(\bfx) \geq \norm*{\bfA\bfx}_\infty
\]
when $\bfx$ is an input point, i.e., $\bfx$ is one of the rows $\bfa_i$ of $\bfA$.
\end{Definition}

We show that the lower bound instance of \cite{LWW2021} is captured by this restriction, and show an algorithm that matches the lower bound up to logarithmic factors.

\subsubsection{Lower Bound}

We will use the following lemma from coding theory, which has previously been used several times in streaming and communication lower bounds for linear algebraic problems \cite{LWW2021, MMWY2021}. 
\begin{Lemma}[\cite{PTB2013}]\label{lem:coding-random-points}
For any $p\geq 1$ and $d = 2^k - 1$ for some integer $k$, there exists a set $S\subseteq\{-1,1\}^d$ and a constant $C_p$ depending only on $p$ which satisfy
\begin{itemize}
    \item $\abs*{S} = d^p$
    \item For any $s,t\in S$ such that $s\neq t$, $\abs*{\angle*{s,t}} \leq C_p\sqrt d$
\end{itemize}
\end{Lemma}

We then have the following:

\begin{Theorem}\label{thm:restricted-subspace-sketch-lb}
    Let $n = d^q$ for some integer $q$. Suppose that a streaming algorithm $\mathcal A$ solves the restricted $\ell_\infty$ subspace sketch problem (Definition \ref{def:restricted-subspace-sketch}) with $\kappa = c\sqrt d$ for some sufficiently small constant $c>0$. Then, $\mathcal A$ must use $\Omega(n)$ bits of space.
\end{Theorem}
\begin{proof}
    We show the result by reduction from the \textsf{INDEX} problem (Theorem \ref{thm:index}). 

    Let $S\subseteq\{-1,1\}^d$ be the set of vectors given by Lemma \ref{lem:coding-random-points} with $n = d^q$. Suppose that Alice has a subset $A\subseteq[n]$. Then, Alice can feed the vectors of $S$ corresponding to her subset $A$, normalized to have norm $\Theta(1)$, and then pass the memory state of $\mathcal A$ to Bob. Now suppose that Bob has the index $b\in [n]$. Then, Bob queries the subspace sketch data structure the vector $\bfx_b\in S$ corresponding to the index $b$.
    
    If $b\in A$, then we have that $Q(\bfx_b) \geq \norm*{\bfA\bfx_b}_\infty = \norm*{\bfx_b}_2^2 = \Theta(1)$. On the other hand, if $b\notin A$, then we have that
    \[
        Q(\bfx_b) \leq \kappa \norm*{\bfA\bfx}_\infty \leq c\sqrt d \cdot \frac1{\Theta(\sqrt d)} = \Theta(c).
    \]
    Thus for $c$ sufficiently small, Bob can distinguish whether $b\in A$ or not and thus $\mathcal A$ must use at least $\Omega(n)$ bits of space.
\end{proof}

\begin{Remark}
By replacing our use of Lemma \ref{lem:coding-random-points} with $n$ random unit vectors in $d$ dimensions, we can instead get a collection of vectors with inner product $\Theta(\sqrt{(\log n)/d})$, which leads to an $\Omega(n)$ bit lower bound for distortions better than $O(\sqrt{d/\log n})$, even for $n$ larger than $\poly(d)$.
\end{Remark}

\subsubsection{Upper Bound}

In this section, we design an algorithm solving the restricted $\ell_\infty$ subspace sketch problem (Definition \ref{def:restricted-subspace-sketch}). Our algorithm is given in Algorithm \ref{alg:l-inf-input-query-unit}.

\begin{algorithm}
	\caption{Restricted $\ell_\infty$ Subspace Sketch}
	\textbf{input:} $\bfA\in\mathbb R^{n\times d}$ in a row arrival stream of $\Theta(1)$ norm rows. \\
	\textbf{output:} Coreset $S\subseteq[n]$.
	\begin{algorithmic}[1] 
        \State $S\gets\varnothing$
        \For{$i\in[n]$}
            \If{there is no $j\in S$ s.t. $\abs*{\angle*{\bfa_j/\norm*{\bfa_j}_2,
            \bfa_i/\norm*{\bfa_i}_2}} \geq 1/\sqrt{2d-1}$}\label{line:keep-row}
                \State $S\gets S \cup \{i\}$
            \EndIf
        \EndFor
	\end{algorithmic}\label{alg:l-inf-input-query-unit}
\end{algorithm}

Our analysis will use the well-known Welch bound from coding theory.

\begin{Theorem}[Inner Product Lower Bound \cite{Wel1974}]\label{thm:inner-product-lb}
    Let $\bfa_1, \bfa_2, \dots, \bfa_M \in\mathbb R^d$ be a set of $M$ unit vectors. Let $k\geq 1$ be an integer. Then,
    \[
        \max_{i\neq j}\abs*{\angle*{\bfa_i,\bfa_j}}^{2k} \geq \frac1{M-1}\bracks*{\frac{M}{\binom{d+k-1}{k}} - 1}
    \]
\end{Theorem}

Using Theorem \ref{thm:inner-product-lb}, we show the following.

\begin{Theorem}\label{thm:restricted-subspace-sketch-alg}
    Let $\bfA\in\mathbb R^{n\times d}$ be a matrix with rows with norm $\Theta(1)$. Then, Algorithm \ref{alg:l-inf-input-query-unit} outputs a coreset $S\subseteq[n]$ such that
    \[
        C\sqrt d\norm*{\bfA_S\bfa_i}_\infty \geq \norm*{\bfA\bfa_i}_\infty
    \]
    for all $i\in[n]$, for some $C>0$ a sufficiently large constant. Furthermore, Algorithm \ref{alg:l-inf-input-query-unit} uses $O(d^2\log n)$ bits of space.
\end{Theorem}

Before proving Theorem \ref{thm:restricted-subspace-sketch-alg}, note that the result implies that Algorithm \ref{alg:l-inf-input-query-unit} solves the restricted $\ell_\infty$ subspace sketch problem, since trivially, we have that
\[
    C\sqrt d\norm*{\bfA_S\bfx}_\infty \leq C\sqrt d\norm*{\bfA\bfx}_\infty
\]
for all $\bfx\in\mathbb R^d$. 

\begin{proof}[Proof of Theorem \ref{thm:restricted-subspace-sketch-alg}]
First note that by assuming that $\bfa_i$ are unit vectors, we only lose $\Theta(1)$ factors in the distortion parameter $\kappa$, so we make this assumption without loss of generality. 

Note that the correctness guarantee is trivial from the construction of the algorithm, since every input point that doesn't satisfy line \ref{line:keep-row} is kept by the coreset. It suffices to argue the space complexity of the algorithm. 

We will argue that the algorithm keeps at most $O(d)$ points in $S$. If we apply Theorem \ref{thm:inner-product-lb} with $k = 1$ and $M = 2d$, we get that for any set of $M = 2d$ unit vectors $\bfa_1, \bfa_2, \dots, \bfa_M \in\mathbb R^d$,
\[
    \max_{i\neq j}\abs*{\angle*{\bfa_i,\bfa_j}}^{2} \geq \frac1{2d-1}\bracks*{\frac{2d}{d} - 1} = \frac1{2d-1}
\]
and thus the algorithm cannot keep more than $2d-1$ points. Storing these points only requires $O(d^2\log n)$ bits of space.
\end{proof}
\section{Applications to Streaming Algorithms for Geometric Problems in High Dimensions}\label{sec:streaming-comp-geo}

We show that our techniques for the streaming subspace sketch problem yield the first one pass $\poly(d)$ space algorithms for a wide variety of geometric approximation problems that are symmetric with respect to the origin. For these problems, the previously known techniques typically only yielded space bounds of the form $\eps^{-\Theta(d)}$ for a $(1+\eps)$ approximation. In contrast, we show how to obtain $\poly(d)$ approximations using $\poly(d)$ bits of space.

\subsection{Directional Width}\label{sec:directional-height}

The most direct application of our results is that of approximating the \emph{directional height} oh a point set, which is a symmetric version of the more well-known \emph{directional width}:
\begin{Definition}[Directional width and height]
    Let $\bfA\in\mathbb R^{n\times d}$. The \emph{directional width} \cite{AHV2005} of $\bfA$ with respect to a unit vector $\bfx$ is defined to be
    \[
        \omega(\bfx,\bfA) = \max_{i\in[n]} \angle*{\bfx,\bfa_i} - \min_{i\in[n]} \angle*{\bfx,\bfa_i}
    \]
    and the \emph{directional height} \cite{IMGR2020, MRWZ2020} of $\bfA$ with respect to a unit vector $\bfx$ is defined to be
    \[
        h(\bfx,\bfA) = \max_{i\in[n]} \abs*{\angle*{\bfx,\bfa_i}}.
    \]
\end{Definition}

The definition of directional height is equivalent to an $\ell_\infty$ subspace sketch data structure, which means that Theorem \ref{thm:subspace-sketch-stream-l-inf} directly yields the result by providing a coreset result for the problem in the high-dimensional regime. Furthermore, Theorem \ref{thm:restricted-subspace-sketch-lb} improves the lower bound of \cite{AS2015} for directional width from $\Omega(d^{1/3})$ to $\Omega(d^{1/2})$. This in turn shows a lower bound of a $\Omega(d^{1/2})$ factor distortion for convex hull as well.

By using the ``peeling'' technique of \cite{AHY2008}, we extend this to \emph{$k$-robust directional width}. We define this for centrally symmetric instances as follows:

\begin{Definition}[Centrally Symmetric $k$-Robust Directional Width \cite{AHY2008}]\label{def:k-robust-width}
Let $\bfA\in\mathbb R^{n\times d}$ be a set of $n$ points in $d$ dimensions. We consider each row $\bfa_i\in\mathbb R^d$ as representing both $\bfa_i$ and $-\bfa_i$, so that the input instance is centrally symmetric. Define the \emph{level} of $\bfa\in\mathbb R^d$ in the direction $\bfx\in\mathbb R^d$ to be
\[
    \abs*{\braces*{i\in[n] : \abs*{\angle*{\bfa_i,\bfx}} > \angle*{\angle*{\bfa,\bfx}}}}
\]
and let $\bfA^\ell[\bfx]$ denote the point (or row) of $\bfA$ at level $\ell$\footnote{For simplicity, we assume that there is at most one vector at a given level, as done in \cite{AHY2008}.}. Then, the $k$-robust directional width is defined to be
\[
    \mathcal E_k(\bfx,\bfA) \coloneqq \abs*{\angle*{\bfA^k[\bfx],\bfx}}.
\]
\end{Definition}

We now turn to showing Theorem \ref{thm:k-robust-width}, which uses the reduction of \cite{AHY2008} to turn coresets for directional width for coresets for $k$-robust directional width, even in one-pass streams.

\kRobustWidth*
\begin{proof}
We follow the reduction described in \cite{AHY2008}. We first discuss an algorithm running in $k+1$ passes, and then describe how this can be implemented in one pass. In $k+1$ iterations, we consider a decreasing sequence of sets of rows $[n] = S_0 \supseteq S_1 \supseteq \ldots \supseteq S_k$, where $S_{i+1} = S_i \setminus \mathcal T_i$, where $\mathcal T_i\subseteq S_i$ is a coreset for directional width as constructed by our Theorem \ref{thm:subspace-sketch-stream-l-inf}. The coreset we output is then $\mathcal T \coloneqq \bigcup_{i=0}^k \mathcal T_i$.

We first argue correctness. Consider an arbitrary direction $\bfx\in\mathbb R^d$. Say that the $i$th iteration is \emph{successful} if $\bfA^j[\bfx]\in\mathcal T_i$ for some $j\in\{0,1,\dots,k\}$, and \emph{unsuccessful} otherwise. Now if $\bfA^j[\bfx]\in\mathcal T$ for every $j$, then we already have that $\mathcal E_k(\bfx,\bfA\vert_{\mathcal T_i}) \geq \mathcal E_k(\bfx,\bfA)$, so we assume that there exists some $j$ such that $\bfA^j[\bfx]\notin\mathcal T$. It then follows that $\bfA^j[\bfx]\notin\mathcal T_i$ for every iteration $i$. Then, let $i$ be any iteration in which the algorithm is unsuccessful in the direction $\bfx$. Then, 
\begin{align*}
    \mathcal E_0(\bfx,\bfA\vert_{\mathcal T_i}) &\geq \frac1\Delta \mathcal E_0(\bfx,\bfA\vert_{S_i}) && \text{since $\mathcal T_i$ is a coreset for $S_i$} \\
    &\geq \frac1\Delta \mathcal E_j(\bfx,\bfA) && \text{since $\bfA^j[\bfx]\in S_i$} \\
    &\geq \frac1\Delta \mathcal E_k(\bfx,\bfA)
\end{align*}
Furthermore, the $\bfA\vert_{\mathcal T_i}^0[\bfx]$ witnessing the above inequality is not one of the $\bfA^j[\bfx]$ of the entire dataset $\bfA$, since this iteration was unsuccessful. Thus, no matter whether the iteration is successful or unsuccessful, the final coreset $\mathcal T$ gains a vector $\bfa\in\mathbb R^d$ with $\abs*{\angle{\bfa,\bfx}} \geq \mathcal E_k(\bfx,\bfA) / \Delta$ in each iteration.

To turn this into a one-algorithm, we can follow Section 2.4 of \cite{AHY2008} and maintain $k+1$ copies of our coreset data structure in parallel, where the $i$th data structure gets inserted with a row $\bfa$ if either the $(i-1)$th copy of the algorithm does not add $\bfa$ to $\mathcal T_{i-1}$. Note that our base coreset algorithm does not delete points, so we do not need to handle this as \cite{AHY2008} does.
\end{proof}

\subsection{Volume Maximization}\label{sec:vol-max}

The work of \cite{MRWZ2020} shows how to convert coresets for directional height into coresets for volume maximization. Using this observation, we show the following result:

\VolMaxStream*
\begin{proof}
    The following is shown in \cite{MRWZ2020}:

    \begin{Lemma}[Lemmas 5.13, 5.14 of \cite{MRWZ2020}]
        Let $r = \Theta((\log n)/C)$. Let $\bfG\in\mathbb R^{d\times r}$ have each entry drawn i.i.d.\ from the Gaussian distribution $\mathcal N(0,1/r)$. Then:
        \begin{itemize}
            \item The volume of the optimal $k$-subset $\bfA\vert_{S_*}$ satisfies 
            \[
                \Pr\braces*{2^k \Vol(\bfA\bfG\vert_{S_*}) \geq \Vol(\bfA\vert_{S_*})} \geq \frac9{10}
            \]
            \item 
            \[
                \Pr\braces*{\forall S\in\binom{[n]}{k}, \Vol(\bfA\bfG\vert_{S}) \leq O(Ck)^{k/2}\Vol(\bfA\vert_{S})} \geq \frac{9}{10}
            \]
        \end{itemize}
    \end{Lemma}

    Thus, up to a $O(Ck)^{k/2}$ factor loss in the approximation factor, we may replace $\bfA$ by the $n\times r$ matrix $\bfA\bfG$. Now applying the observation of Section \ref{sec:directional-height}, we can obtain a directional height coreset for $\bfA\bfG$ from our Theorem \ref{thm:subspace-sketch-stream-l-inf}, which produces a set $T\subseteq[n]$ of size $\abs*{T} \leq O(r\log n)$ with distortion $\kappa = O(\sqrt{r\log n})$. Observation 5.9 of \cite{MRWZ2020} and Lemma 3.3 of \cite{IMGR2019} then shows that the maximum volume subset of the directional height coreset approximates the maximum volume subset of $\bfA$ up to a factor of $\kappa^{2k} = (\kappa^4)^{k/2}$. The total approximation factor is thus $O(\kappa^4 Ck)^{k/2}$, while the space complexity is $O(dr\log n + \abs*{T}d\log n) = O(\abs*{T}d\log n)$ for storing $\bfG$ and the coreset.
    
    If we only need to output the indices of the coreset, then we can first replace the Gaussian matrix $\bfG$ with a subspace embedding with a small seed as in, e.g., \cite{KMN2011}. Then by setting $\delta = \exp(-\Theta(k^2\log n))$, we have an $O(\log\frac1\delta)\times d$ matrix $\bfS$ such that, with probability at least $1 - n^{-k}$, for any fixed $d\times k$ matrix $\bfR$, $\norm*{\bfS\bfR\bfx}_2 = \Theta(1)\norm*{\bfR\bfx}_2$ for all $\bfx\in\mathbb R^k$, where $\bfS$ can be generated from a seed of length $\tilde O(\log k + \log\frac1\delta)$. In particular, $\norm*{\bfS\bfR}_2 = \Theta(1)\norm*{\bfR}_2$ under this event. We now consider any subset $S\in\binom{[n]}{k}$. Then by the same reasoning as in \cite{MRWZ2020}, we have that the volume spanned by $\bfA\vert_S = \bfU\bfSigma\bfV^\top$ written in the SVD is 
    \[
        \sqrt{\det(\bfA\vert_S\bfA\vert_S^\top)} = \sqrt{\det(\bfSigma^2)}
    \]
    while the volume of the embedded matrix $\bfA\vert_S\bfS^\top$ is at most
    \[
        \sqrt{\det(\bfA\vert_S\bfS^\top\bfS\bfA\vert_S^\top)} = \sqrt{\det(\bfSigma\bfV^\top\bfS^\top\bfS\bfV\bfSigma)} \leq \norm*{\bfS\bfV}_2^k \sqrt{\det(\bfSigma^2)}.
    \]
    Conditioned on the operator norm preservation of $\bfV$ by $\bfS$ for all $\binom{n}{k}$ subsets $S$, this is at most
    \[
        O(\norm*{\bfV}_2)^k \sqrt{\det(\bfSigma^2)} \leq \exp(O(k))\sqrt{\det(\bfSigma^2)}.
    \]
    The fact that the volume of the maximal volume subset $S_*$ does not shrink by more than $\exp(O(k))$ follows similarly as in \cite{MRWZ2020}. Then, we repeat the reasoning as before with $r = O(k^2\log n)$ on using the directional height coresets, so that our total space usage is just the seed length for the subspace embedding and the storage of the directional height coreset, which is $O(\abs*{T}\log n) = O(r\log^2 n) = O(k^2\log^3 n)$. The total distortion is $O(\sqrt{r\log n})^{2k} = O(k\log n)^k$.
\end{proof}

\subsection{Convex Hulls}\label{sec:convex-hull}

In this section, we note that our coreset results for the streaming $\ell_\infty$ subspace sketch problem in fact immediately yield streaming algorithms for computing coresets for convex hulls. This follows from the observation that for a matrix $\bfA\in\mathbb R^{n\times d}$ the \emph{polar body} of the polytope $\braces*{\bfx\in\mathbb R^d : \norm*{\bfA\bfx}_\infty \leq 1}$ is the symmetric convex hull $\conv(\braces*{\pm\bfa_1,\pm\bfa_2,\dots,\pm\bfa_n})$ of the rows. 

The following are standard elementary facts about polars that we will need:
\begin{Lemma}[Polars and their properties, Exercises 1.1.14, 2.3.2 of \cite{HW2020}, Section 3.5 of \cite{Tod2016}]\label{lem:polars}
    Let $K\subset\mathbb R^d$ be a convex body and define the polar $K^\circ$ as
    \[
        K^\circ \coloneqq \braces*{\bfx\in\mathbb R^d : \forall \bfx'\in \mathbb R^d, \angle*{\bfx,\bfx'} \leq 1}.
    \]
    Then, the following hold:
    \begin{itemize}
        \item if $K\subset L$, then $K^\circ \supset L^\circ$
        \item for $r>0$, $(r\cdot K)^\circ = r^{-1}\cdot K^\circ$
        \item if $0\in \mathop{\mathrm{int}}K$, then $(K^\circ)^\circ = K$
        \item for $\bfA\in\mathbb R^{n\times d}$, $\conv(\{\bfa_1,\bfa_2,\dots,\bfa_n\})^\circ = \braces*{\bfx\in\mathbb R^d:\norm*{\bfA\bfx}_\infty\leq 1}$
        \item for an ellipsoid $E = \braces*{\bfx\in\mathbb R^d : \bfx^\top\bfH\bfx\leq 1}$, $E^\circ$ is the ellipsoid $E^\circ = \braces*{\bfx\in\mathbb R^d : \bfx^\top\bfH^{-1}\bfx\leq 1}$
    \end{itemize}
\end{Lemma}

This observation, combined with Theorem \ref{thm:subspace-sketch-stream-l-inf}, yields the first polynomial space algorithm for approximating convex hulls in the worst case:

\ConvexHullStream*
\begin{proof}
    Let $S\subseteq[n]$ be the coreset computed by Algorithm \ref{alg:subspace-sketch-stream-l-inf}. Let $K = \braces*{\bfx\in\mathbb R^d : \norm*{\bfA\bfx}_\infty\leq 1}$ and let $K_S = \braces*{\bfx\in\mathbb R^d : \norm*{\bfA\vert_S\bfx}_\infty\leq 1}$. By Theorem \ref{thm:subspace-sketch-stream-l-inf}, we are guaranteed that
    \[
        K \subseteq K_S \subseteq \Delta K
    \]
    for $\Delta = O(\sqrt{d\log n})$ in the row arrival streaming model and $\Delta = O(\sqrt{d\log(n\kappa^\OL)})$ in the online coreset model. Then by Lemma \ref{lem:polars}, we may take polars on this chain of inclusions to conclude that
    \[
        K^\circ \supseteq K_S^\circ \supseteq \frac1\Delta K^\circ.
    \]
    Since $K^\circ = \conv(\{\pm\bfa_i\}_{i=1}^n)$ and $K_S^\circ = \conv(\{\pm\bfa_i\}_{i\in S})$, we conclude.
\end{proof}

\subsection{L\"owner--John Ellipsoids}

We consider the problem of computing an approximate L\"owner--John ellipsoid of a convex symmetric polytope, also known as the problem of minimum volume enclosing ellipsoid (MVEE). We define our notion of approximation of L\"owner--John ellipsoids as follows:

\begin{Definition}
Let $K\subseteq \mathbb R^d$ be a convex body and let $E$ be the L\"owner--John ellipsoid of $K$. We say that an ellipsoid $E'$ is an \emph{$\alpha$-approximate L\"owner--John ellipsoid} for $K$ if
\[
    E\subseteq E'\subseteq \alpha E.
\]
\end{Definition}

\subsubsection{Upper Bound}

In the literature, there are two closely related variations to this problem (see Equations (1.1.1) and (1.1.2) of \cite{Tod2016}). In one, more common in the computational geometry community, the input data set $\bfA\in\mathbb R^{n\times d}$ is interpreted as the convex hull of the $n$ rows, i.e., $K = \conv(\braces*{\pm\bfa_1,\pm\bfa_2,\dots,\pm\bfa_n})$. In the other, more common in the optimization community, the $\bfA$ is interpreted as a set of $n$ linear constraints, and the input polytope is $K = \braces*{\bfx\in\mathbb R^d : \norm*{\bfA\bfx}_\infty \leq 1}$. As noted in Section \ref{sec:convex-hull}, these two interpretations are polars of each other.

Our results for the streaming $\ell_\infty$ subspace sketch problem in Theorem \ref{thm:subspace-sketch-stream-l-inf} apply most readily to the latter interpretation, i.e.\ the linear inequalities interpretation, and we immediately obtain the following:

\EllipsoidStream*


The proof is sketched in Remark \ref{rem:ellipsoid}.

We also show that we can also get results in the convex hull interpretation, by using the fact that these two interpretations of the input matrix $\bfA$ are polars of each other. See Section 3.5 of \cite{Tod2016} for a discussion on polars and L\"owner--John ellipsoids. 

Using basic facts about polars (Lemma \ref{lem:polars}), we obtain the following:

\begin{Corollary}\label{cor:ellipsoids-convex-hull}
    Let $\bfA$ be an $n\times d$ matrix presented in one pass over a row arrival stream. Define the polytope $K = \conv(\{\pm\bfa_1,\pm\bfa_2,\dots,\pm\bfa_n\})$. There is an algorithm $\mathcal A$ which maintains a coreset $S\subseteq[n]$ from which we can compute an ellipsoid $E'$ such that
    \[
        E' \subseteq K \subseteq \Delta E'.
    \]
    where
    \begin{itemize}
        \setlength\itemsep{-0.1em}
        \item in the streaming model, $\Delta = O(\sqrt{d\log n})$, $\abs*{S} = O(d\log n)$, and $\mathcal A$ uses $O(d^2\log^2 n)$ bits of space.
        \item in the online coreset model, $\Delta = O(\sqrt{d\log(n\kappa^\OL)})$ and $\abs*{S} = O(d\log(n\kappa^\OL))$.
    \end{itemize}
    Since $K\subseteq E\subseteq \sqrt d K$, $E'$ is an $O(\Delta\sqrt d)$-approximate L\"owner--John ellipsoid.
\end{Corollary}
\begin{proof}
    We claim that we can just interpret the row arrival stream as in Theorem \ref{thm:john-ellipsoid-streaming-ub}, and then simply invert the quadratic form of the ellipsoid. Using Theorem \ref{thm:john-ellipsoid-streaming-ub} and Lemma \ref{lem:polars}, we obtain some ellipsoid $E$ such that
    \[
        E \subseteq K^\circ \subseteq \lambda E
    \]
    for some $\lambda$. Then, the ellipsoid with the inverse quadratic form of $E$ is $E^\circ$ and satisfies
    \[
        E^\circ \supseteq K \supseteq \frac1\lambda E^\circ
    \]
    by Lemma \ref{lem:polars}. Scaling by $\lambda$ gives the desired conclusion.
\end{proof}

\subsubsection{Lower Bound}

In this section, we show the negative result that approximate L\"owner--John ellipsoids cannot be maintained in the row arrival model with small space, if the desired approximation is much smaller than $\sqrt d$.

Our main result of the section is the following.

\begin{Theorem}\label{thm:ellipsoid-lb}
Let $n = d^c$, where $c\geq 1$ is any constant integer. Suppose an algorithm $\mathcal A$ computes an $\alpha$-approximate L\"owner--John ellipsoid of any $n\times d$ matrix $\bfA$ with probability at least $2/3$, for $\alpha = c'\sqrt d$ for a sufficiently small constant $c'$, in one pass over a row arrival stream. Then, $\mathcal A$ must use $\Omega(n)$ bits of space.
\end{Theorem}
\begin{proof}
We show the result by reduction from the \textsf{INDEX} problem (Theorem \ref{thm:index}).

Let $S$ be the set constructed in Lemma \ref{lem:coding-random-points} with $p$ in the lemma set to $c$, so that $\abs*{S} = d^c = n$. Then, Alice constructs an $\abs*{A}\times d$ matrix $\bfA$ by choosing the vectors of $S$ corresponding to the indices $i\in A$. Alice then runs the algorithm $\mathcal A$ on the rows of $\bfA$, then passes the working memory of the algorithm to Bob. 

Let $i_*\in[n]$ be the index given to Bob. We claim that Bob can then figure out whether $i_*\in A$ or not using this working memory. Let $\bfb\in S$ be the vector in $S$ indexed by $i_*$. Let $\bfu_1, \bfu_2, \dots, \bfu_{d-1}\in\mathbb R^d$ be an orthonormal basis to the orthogonal complement $\braces*{\bfx\in\mathbb R^d : \angle*{\bfb,\bfx} = 0}$ of $\bfb$. Then, Bob inserts the following rows into the working memory of $\mathcal A$:
\begin{itemize}
    \item $4(d-1)$ rows $\pm d\cdot \bfu_i \pm \bfb/\sqrt d$ for $i\in[d-1]$
    \item $2(d-1)$ rows $\pm R\cdot \bfu_i$ for $i\in[d-1]$, for a large $R = \poly(d)$ to be determined
\end{itemize}
Bob will then report that $i_*\in A$ if and only if $\bfb$ belongs to the $\alpha$-approximate L\"owner--John ellipsoid that is output by $\mathcal A$.

If $i_*\in A$, then it is obvious that $\bfb$ must be in the L\"owner--John ellipsoid, so suppose that $i_*\notin A$. By rotating, we assume without loss of generality that $\bfb = \sqrt d\cdot\bfe_1$ and $\bfu_i = \bfe_{i+1}$ for $i\in[d-1]$. Now consider the exact L\"owner--John ellipsoid $E$ of the input dataset including all rows added by both Alice and Bob, and let $g$ be the largest magnitude achieved by a point $\bfg\in E$, in the direction of $\bfb$. Suppose for contradiction that $g \geq 2$.

\paragraph{Replacing Alice's points by a box.}
Let $V = \{\pm d\cdot \bfu_i \pm \bfb/\sqrt d : i\in[d-1]\}$ be the rows added by Bob. We first show that the L\"owner--John ellipsoid does not change if we remove all of Alice's points, by showing that the convex hull of $V$ must contain Alice's points. Note that Alice's points all have $\ell_2$ norm at most $\sqrt d$ and $\bfe_1$ component at most $1$. If $\bfx$ is any point with $\bfx_1 = 0$, then $d\cdot \bfx / \norm*{\bfx}_1$ is a convex combination of $\pm d\cdot\bfe_i$. The $\ell_2$ norm of this point is at least
\[
    \norm*{d\cdot \frac{\bfx}{\norm*{\bfx}_1}}_2 = d\frac{\norm*{\bfx}_2}{\norm*{\bfx}_1} \geq \sqrt d.
\]
Thus, applying this to any of Alice's points $\bfx$ with the first coordinate removed, these points must lie in $\conv(V)$, since $d\cdot \bfx / \norm*{\bfx}_1\in\conv(V)$ is a vector in the same direction with a greater magnitude as $\bfx$ that is also in $\conv(V)$. It follows that $\bfx$ must lie in $\conv(V)$ as well. 

\paragraph{Reduction to a two-dimensional ellipse.}
Note that $V$ is symmetric with respect to flipping signs on coordinates, and so is $\conv(V)$, and thus so is the L\"owner--John ellipsoid of $\conv(V)$. Now let $\bfv\in V$ be any vertex of $\conv(V)$, and consider the two-dimensional ellipse $E'$ obtained by intersecting $E$ with the plane spanned by $\bfv$, and $\bfb$. Write this ellipse as $E' = \braces*{(x,y) : ax^2 + by^2 \leq 1}$, where the cross term disappears due to symmetry of the ellipse. We will think of the $x$ direction as the $\bfb$ direction, and refer to this coordinate system as the $E'$ coordinate system.

\paragraph{Bounds on the ellipse.}
If the $V$ vertices do not contact the ellipsoid, then they can be removed from the L\"owner--John ellipsoid, which means that the L\"owner--John ellipsoid would be degenerate since it would lie on a $(d-1)$-dimensional space. Thus, the vertices of $V$ must contact the ellipsoid. Similarly, for large enough $R$, the points $\pm R\cdot\bfu_i$ must also contact the ellipsoid, since otherwise removing them would lead to a John ellipse of bounded radius. Note that the $\ell_2$ diameter of $\conv(V)$ is at most $O(d)$, so a L\"owner--John ellipsoid of $\conv(V)$ would have radius at most $O(d^{3/2})$, which means the above holds when $R$ is chosen to be larger than some $O(d^{3/2})$. Also, we have a point $(g,0)\in E'$ for $g > 2$. Then, we have that $a = 1/g^2$ and $b = 1/R^2$ so that
\[
    E' = \braces*{(x,y) : \frac1{g^2} x^2 + \frac1{R^2} y^2 \leq 1}.
\]
Note that the $V$ vertices have the form $(\pm d, \pm 1)$ in the $E'$ coordinate system. However, we then have that
\[
    \frac1{g^2} + \frac{d^2}{R^2} \ll 1
\]
so they in fact cannot contact the ellipse. We conclude that $g > 2$ is impossible.

Finally, even if we have an $\alpha$-approximate L\"owner--John ellipsoid, $\bfb$ will still not be contained in the ellipsoid, so the Bob will still output the correct answer.
\end{proof}

\begin{Remark}
By replacing our use of Lemma \ref{lem:coding-random-points} with $n$ random unit vectors in $d$ dimensions, we can instead get a collection of vectors with inner product $\Theta(\sqrt{(\log n)/d})$, which leads to an $\Omega(n)$ bit lower bound for distortions better than $O(\sqrt{d/\log n})$, even for $n$ larger than $\poly(d)$. 
\end{Remark}

\begin{Remark}
Note that the above lower bound holds even if Alice and Bob compute a general convex body $K$ such that
\[
    E \subseteq K \subseteq \alpha E,
\]
since such a $K$ can still detect whether Bob's point is in Alice's point set or not. 
\end{Remark}

\subsection{Minimum-Width Spherical Shell}\label{sec:spherical-shell}

As mentioned earlier, our proof of Theorem \ref{thm:spherical-shell} for minimum-width spherical shells requires additional care to handle general instances, rather than just centrally symmetric instances.

\SphericalShell*
\begin{proof}
We will always store the first point $\bfa_1$ in order to translate our input instance to the origin. Now for each $i\in[n]$, define the vector $\bfb_i\in\mathbb R^{d+1}$ by setting the first $d$ coordinates to be $-2(\bfa_i-\bfa_1)$ and the last coordinate to be $\norm*{\bfa_i-\bfa_1}_2^2$. Given $\bfa_1$, we can always compute $\bfb_i$ if we have stored $\bfa_i$. Similarly, define $\bfb_i''\in\mathbb R^{d+2}$ to be $\bfb_i$ with an additional $1$ appended as the $(d+2)$th coordinate.

We now proceed by a variation on the standard linearization trick \cite{AHV2004}. Suppose that we wish to compute the width $R - r$ of the minimum-width spherical shell $\sigma(\bfc,r,R)$ containing $\{\bfa_i\}_{i=1}^n$, centered at some arbitrary $\bfc\in\mathbb R^d$. Note that the inner radius is given by $r = \min_{j=1}^n \norm*{\bfc-\bfa_j}_2$ while the outer radius is given by $R = \max_{i=1}^n \norm*{\bfc-\bfa_i}_2$. Now note that
\begin{align*}
    R^2 - r^2 &= \max_{i=1}^n \norm*{\bfc-\bfa_i}_2^2 -\min_{j=1}^n \norm*{\bfc-\bfa_j}_2^2 \\
    &= \max_{i=1}^n \norm*{\bfc}_2^2 -2\angle*{\bfc,\bfa_i} + \norm*{\bfa_i}_2^2 -\min_{j=1}^n \norm*{\bfc}_2^2 -2\angle*{\bfc,\bfa_j} + \norm*{\bfa_j}_2^2 \\
    &= \max_{i=1}^n -2\angle*{\bfc,\bfa_i} + \norm*{\bfa_i}_2^2 -\min_{j=1}^n -2\angle*{\bfc,\bfa_j} + \norm*{\bfa_j}_2^2 \\
    &= \max_{i=1}^n \angle*{\bfb_i,\bfc'} -\min_{j=1}^n \angle*{\bfb_j,\bfc'}
\end{align*}
where $\bfc' = [\bfc, 1]$. Then by the discussion in Section \ref{sec:high-dim-cg}, our $\ell_\infty$ subspace sketch coreset result of Theorem \ref{thm:subspace-sketch-stream-l-inf} can estimate this up to a factor of $\Delta$ both in the row arrival streaming model and the online coreset model. Similarly, note that
\begin{align*}
    R^2 &= \max_{i=1}^n \norm*{\bfc-\bfa_i}_2^2 \\
    &= \max_{i=1}^n \norm*{\bfc}_2^2 -2\angle*{\bfc,\bfa_i} + \norm*{\bfa_i}_2^2 \\
    &= \max_{i=1}^n \angle*{\bfb_i'',\bfc''}
\end{align*}
where $\bfc'' = [\bfc,1,\norm*{\bfc}_2^2]$. We estimate this quantity up to a $\Delta$ factor using Theorem \ref{thm:subspace-sketch-stream-l-inf} as well. Note then that
\[
    R - r = \frac{R^2 - r^2}{R+r} = \Theta\parens*{\frac{R^2 - r^2}{R}}
\]
and we obtain a $\Delta$ factor approximation to the numerator, while we obtain a $\sqrt\Delta$ factor approximation to the denominator. Thus, overall, we obtain a $\Delta^{3/2}$-approximation to the entire quantity.
\end{proof}
\section{An Elementary Proof of the Lewis--Tomczak-Jaegermann \texorpdfstring{\cite{LT1980}}{[LT80]} Change of Density}\label{sec:LTJ80}

In this section, we obtain a short and elementary proof of Theorem \ref{thm:lp-lq-change-of-density} via Lewis weights, by proving Theorem \ref{thm:lp-lq-change-of-density-lewis}.

\subsection{Lewis Weight Switching}

We first show the following crucial identity, which shows that one can reweight a matrix by $\ell_p$ Lewis weights, so that the $\ell_q$ Lewis weights of the resulting matrix coincides with the $\ell_p$ Lewis weights of the original matrix. Note that this identity is true by definition for $q = 2$ (Definition \ref{def:lewis-weights}), since $\ell_2$ Lewis weights are just leverage scores. Thus, this result shows that although Lewis weights are defined by normalizing a change of density with respect to $\ell_2$ (see \cite{CP2015}), they actually simultaneously satisfy the analogous property for all $\ell_q$ as well.

\begin{Lemma}[Lewis weight switching]\label{lem:lewis-weight-switching}
Let $\bfA\in\mathbb R^{n\times d}$ and let $p,q>0$. Let $\bfB \coloneqq \bfW^p(\bfA)^{1/q - 1/p}\bfA$. Then, for each $i\in[n]$,
\[
    \bfw_i^q(\bfB) = \bfw_i^p(\bfA).
\]
Furthermore, the two Lewis bases coincide.
\end{Lemma}
\begin{proof}
We have that
\begin{align*}
    \bftau_i(\bfW^p(\bfA)^{1/2 - 1/q}\bfB) &= \bftau_i(\bfW^p(\bfA)^{1/2 - 1/q}\cdot \bfW^p(\bfA)^{1/q - 1/p}\bfA) \\
    &= \bftau_i(\bfW^p(\bfA)^{1/2 - 1/p}\bfA) = \bfw_i^p(\bfA)
\end{align*}
so
\[
    \bfw_i^q(\bfW^p(\bfA)^{1/q - 1/p}\bfA) = \bfw_i^p(\bfA)
\]
by uniqueness of Lewis weights \cite{CP2015}. The Lewis bases coincide since
\[
    \bfW^p(\bfA)^{1/2-1/p}\bfA\bfR = \bfW^p(\bfA)^{1/2-1/q}\bfW^p(\bfA)^{1/q-1/p}\bfA\bfR = \bfW^p(\bfA)^{1/2-1/q}\bfB\bfR.
\]
\end{proof}

In fact, given only one-sided $\ell_p$ Lewis weights (Definition \ref{def:one-sided-lewis}), we can prove a similar inequality:

\begin{Lemma}[One-sided Lewis weight switching]\label{lem:one-sided-lewis-weight-switching}
Let $\bfA\in\mathbb R^{n\times d}$ and let $p,q>0$. Let $\bfv\in\mathbb R^n$ be one-sided $\ell_p$ Lewis weights for $\bfA$ and let $\bfR$ be the corresponding one-sided $\ell_p$ Lewis basis. Let $\bfV = \diag(\bfv)$ and $\bfB \coloneqq \bfV^{1/q-1/p}\bfA$. Then, $\bfv$ are one-sided $\ell_q$ Lewis weights and $\bfR$ is a one-sided $\ell_q$ Lewis basis for $\bfB$, i.e.,
\[
    \bftau_i(\bfV^{1/2-1/q}\bfB) \leq \bfv_i.
\]
\end{Lemma}
\begin{proof}
We have that $\bfV^{1/2-1/p}\bfA\bfR$ is orthonormal, which means
\[
    \bfV^{1/2-1/q}\bfB\bfR = \bfV^{1/2-1/q}\bfV^{1/q-1/p}\bfA\bfR = \bfV^{1/2-1/p}\bfA\bfR
\]
is as well. Then,
\[
    \bftau_i(\bfV^{1/2-1/q}\bfB) = \norm*{\bfe_i^\top \bfV^{1/2-1/q}\bfB\bfR}_2^2 = \norm*{\bfe_i^\top \bfV^{1/2-1/p}\bfA\bfR}_2^2 = \bftau_i(\bfV^{1/2-1/p}\bfA) \leq \bfv_i
\]
as desired.
\end{proof}

\subsection{Proof of Change of Density}

Using the above, we show that reweighting the rows of $\bfA$ by a scalar multiple of the $\ell_p$ Lewis weights provide optimal approximations of $\ell_p$ by $\ell_q$. The following lemmas show the upper bounds and lower bounds. The proofs roughly follow, but are still slightly different from, the estimates in Lemma 2.6 of \cite{JLS2021} and Lemma 8 in Chapter III.B of \cite{Woj1991}, which show the analogous results for $q = 2$. The estimates are an elementary combination of Lewis weight switching (Lemma \ref{lem:lewis-weight-switching}/Lemma \ref{lem:one-sided-lewis-weight-switching}), sensitivity bounds (Lemma \ref{lem:lewis-weights-bound-sensitivities}/Lemma \ref{lem:one-sided-lewis-weights-bound-sensitivities}), and H\"older's inequality. 

\begin{Lemma}[Upper bound, $p\geq q$]\label{lem:lp-lq-ub-gt}
Let $\bfA\in\mathbb R^{n\times d}$ and $p\geq q>0$. Let $\bfv\in\mathbb R^n$ be one-sided $\ell_p$ Lewis weights for $\bfA$. Let $\bfV = \diag(\bfv)$. For all $\bfx\in\mathbb R^d$,
\[
    \norm*{\bfA\bfx}_p \leq \norm*{\bfv}_1^{[0\lor(1/2 - 1/q)](p-q)/p}\norm*{\bfV^{1/q-1/p}\bfA\bfx}_q
\]
\end{Lemma}
\begin{proof}
For $i\in[n]$, we have that
\begin{align*}
    \abs*{[\bfA\bfx](i)} &= \bfv_i^{1/p - 1/q}\cdot [\bfV^{1/q - 1/p}\bfA\bfx](i) \\
    &\leq \bfv_i^{1/p - 1/q}\cdot \left[\norm*{\bfv}_1^{0\lor (q/2 - 1)}\cdot \bfv_i\cdot \norm*{\bfV^{1/q - 1/p}\bfA\bfx}_q^q\right]^{1/q} && \text{Lemmas \ref{lem:one-sided-lewis-weights-bound-sensitivities}, \ref{lem:one-sided-lewis-weight-switching}} \\
    &= \norm*{\bfv}_1^{0\lor (1/2 - 1/q)}\cdot \bfv_i^{1/p}\norm*{\bfV^{1/q - 1/p}\bfA\bfx}_q.
\end{align*}
Then, 
\begin{align*}
    \norm*{\bfA\bfx}_p^p &= \sum_{i=1}^n \abs*{[\bfA\bfx](i)}^p = \sum_{i=1}^n \abs*{[\bfA\bfx](i)}^{p-q}\cdot \abs*{[\bfA\bfx](i)}^{q} \\
    &\leq \sum_{i=1}^n \norm*{\bfv}_1^{[0\lor(1/2 - 1/q)](p-q)}\cdot \bfv_i^{(p-q)/p}\norm*{\bfV^{1/q - 1/p}\bfA\bfx}_q^{p-q}\cdot \abs*{[\bfA\bfx](i)}^{q} \\
    &= \norm*{\bfv}_1^{[0\lor(1/2 - 1/q)](p-q)}\norm*{\bfV^{1/q - 1/p}\bfA\bfx}_q^{p-q}\sum_{i=1}^n \bfv_i^{q(1/q-1/p)}\cdot \abs*{[\bfA\bfx](i)}^{q} \\
    &= \norm*{\bfv}_1^{[0\lor(1/2 - 1/q)](p-q)}\norm*{\bfV^{1/q - 1/p}\bfA\bfx}_q^{p-q} \cdot \norm*{\bfV^{1/q - 1/p}\bfA\bfx}_q^q \\
    &= \norm*{\bfv}_1^{[0\lor(1/2 - 1/q)](p-q)}\norm*{\bfV^{1/q - 1/p}\bfA\bfx}_q^{p}.
\end{align*}
Taking $p$th roots on both sides gives the desired result.
\end{proof}

\begin{Lemma}[Upper bound, $q\geq p$]\label{lem:lp-lq-ub-lt}
Let $\bfA\in\mathbb R^{n\times d}$ and $q\geq p>0$. Let $\kappa \geq 1$ and let $\bfw\in\mathbb R^n$ be $\alpha$-approximate $\ell_p$ Lewis weights for $\bfA$. Let $\bfW = \diag(\bfw)$. For all $\bfx\in\mathbb R^d$,
\[
    \norm*{\bfA\bfx}_p \leq (\alpha d)^{1/p-1/q}\norm*{\bfW^{1/q-1/p}\bfA\bfx}_q
\]
\end{Lemma}
\begin{proof}
We have
\begin{align*}
    \norm*{\bfA\bfx}_p^p &= \sum_{i=1}^n \bfw_i^{1-p/q} \cdot [\bfW^{1/q-1/p}\bfA\bfx](i)^p \\
    &= \bracks*{\sum_{i=1}^n \bfw_i^{(1-p/q)/(1-p/q)}}^{1-p/q}\bracks*{\sum_{i=1}^n [\bfW^{1/q-1/p}\bfA\bfx](i)^q}^{p/q} && \text{H\"older's inequality} \\
    &= (\alpha d)^{1-p/q}\norm*{\bfW^{1/q-1/p}\bfA\bfx}_q^p.
\end{align*}
Taking $p$th roots on both sides gives the desired result.
\end{proof}

\begin{Lemma}[Lower bound, $p\geq q$]\label{lem:lp-lq-lb-gt}
Let $\bfA\in\mathbb R^{n\times d}$ and $p\geq q>0$. Let $\kappa\geq1$ and let $\bfw\in\mathbb R^n$ be $\alpha$-approximate $\ell_p$ Lewis weights for $\bfA$. Let $\bfW = \diag(\bfw)$. For all $\bfx\in\mathbb R^d$,
\[
    \norm*{\bfW^{1/q-1/p}\bfA\bfx}_q \leq (\alpha d)^{1/q-1/p}\norm*{\bfA\bfx}_p
\]
\end{Lemma}
\begin{proof}
We have
\begin{align*}
    \norm*{\bfW^{1/q-1/p}\bfA\bfx}_q^q &= \sum_{i=1}^n \bfw_i^{1-q/p}[\bfA\bfx](i)^q \\
    &\leq \bracks*{\sum_{i=1}^n \bfw_i^{(1-q/p)/(1-q/p)}}^{1-q/p} \bracks*{\sum_{i=1}^n [\bfA\bfx](i)^p}^{q/p} && \text{H\"older's inequality} \\
    &\leq (\alpha d)^{1-q/p}\norm*{\bfA\bfx}_p^q && \text{Definition \ref{def:approximate-lewis-overestimate}}.
\end{align*}
Taking $q$th roots on both sides gives the desired result.
\end{proof}

\begin{Lemma}[Lower bound, $q\geq p$]\label{lem:lp-lq-lb-lt}
Let $\bfA\in\mathbb R^{n\times d}$ and $q\geq p>0$. Let $\bfv\in\mathbb R^n$ be one-sided $\ell_p$ Lewis weights for $\bfA$. Let $\bfW = \diag(\bfw)$. For all $\bfx\in\mathbb R^d$,
\[
    \norm*{\bfV^{1/q-1/p}\bfA\bfx}_q \leq \norm*{\bfv}_1^{[0\lor(1/2-1/p)](q-p)/q}\norm*{\bfA\bfx}_p
\]
\end{Lemma}
\begin{proof}
We have
\begin{align*}
    \norm*{\bfV^{1/q-1/p}\bfA\bfx}_q^q &= \sum_{i=1}^n \bfv_i^{1-q/p}[\bfA\bfx](i)^q = \sum_{i=1}^n \bfv_i^{1-q/p}[\bfA\bfx](i)^{q-p}[\bfA\bfx](i)^p \\
    &\leq \sum_{i=1}^n \bfv_i^{1-q/p}\bracks*{\norm*{\bfv}_1^{0\lor(p/2-1)}\bfv_i\norm*{\bfA\bfx}_p^p}^{(q-p)/p}[\bfA\bfx](i)^p && \text{Lemma \ref{lem:one-sided-lewis-weights-bound-sensitivities}} \\
    &= \norm*{\bfA\bfx}_p^{q-p} \norm*{\bfv}_1^{[0\lor(p/2-1)](q-p)/p}\sum_{i=1}^n [\bfA\bfx](i)^p = \norm*{\bfv}_1^{[0\lor(p/2-1)](q-p)/p}\norm*{\bfA\bfx}_p^q.
\end{align*}
Taking $q$th roots on both sides gives the desired result.
\end{proof}

Combining the above lemmas yields the following conclusion.

\ChangeOfDensity*
\begin{proof}
Let $\bfv$ be the one-sided $\ell_p$ Lewis weights such that $\bfw\geq \bfv$. First consider the case of $p\geq q>0$. Note that in this case, $1/q - 1/p \geq 0$ so Lemmas \ref{lem:lp-lq-ub-gt} and \ref{lem:lp-lq-lb-gt} yield that
\[
    \norm*{\bfA\bfx}_p \leq \lambda_{d,p,q} \norm*{\bfV^{1/q-1/p}\bfA\bfx}_q \leq \lambda_{d,p,q} \norm*{\bfW^{1/q-1/p}\bfA\bfx}_q \leq \kappa_{d,p,q}\lambda_{d,p,q} \norm*{\bfA\bfx}_p.
\]
On the other hand, if $q\geq p>0$, then $1/q - 1/p \leq 0$ so Lemmas \ref{lem:lp-lq-ub-lt} and \ref{lem:lp-lq-lb-lt} yield that
\[
    \norm*{\bfA\bfx}_p \leq \kappa_{d,p,q}\norm *{\bfW^{1/q-1/p}\bfA\bfx}_q \leq \kappa_{d,p,q} \norm*{\bfV^{1/q-1/p}\bfA\bfx}_q \leq \kappa_{d,p,q}\lambda_{d,p,q} \norm*{\bfA\bfx}_p.
\]

For $p\geq q$, we have that the total distortion is $(\alpha d)^\beta$, for
\[
    \beta = \bracks*{\frac12 - \frac1q}\frac{p-q}{p} + \bracks*{\frac1q - \frac1p} = \bracks*{\frac12 - \frac1q}\frac{p-q}{p} + \frac1q \frac{p-q}{p} = \frac{p-q}{2p} = \frac12\parens*{1-\frac{q}{p}}
\]
if $q\geq 2$, and $(\alpha d)^{\frac1q - \frac1p}$ if $q\leq 2$. Next, when $q\geq p>0$, then we have that the total distortion is $(\alpha d)^\beta$ for
\[
    \beta = \bracks*{\frac12 - \frac1p}\frac{q-p}{q} + \bracks*{\frac1p - \frac1q} = \bracks*{\frac12 - \frac1p}\frac{q-p}{q} + \frac1p \frac{q-p}{q} = \frac{q-p}{2q} = \frac12\parens*{1-\frac{p}{q}}
\]
if $p\geq 2$, and $(\alpha d)^{\frac1p - \frac1q}$ if $p\leq 2$. These yield the claimed bounds.

\end{proof}

\section{Applications to Numerical Linear Algebra}\label{sec:app-lin-alg}

In this section, we show applications of the embeddings in Section \ref{sec:LTJ80} to various problems in numerical linear algebra. In general, we will obtain smaller space/dimension compared to existing results by trading off for larger distortions.

\subsection{Subspace Embeddings from \texorpdfstring{$\ell_p$}{lp} to \texorpdfstring{$\ell_q$}{lq}}

By composing the embedding of Theorem \ref{thm:lp-lq-change-of-density-lewis} with subspace embeddings for $\ell_q$ \cite{BLM1989, CP2015}, we obtain new subspace embeddings from $\ell_p$ into $\ell_q$. Our results add to an old and long line of work in geometric functional analysis on embedding subspaces of $L_p$ into $\ell_q^r$, including \cite{JS1982, Sch1985, Sch1987,BLM1989, Tal1990, Tal1995, Zva2000, SZ2001, Sch2011, LT2011, LWW2021}. See also the surveys of \cite{JS2001, KKLT2021} on this topic.

\begin{Theorem}\label{thm:lp-lq-subspace-embedding}
    Let $\bfA\in\mathbb R^{n\times d}$ and $2\leq q \leq p < \infty$. Then, there is an algorithm which outputs a matrix $\bfS\in\mathbb R^{s\times n}$ such that
    \[
        \Pr\braces*{\forall \bfx\in\mathbb R^d, \norm*{\bfA\bfx}_p \leq \norm*{\bfS\bfA\bfx}_q \leq \kappa\norm*{\bfA\bfx}_p} \geq 1 - \delta
    \]
    for
    \[
        \kappa = O\parens*{d^{\frac12\parens*{1-\frac{q}{p}}}},\qquad s = d^{q/2}\poly(\log d, \log\delta^{-1})
    \]
    with running time $\tilde O(\nnz(\bfA) + d^\omega)$.
\end{Theorem}
\begin{proof}
    By Theorem \ref{thm:lp-lq-change-of-density-lewis}, one can use the $\ell_p$ Lewis weights of $\bfA$ to compute a diagonal matrix $\bfD$ such that
    \[
        \norm*{\bfA\bfx}_p \leq \norm*{\bfD\bfA\bfx}_q \leq \kappa\norm*{\bfA\bfx}_p
    \]
    for $\kappa = O\parens*{d^{\frac12\parens*{1-\frac{q}{p}}}}$. Then by Theorem \ref{thm:lewis-weight-subspace-embedding}, one can use the $\ell_q$ Lewis weights of $\bfA$ to sample an $s\times n$ matrix $\bfR$ such that
    \[
        \Pr\braces*{\forall \bfx\in\mathbb R^d, \norm*{\bfD\bfA\bfx}_p \leq \norm*{\bfR\bfD\bfA\bfx}_q \leq O(1)\norm*{\bfD\bfA\bfx}_p} \geq 1-\delta
    \]
    with $s = d^{q/2}\poly(\log d, \log\delta^{-1})$. Chaining together these estimates leads to the conclusion of the theorem, with $\bfS = \bfR\bfD$.

    The running time is dominated by the computation of $\ell_p$ and $\ell_q$ Lewis weights, which is $\tilde O(\nnz(\bfA) + d^\omega)$ by Theorem \ref{thm:cohen-peng-fast-lewis-weights}.
\end{proof}

Note that subspace embeddings immediately imply algorithms for the subspace sketch problem (see Section \ref{sec:subspace-sketch}, as well as \cite{LWW2021}). Thus, lower bounds for the subspace sketch problem imply lower bounds for subspace embeddings. It then follows from Theorem \ref{thm:lww2021-subspace-sketch-lower-bound} that the tradeoff between the distortion $\kappa$ and the sketching dimension $s$ in Theorem \ref{thm:lp-lq-subspace-embedding} is optimal up to constant factors.

\subsection{\texorpdfstring{$\ell_p$}{lp} Regression}

It is well-known that subspace embeddings can be used to speed up algorithms for $\ell_p$ regression \cite{Sar2006,DDHKM2009, CW2013, NN2013, WZ2013,MM2013, CP2015, CDMMMW2016}. See also the surveys \cite{Mah2011, Woo2014, MT2020}. By using the $\ell_p$-$\ell_q$ subspace embeddings from the previous section, we obtain faster algorithms for $\ell_p$ linear regression, if one is willing to accept larger multiplicative errors.

In the high precision regime, the following is known for solving $\ell_p$ regression:

\begin{Theorem}[Fast $\ell_p$ Regression Solvers]\label{thm:lp-regression-solver}
    Let $\bfA\in\mathbb R^{n\times d}$ and $\bfb\in\mathbb R^n$ and let $2\leq p<\infty$. Let $\eps>0$ be an accuracy parameter. Then, there is an algorithm which outputs $\hat\bfx\in\mathbb R^d$ such that
    \[
        \norm*{\bfA\hat\bfx - \bfb}_p \leq (1+\eps)\min_{\bfx\in\mathbb R^d}\norm*{\bfA\bfx-\bfb}_p,
    \]
    running in time $\tilde O(n^{\max\{\omega, 7/3\}}\log\eps^{-1})$ \cite{AKPS2019}. For the current value of $\omega \approx 2.37286$ \cite{AW2021}, this algorithm runs in time $\tilde O(n^{\omega}\log\eps^{-1})$. 
\end{Theorem}

Composing the above result with the subspace embedding in Theorem \ref{thm:lp-lq-subspace-embedding} yields the following:

\begin{Theorem}\label{thm:lp-regression-large-distortion}
    Let $\bfA\in\mathbb R^{n\times d}$ and $\bfb\in\mathbb R^n$ and let $2\leq p<\infty$. Let $2\leq q\leq p$. Then, there is an algorithm which outputs $\hat\bfx\in\mathbb R^d$ such that
    \[
        \norm*{\bfA\hat\bfx - \bfb}_p \leq O(d^{\frac12\parens*{1-\frac{q}{p}}})\min_{\bfx\in\mathbb R^d}\norm*{\bfA\bfx-\bfb}_p,
    \]
    running in time $\tilde O(\nnz(\bfA) + d^{\max\{\omega,7/3\}\cdot q/2})$.
\end{Theorem}
\begin{proof}
We use the standard sketch-and-solve framework for $\ell_p$ regression. Let $\bfA' \coloneqq [\bfA\ \bfb]$ be the $n\times(d+1)$ matrix obtained by concatenating $\bfA$ with $\bfb$. Let $\bfS\in\mathbb R^{s\times n}$ be the subspace embedding sketch of Theorem \ref{thm:lp-lq-subspace-embedding} for the matrix $\bfA'$. Then by using Theorem \ref{thm:lp-regression-solver}, we obtain $\hat\bfx\in\mathbb R^d$ such that
\[
    \norm*{\bfS\bfA\hat\bfx - \bfS\bfb}_q \leq O(1) \min_{\bfx\in\mathbb R^d} \norm*{\bfS\bfA\bfx - \bfS\bfb}_q
\]
in time $\tilde O(s^{\max\{\omega,7/3\}}) = \tilde O(d^{\max\{\omega,7/3\}\cdot q/2})$. Now let $\bfx_* \coloneqq \arg\min_{\bfx\in\mathbb R^d}\norm*{\bfA\bfx-\bfb}_p$. Then, this $\hat\bfx$ satisfies
\begin{align*}
    \norm*{\bfA\hat\bfx - \bfb}_p &\leq \norm*{\bfS\bfA\hat\bfx - \bfS\bfb}_q && \text{Theorem \ref{thm:lp-lq-subspace-embedding}} \\
    &\leq O(1) \min_{\bfx\in\mathbb R^d} \norm*{\bfS\bfA\bfx - \bfS\bfb}_q \\
    &\leq O(1) \norm*{\bfS\bfA\bfx_* - \bfS\bfb}_q \\
    &\leq O(d^{\frac12\parens*{1-\frac{q}{p}}})\norm*{\bfA\bfx_* - \bfb}_p && \text{Theorem \ref{thm:lp-lq-subspace-embedding}} \\
    &= O(d^{\frac12\parens*{1-\frac{q}{p}}})\min_{\bfx\in\mathbb R^d}\norm*{\bfA\bfx - \bfb}_p
\end{align*}
as claimed.
\end{proof}

\subsection{\texorpdfstring{$\ell_p$}{lp} Low Rank Approximation}

A useful application of fast algorithms for $\ell_p$ regression with large errors is in $\ell_p$ low rank approximation via column subset selection, for which algorithms already provably must have $\poly(k)$ factor error, where $k$ is the desired rank parameter \cite{CGKLPW2017,DWZZR2019}.

More formally, in the $\ell_p$ column subset selection problem, we are given an $n\times d$ matrix $\bfA$, $p>0$, and a rank parameter $k$, and we seek a rank $k$ approximation of $\bfA$ by selecting a set of $k$ columns $S$ of $\bfA$ that minimizes
\[
    \norm*{\bfA - \bfA_S\cdot \bfV}_p
\]
where $\bfA_S$ is the $k$ selected columns, $\bfV\in\mathbb R^{k\times d}$, and $\norm*{\cdot}_p$ is the entrywise $\ell_p$ norm for matrices.

The work of \cite{CGKLPW2017} designed a bicriteria algorithm for $\ell_p$ column subset selection which repeatedly samples $2k$ random columns $R$ of $\bfA$, solves $\ell_p$ regression with the columns of $\bfA$ outside of $R$ as the target vectors and the columns inside $R$ as the design matrix, and selects the columns with low fitting error. This algorithm selects a total of $O(k\log d)$ columns $S$ of $\bfA$ with high probability, and gives an $O(k)$ factor approximation relative to the best rank $k$ approximation, that is,
\[
    \norm*{\bfA - \bfA_S}_p \leq (k+1)\min_{\text{rank $k$ $\bfX$}}\norm*{\bfA - \bfX}_p
\]
This algorithm was re-analyzed in \cite{DWZZR2019} with an improved guarantee, giving a $O(k^{1/p \lor (1-1/p)})$ factor approximation.

We now show an improved running time for this algorithm at a cost of a slightly larger distortion, by using Theorem \ref{thm:lp-regression-large-distortion}.

Following the notation of \cite{CGKLPW2017, DWZZR2019}, we write
\[
    \bfDelta \coloneqq \bfA - \bfA_*
\]
where $\bfA_* = \arg\min_{\text{rank $k$ $\bfX$}} \norm*{\bfA - \bfX}_p$. As in the proof of Theorem 7 of \cite{CGKLPW2017}, we assume that our algorithm knows $\norm*{\bfDelta}_p^p$ up to a factor of $2$, by paying a factor of $O(\log n)$ in the running time.

\begin{Definition}[Approximate coverage, Definition 5 \cite{CGKLPW2017}, Definition D.1 \cite{DWZZR2019}]
    Let $S$ be a subset of $k$ column indices. We say that column $\bfA_i$ is $(\kappa,p)$-approximately covered by $S$ if for $p\in[1,\infty)$ we have
    \[
        \min_{\bfx\in\mathbb R^k}\norm*{\bfA_S\bfx - \bfA_i}_p^p \leq \kappa\frac{(k+1)^{1\lor (p-1)}}{d}\norm*{\bfDelta}_p^p
    \]
    or for $p = \infty$ we have
    \[
        \min_{\bfx\in\mathbb R^k}\norm*{\bfA_S\bfx - \bfA_i}_\infty \leq \kappa(k+1)\norm*{\bfDelta}_\infty
    \]
\end{Definition}

\begin{algorithm}
	\caption{$\ell_p$ Column Subset Selection (Algorithm 3 of \cite{CGKLPW2017}, Algorithm 2 of \cite{DWZZR2019})}
	\textbf{input:} $\bfA\in\mathbb R^{n\times d}$, $p>2$, rank $k$, $2\leq q \leq p$. \\
	\textbf{output:} $O(k\log d)$ columns of $\bfA$. \\
    $\textsc{SelectColumns}(\bfA,p,k)$
	\begin{algorithmic}[1] 
        \If{$d\leq 2k$}
            \State \Return all columns of $\bfA$
        \EndIf
        \State $\kappa \gets O(k^{\frac12\parens*{1-\frac{q}{p}}})$
        \Repeat
            \State Let $R$ be a uniformly random set of $2k$ columns of $\bfA$
        \Until{$1/10$ fraction of columns of $\bfA$ are $(\kappa^p,p)$-approximately covered (using Theorem \ref{thm:lp-regression-large-distortion})}\label{line:approx-cover}
        \State Let $\bfA_{\overline R}$ be the columns not covered by $R$
        \State \Return $\bfA_R\cup \textsc{SelectColumns}(\bfA_{\overline R},p,k)$
	\end{algorithmic}\label{alg:lp-css}
\end{algorithm}

\begin{Theorem}\label{thm:lp-css}
    Let $\bfA\in\mathbb R^{n\times d}$, $p>2$, and $2\leq q\leq p$. Algorithm \ref{alg:lp-css} outputs a set of $O(k\log d)$ columns $S\subseteq[d]$ of $\bfA$ such that
    \[
        \norm*{\bfA - \bfA_S\cdot\bfV}_p \leq O(k^{1-\frac1p + \frac12\parens*{1-\frac{q}{p}}})\norm*{\bfDelta}_p = O(k^{\frac32-\frac{1+q/2}p})\norm*{\bfDelta}_p
    \]
    and runs in time
    \[
        \tilde O(\nnz(\bfA)d + k^{\max\{\omega,7/3\}\cdot q/2}d).
    \]
\end{Theorem}
\begin{proof}
    We only briefly outline the proof of \cite{CGKLPW2017,DWZZR2019}, leaving rest of the details to these works. The work of \cite{CGKLPW2017} first shows the existence of $k$ columns which achieves a distortion of $O(k)$, and that $O(k\log d)$ columns achieving this value can be identified using Algorithm \ref{alg:lp-css}. Their argument roughly proceeds as follows. If we select $2k$ columns of $\bfA$, then a random column has a regression cost of roughly $O(k^p)\norm*{\Delta}_p^p / d$ on average, which means that a constant fraction of the columns have regression cost roughly $O(k^p)\norm*{\Delta}_p^p / d$. These columns can be identified by using an $\ell_p$ regression solver. The work of \cite{DWZZR2019} improves upon this result by showing the existence of $k$ columns achieving a distortion of $O(k^{1/p \lor (1-1/p)}) = O(k^{1-1/p})$ for $p>2$, which immediately yields the improved distortion result using the same algorithm. 
    
    When we replace the use of exact $\ell_p$ regression solvers with our Theorem \ref{thm:lp-regression-large-distortion}, we distort the regression costs by a factor of $O(k^{\frac12\parens*{1-\frac{q}{p}}})$, which means we must consider larger errors to be covered, which is reflected in line \ref{line:approx-cover}. Finally, the running time follows from solving $\ell_p$ regression with $O(k)$ columns for a total of $\tilde O(d)$ times.
\end{proof}

\section{The Subspace Sketch Problem with Large Approximation}\label{sec:subspace-sketch}

Using our embedding result from Section \ref{sec:LTJ80}, we show new tight upper bounds on the subspace sketch problem in the large approximation regime.

The following lower bound is known, due to results given in Section 4 of \cite{LWW2021}\footnote{\cite{LWW2021} instantiates the result with $N = \Theta(d^{p/2})$, but the proof immediately gives the claimed result.}:

\begin{Theorem}[Section 4 of \cite{LWW2021}]\label{thm:lww2021-subspace-sketch-lower-bound}
Let $\bfA\in\mathbb R^{n\times d}$, $p\geq 2$, and let $Q_p$ be a data structure which compresses $\bfA$ to support queries of the form $Q_p(\bfx)$ for $\bfx\in\mathbb R^d$. Then:
\begin{itemize}
    \item \textbf{For each guarantee}: If
    \[
        \Pr\braces*{Q_p(\bfx) \leq \norm*{\bfA\bfx}_p \leq \frac{\sqrt d}{N^{1/p}}Q_p(\bfx)} \geq \frac23
    \]
    for each $\bfx\in\mathbb R^d$, then $Q_p$ must use $\Omega(N)$ bits of space.
    \item\textbf{For all guarantee}: If
    \[
        \Pr\braces*{\forall \bfx\in\mathbb R^d, Q_p(\bfx) \leq \norm*{\bfA\bfx}_p \leq \frac{\sqrt d}{N^{1/p}}Q_p(\bfx)} \geq \frac23
    \]
    then $Q_p$ must use $\Omega(Nd)$ bits of space.
\end{itemize}
\end{Theorem}

By letting $N = \Theta(d^{q/2})$ for $q \leq p$, this gives the following tradeoff:
\begin{itemize}
    \item \textbf{For each guarantee}: If
    \[
        \Pr\braces*{Q_p(\bfx) \leq \norm*{\bfA\bfx}_p \leq d^{\frac{p-q}{2p}}Q_p(\bfx)} \geq \frac23
    \]
    for each $\bfx\in\mathbb R^d$, then $Q_p$ must use $\Omega(d^{q/2})$ bits of space.
    \item\textbf{For all guarantee}: If
    \[
        \Pr\braces*{\forall \bfx\in\mathbb R^d, Q_p(\bfx) \leq \norm*{\bfA\bfx}_p \leq d^{\frac{p-q}{2p}}Q_p(\bfx)} \geq \frac23
    \]
    then $Q_p$ must use $\Omega(d^{q/2+1})$ bits of space.
\end{itemize}

We use our new embedding result to show the following upper bounds that match the lower bounds of Theorem \ref{thm:lww2021-subspace-sketch-lower-bound} up to constant factors:

\begin{Theorem}\label{thm:subspace-sketch-lp-lq}
Let $\bfA\in\mathbb R^{n\times d}$ and $p\geq q\geq 2$. Then:
\begin{itemize}
    \item \textbf{For each guarantee}: There exists a data structure $Q_p$ such that
    \[
        \Pr\braces*{Q_p(\bfx) \leq \norm*{\bfA\bfx}_p \leq O\parens*{d^{\frac{p-q}{2p}}}Q_p(\bfx)} \geq \frac23
    \]
    for each $\bfx\in\mathbb R^d$, using at most $O(d^{q/2}\poly\log n)$ bits of space.
    \item\textbf{For all guarantee}: There exists a data structure $Q_p$ such that
    \[
        \Pr\braces*{\forall \bfx\in\mathbb R^d, Q_p(\bfx) \leq \norm*{\bfA\bfx}_p \leq O\parens*{d^{\frac{p-q}{2p}}}Q_p(\bfx)} \geq \frac23
    \]
    using at most $O(d^{q/2+1}\poly\log n)$ bits of space.
\end{itemize}
\end{Theorem}
\begin{proof}
In the for all model, our result follows from Theorem \ref{thm:lp-lq-subspace-embedding}. In the for each model, we further compose with a frequency moments sketch as in Remark 4.4 of \cite{LWW2021} (using Theorem \ref{thm:frequency-moment-sketch}) to further reduce the number of rows to $r^{1-2/q} = \tilde O(d^{q/2 - 1})$, so that the final space bound is $O(d^{q/2}\poly\log n)$ bits of space. Note that in order to ensure that our space complexity only uses a bit complexity of $O(\log n)$, we may add $1/n$ to Lewis weights and round them, since Theorem \ref{thm:lp-lq-subspace-embedding} only requires $O(1)$-approximate $\ell_p$ Lewis weights.
\end{proof}

\section{Streaming Computation of Lewis Weights}

\subsection{Rounding and Approximation}

We record some easy definitions and lemmas for rounding and approximation that will be useful for our streaming algorithms.

\begin{Definition}\label{def:round}
Let $\delta>0$. We define $\round_\delta(a) \coloneqq \delta\ceil*{a/\delta}$. 
\end{Definition}

Note the following:
\begin{Lemma}
We have that
\[
    a\leq \round_\delta(a) < a + \delta.
\]
\end{Lemma}

\begin{Definition}
Let $\alpha\geq1$. If $x,y\geq0$, we write $x \approx_\alpha y$ if
\[
    \frac1\alpha x \leq y \leq \alpha x.
\]
If $\bfx,\bfy\in\mathbb R^n$ have nonnegative entries, we write $\bfx\approx_\alpha\bfy$ if $\bfx_i\approx_\alpha\bfy_i$ for every $i\in[n]$. If $\bfX,\bfY\in\mathbb R^{n\times n}$ are PSD matrices, we write $\bfX\approx_\alpha\bfY$ if
\[
    \frac1\alpha \bfX \preceq \bfY \preceq \alpha\bfX
\]
in the usual L\"owner order. 
\end{Definition}

\begin{Lemma}[Fact 3.1 \cite{CP2015}]
    Let $\bfX,\bfY\in\mathbb R^{d\times d}$ and $\bfA\in\mathbb R^{n\times d}$. The following hold:
    \begin{itemize}
    \item If $\bfX,\bfY$ are positive definite, then $\bfX\approx_\alpha \bfY \implies \bfX^{-1} \approx_\alpha \bfY^{-1}$. 
    \item $\bfX\approx_\alpha\bfY \implies \bfA^\top\bfX\bfA \approx_\alpha \bfA^\top\bfY\bfA$.
    \end{itemize}
\end{Lemma}

The following lemma shows that rounding large numbers gives relative error approximations. 

\begin{Lemma}\label{lem:round-approx}
Let $\alpha \geq 1$ and $\delta>0$. Let $x,y \geq L$ for some $L > 2\delta$. If $x\approx_\alpha y$, then
\[
    \round_\delta(x) \approx_{\alpha(1+2\delta/L)} \round_\delta(y)
\]
\end{Lemma}
\begin{proof}
Note that $1/(1+2x) \leq 1-x$ on $x\in[0,1/2]$. We then have that
\begin{align*}
    \round_\delta(y) &\geq y \geq \frac1\alpha x \geq \frac1\alpha (\round_\delta(x) - \delta) \geq \frac1\alpha \parens*{\round_\delta(x) - \delta \frac{\round_\delta(x)}{L}} \\
    &= \frac1\alpha \parens*{1 - \frac{\delta}{L}}\round_\delta(x) \geq \frac1{\alpha(1-2\delta/L)}\round_\delta(x)
\end{align*}
and
\begin{align*}
    \round_\delta(y) &\leq y + \delta \leq \alpha x + \delta \leq \alpha \round_\delta(x) + \delta \frac{\round_\delta(x)}{L} \\
    &= \alpha \parens*{1 + \frac{\delta}{L}}\round_\delta(x) \leq \alpha(1+2\delta/L)\round_\delta(x).
\end{align*}
\end{proof}

\subsection{\texorpdfstring{$O(\log\log n)$}{O(log log n)} Pass Algorithm for \texorpdfstring{$0 < p < 4$}{0 < p < 4}}
We first note for $0 < p < 4$, by carefully implementing the fast iterative algorithm of \cite{CP2015} in a stream, one can obtain an $O(\log\log n)$ pass algorithm. In Lemma 2.4, \cite{CP2015} show that repeatedly performing the update
\[
    \bfw_i' \gets \parens*{\bfa_i^\top(\bfA^\top\bfW^{1-2/p}\bfA)^{-1}\bfa_i}^{p/2}
\]
for all $i\in[n]$ quickly yields constant factor solutions in $O(\log\log n)$ iterations. Naively, this requires $\tilde\Theta(n)$ space even to just maintain the weights $\bfw'$. We show that by carefully just maintaining the quadratic form $\bfA^\top\bfW^{1-2/p}\bfA$, we obtain an algorithm using only $O(d^2\log n)$ bits of space.

\begin{algorithm}
	\caption{Lewis quadratic in a stream}
	\textbf{input:} $\bfA\in\mathbb R^{n\times d}$ in a row arrival stream, $0<p<4$, number of passes $T$. \\
	\textbf{output:} Lewis quadratic $\bfA^\top\bfW^{1-2/p}\bfA$.
	\begin{algorithmic}[1] 
        \State $\delta\gets \Theta(\min\braces*{1, \Theta(d/n)^{1-2/p}})$ an integer power of $2$
	    \State $\bfM\gets 0$ \Comment{$\bfM$ will maintain $\bfA^\top\bfW^{1-2/p}\bfA$}
	    \For{$i\in[n]$}
	        \State $\bfM \gets \bfM + \bfa_i\bfa_i^\top$ \Comment{Compute $\bfM = \bfA^\top\bfA$, i.e., $\bfW = \diag(\mathbf{1}_n)$}
	    \EndFor
	    \For{$t\in [T]$}
	        \State $\bfQ\gets 0$
	        \For{$i\in[n]$}
	            \State $w\gets \round_{\delta}\parens*{\max\braces*{\Theta(d/n),\bracks*{\bfa_i^\top\bfM^{-1}\bfa_i}^{p/2}}^{1-2/p}}$ \Comment{Compute $\bfW^{1-2/p}$ with new $\bfW$}\label{line:round-w}
	            \State $\bfQ\gets\bfQ + w\cdot \bfa_i\bfa_i^\top$ \Comment{Compute $\bfQ = \bfA^\top\bfW^{1-2/p}\bfA$ with new $\bfW$}
	        \EndFor
	        \State $\bfM\gets\bfQ$
	    \EndFor
        \State \Return $\bfM$
	\end{algorithmic}\label{alg:lewis-quadratic-stream}
\end{algorithm}

\begin{Lemma}[Lewis quadratic in a stream]\label{lem:lewis-quadratic-stream}
Let $\bfA\in\mathbb R^{n\times d}$ and let $0 < p < 4$ be a constant. Then, there is an $O(\log\log n)$ pass $O(d^2\log n)$ bit space algorithm (Algorithm \ref{alg:lewis-quadratic-stream}) in the row arrival model, which computes $\bfM = \bfA^\top\bfW^{1-2/p}\bfA\in\mathbb R^{d\times d}$, where $\bfW$ are $O(1)$-approximate $\ell_p$ Lewis weights of $\bfA$. Furthermore, we may recover $O(1)$-approximations to $\bfW$ given $\bfM$ by computing
\begin{equation}\label{eq:recover-weight-from-quadratic}
    \tilde\bfw_i = \max\braces*{\gamma, \parens*{\bfa_i^\top \bfM^{-1}\bfa_i}^{p/2}}
\end{equation}
for some $\gamma = 1/\poly(n)$.
\end{Lemma}

Before proving Lemma \ref{lem:lewis-quadratic-stream} in Section \ref{sec:lewis-quadratic-stream-proof}, we show the theorem which follows for solving the subspace sketch problem in a stream.

\begin{Theorem}[Subspace sketch in a stream, $2 < p < 4$]\label{thm:subspace-sketch-stream-2<p<4}
    Let $\bfA\in\mathbb R^{n\times d}$, $2 < p < 4$. Let $2 \leq q \leq p$. Then, there is a $O(\log\log n)$ pass streaming algorithm which uses $O(d^{q/2+1}\poly(\log n,\eps^{-1}))$ space and outputs a data structure $Q_p$ such that
    \[
        \Pr\braces*{\forall\bfx\in\mathbb R^d, \norm*{\bfA\bfx}_p \leq Q_p(\bfx) \leq O(d^{\frac12\parens*{1-\frac{q}{p}}})\norm*{\bfA\bfx}_p} \geq \frac23
    \]
\end{Theorem}
\begin{proof}
    We first obtain the Lewis quadratic $\bfM = \bfA^\top\bfW^{1-2/p}\bfA$ for $O(1)$-approximate $\ell_p$ weights $\bfw$ by Lemma \ref{lem:lewis-quadratic-stream} using $O(\log\log n)$ passes and $O(d^2\log n)$ bits of space. Furthermore, given the row $\bfa_i$, we recover the corresponding weight, up to constant factors, by using the formula of Equation \eqref{eq:recover-weight-from-quadratic}. Thus, given a row arrival stream of $\bfA$ and $\bfM$, we can simulate a row arrival stream of a matrix $\bfB$ such that
    \[
        \forall \bfx\in\mathbb R^d, \norm*{\bfA\bfx}_p \leq \norm*{\bfB\bfx}_q \leq O(d^{\frac12\parens*{1-\frac{q}{p}}})\norm*{\bfA\bfx}_p
    \]
    by Theorem \ref{thm:lp-lq-change-of-density-lewis}.
    
    It then suffices to apply Theorem \ref{thm:1+eps-merge-reduce-subspace-sketch} for $\ell_q$ to obtain a subspace sketch of $\bfB$ with $O(1)$ distortion, as in the proof of the ``for all'' model algorithm of Theorem \ref{thm:subspace-sketch-lp-lq}. 
\end{proof}

\subsubsection{Proof of Lemma \ref{lem:lewis-quadratic-stream}}\label{sec:lewis-quadratic-stream-proof}

Our proof will follow the analysis of Section 3 in \cite{CP2015}, while making appropriate modifications to handle bit complexity. 

\paragraph{Thresholded Lewis weights.}
We first analyze the limit of the process which repeatedly performs the update
\[
    \bfw_i' \gets \max\braces*{\gamma, \parens*{\bfa_i^\top(\bfA^\top\bfW^{1-2/p}\bfA)^{-1}\bfa_i}^{p/2}}
\]
for some $\gamma>0$, for each $i\in[n]$. Note that the thresholding by $\gamma = 1/\poly(n)$ is necessary in order to get bounded bit complexity, so that rounding to the nearest $1/\poly(n)$ will be a relative error approximation. 

\begin{Lemma}\label{lem:continuous-contraction-limit}
    Let $\bfA\in\mathbb R^{n\times d}$, $0 < p < 4$, and $\gamma\geq0$. Define $f:\mathbb R^n\to\mathbb R^n$ via
    \begin{equation}\label{eq:thresholded-contraction-mapping}
        f_i(\bfw) = \max\braces*{\gamma, \parens*{\bfa_i^\top(\bfA^\top\bfW^{1-2/p}\bfA)^{-1}\bfa_i}^{p/2}}.
    \end{equation}
    Then for any $\bfv,\bfw\in\mathbb R^n$ with nonnegative entries such that $\bfv\approx_\alpha\bfw$, we have that $f(\bfv)\approx_{\alpha^{\abs*{p/2-1}}} f(\bfw)$. 
\end{Lemma}
\begin{proof}
The result was shown in Lemma 3.2 of \cite{CP2015} for $\gamma = 0$. Note then that the thresholding by $\gamma$ can only move points closer together. That is,
\begin{equation*}\label{eq:thresholded-contraction}
\begin{aligned}
    \bfv \approx_\alpha \bfw &\implies \parens*{\bfa_i^\top(\bfA^\top\bfV^{1-2/p}\bfA)^{-1}\bfa_i}^{p/2} \approx_{\alpha^{\abs*{p/2-1}}} \parens*{\bfa_i^\top(\bfA^\top\bfW^{1-2/p}\bfA)^{-1}\bfa_i}^{p/2} \\
    &\implies \max\braces*{\gamma, \parens*{\bfa_i^\top(\bfA^\top\bfV^{1-2/p}\bfA)^{-1}\bfa_i}^{p/2}} \approx_{\alpha^{\abs*{p/2-1}}} \max\braces*{\gamma, \parens*{\bfa_i^\top(\bfA^\top\bfW^{1-2/p}\bfA)^{-1}\bfa_i}^{p/2}}
\end{aligned}
\end{equation*}
\end{proof}

From the above result, the Banach fixed point theorem applied to the log weights yields a unique solution $\bfw\in\mathbb R^n$ which satisfies
\begin{equation}\label{eq:fixed-point-streaming-lewis-weights}
    \bfw_i = \max\braces*{\gamma, \parens*{\bfa_i^\top(\bfA^\top\bfW^{1-2/p}\bfA)^{-1}\bfa_i}^{p/2}}
\end{equation}
for each $i\in[n]$. This fixed point satisfies the following bounds:

Note then that the $\bfw$ are one-sided $\ell_p$ Lewis weights (Definition \ref{def:one-sided-lewis}). Furthermore, we have the following bound on the sum of these one-sided Lewis weights:

\begin{Lemma}
Let $\bfw\in\mathbb R^n$ be the fixed point of the map of Equation \ref{eq:thresholded-contraction-mapping}. Then,
\[
    \bfw_i = \max\{\bftau_i(\bfW^{1/2-1/p}\bfA), \gamma\} \geq \bftau_i(\bfW^{1/2-1/p}\bfA)
\]
for each $i\in[n]$. In particular,
\[
    d \leq \norm*{\bfw}_1 \leq d + \gamma n
\]
\end{Lemma}
\begin{proof}
For each $i\in[n]$, we have that
\[
    \bfw_i = \max\braces*{\gamma, \parens*{\bfa_i^\top(\bfA^\top\bfW^{1-2/p}\bfA)^{-1}\bfa_i}^{p/2}} \geq \parens*{\bfa_i^\top(\bfA^\top\bfW^{1-2/p}\bfA)^{-1}\bfa_i}^{p/2}
\]
Rearranging then yields that
\[
    \bfw_i \geq (\bfw_i^{1/2-1/p}\bfa_i^\top)(\bfA^\top\bfW^{1-2/p}\bfA)^{-1}(\bfw_i^{1/2-1/p}\bfa_i) = \bftau_i(\bfW^{1/2-1/p}\bfA)
\]
Furthermore, for every $i\in[n]$ such that the max is achieved by the latter term of the max, we have that
\[
    \bfw_i = (\bfw_i^{1/2-1/p}\bfa_i^\top)(\bfA^\top\bfW^{1-2/p}\bfA)^{-1}(\bfw_i^{1/2-1/p}\bfa_i) = \bftau_i(\bfW^{1/2-1/p}\bfA)
\]
Thus, $\bfw_i = \max\{\bftau_i(\bfW^{1/2-1/p}\bfA),\gamma\}$. Summing over $i\in[n]$ gives that
\[
    d \leq \sum_{i=1}^n \bftau_i(\bfW^{1/2-1/p}\bfA) \leq \norm*{\bfw}_1 \leq \sum_{i=1}^n \bracks*{\bftau_i(\bfW^{1/2-1/p}\bfA) + \gamma} = d + \gamma n
\]
by Lemma \ref{lem:lewis-weights-sum-to-d}.

\end{proof}

\paragraph{Approximating thresholded Lewis weights with bounded bit complexity.} 
Next, we show that the rounding done in Algorithm \ref{alg:lewis-quadratic-stream} will give us relative error estimates on the thresholded Lewis weights. We start with the initial distortion, by adapting Lemma 3.5 of \cite{CP2015}.

\begin{Lemma}\label{lem:contraction-initial}
    Let $\bfw\in\mathbb R^n$ be the fixed point of the map of Equation \ref{eq:thresholded-contraction-mapping}. Then,
    \[
        \max\{\gamma, [\bftau_i(\bfA)]^{p/2}\} \approx_{n^{\abs*{p/2-1}}} \bfw_i
    \]
\end{Lemma}
\begin{proof}
    The proof of Lemma 3.5 of \cite{CP2015} essentially goes through. Let $\bfU = \bfA\bfR\in\mathbb R^{n\times d}$ be the one-sided Lewis basis for $\bfw$. We will prove the result with $\bfA$ replaced by $\bfU$, since leverage scores depend only on the column span of a matrix and not on any particular basis for the space, and $\bfw$ is only defined in terms of leverage scores. 
    
    Let $\bfB = \bfW^{1/2-1/p}\bfU$, so that
    \[
        \bfB^\top\bfB = (\bfW^{1/2-1/p}\bfU)^\top (\bfW^{1/2-1/p}\bfU) = \bfI
    \]
    We first show that $\bfU^\top\bfU = (\bfW^{1/p-1/2}\bfB)^\top(\bfW^{1/p-1/2}\bfB) \approx_{n^{\abs*{1-2/p}}} \bfI$. For a unit vector $\bfx\in\mathbb R^d$,
    \[
        1 = \norm*{\bfx}_2^2 = \norm*{\bfB\bfx}_2^2 = \sum_{i=1}^n (\bfx^\top\bfb_i)^2 = \sum_{i=1}^n \bfw_i \frac{(\bfx^\top\bfb_i)^2}{\bfw_i},
    \]
    and
    \[
        \norm*{\bfU\bfx}_2^2 = \norm*{\bfW^{1/p-1/2}\bfB\bfx}_2^2 = \sum_{i=1}^n \bfw_i^{2/p-1}(\bfx^\top\bfb_i)^2 = \sum_{i=1}^n \bfw_i^{2/p}\frac{(\bfx^\top\bfb_i)^2}{\bfw_i}.
    \]
    Then by the Cauchy--Schwarz inequality,
    \[
        \frac{(\bfx^\top\bfb_i)^2}{\bfw_i} \leq \frac{\norm*{\bfx}_2^2\norm*{\bfb_i}_2^2}{\bfw_i} = \frac{\bftau_i(\bfB)}{\bfw_i} \leq \frac{\bfw_i}{\bfw_i} = 1
    \]
    so the distortion between $\norm*{\bfB\bfx}_2^2$ and $\norm*{\bfU\bfx}_2^2$ is bounded by the distortion between $\norm*{\bfw}_1$ and $\norm*{\bfw}_{2/p}^{2/p}$, which is $n^{\abs*{1-2/p}}$. 
    
    By inverting the matrices and applying the quadratic form to $\bfu_i$ and raising to the $p/2$th power, we find that
    \[
        [\bftau_i(\bfU)]^{p/2} = [\bfu_i^\top(\bfU^\top\bfU)^{-1}\bfu_i]^{p/2} \approx_{n^{\abs*{p/2-1}}} [\bfu_i^\top(\bfU^\top\bfW^{1-2/p}\bfU)^{-1}\bfu_i]^{p/2}
    \]
    Since thresholding by $\gamma$ can only decrease the distortion, we have that
    \[
        \max\{\gamma,\bftau_i(\bfU)^{p/2}\} \approx_{n^{\abs*{p/2-1}}} \max\braces*{\gamma, [\bfu_i^\top(\bfU^\top\bfW^{1-2/p}\bfU)^{-1}\bfu_i]^{p/2}} = \bfw_i
    \]
    as claimed.
\end{proof}

We are now ready to prove Lemma \ref{lem:lewis-quadratic-stream} by putting the previous pieces together. 

\begin{proof}[Proof of Lemma \ref{lem:lewis-quadratic-stream}]
We follow Lemma 2.4 of \cite{CP2015}. Note first that before rounding, $w$ on line \ref{line:round-w} is in $[1, \Theta(\eps d/n)^{1-2/p}]$ for $p < 2$ and $[\Theta(d/n)^{1-2/p}, 1]$ for $p > 2$. Let $L = \min\{1, \Theta(d/n)^{1-2/p}\}$. Then if we round $\bfw_i^{1-2/p}$ to the nearest multiple of $\delta$ in line \ref{line:round-w}, then this is only a $(1+\beta)$-factor distortion, for $\beta = 2\delta/L$. This translates to a $(1+\beta)^{\abs*{1-2/p}}$-factor distortion of $\bfw_i$. Then by Lemma \ref{lem:continuous-contraction-limit}, if $\bfv\approx_\alpha\bfw$, then the resulting $f(\bfv)$ and $f(\bfw)$ after this rounding process are distorted by a factor of
\[
    (1+\beta)^{\abs*{1-2/p}}\alpha^{\abs*{p/2-1}}.
\]
Note also that by Lemma \ref{lem:contraction-initial}, we initially have an $n^{\abs*{p/2-1}}$ factor approximation. Then, the distortion from the initial approximation is $n^{\abs*{p/2-1}^T}$ while the distortion from the rounding is at most $(1+\beta)^{\abs*{1-2/p}/(1-\abs*{p/2-1})}$. Then, with $T = O(\log_{\abs*{p/2-1}} \log n) = O(\log\log n)$, the distortion from the initial approximation is at most $n^{O(1/\log n)} = O(1)$, and with
\[
    \delta = O\parens*{\frac{1-\abs*{p/2-1}}{\abs*{1-2/p}} L} = O(L),
\]
the distortion from the rounding is at most $O(1)$. Since $L = 1/\poly(n)$, we can choose $\delta$ to be an integer power of $2$ so that the bit complexity is $O(\log n)$. 

We only store $\bfM$ which is a $d\times d$ matrix with entries bounded by $O(\log n)$ bits, so the total space usage is $O(d^2\log n)$. The number of passes is $T = O(\log\log n)$. 
\end{proof}

\subsection{\texorpdfstring{$O(\log\log n)$}{O(log log n)} Pass Algorithm for \texorpdfstring{$p\geq 4$}{p >= 4}}

We now extend the $O(\log\log n)$-pass algorithm from the previous section to arbitrary constant $p$. To do this, we use the observation of \cite{MMMWZ2022} of using the $\ell_{p/k}$ Lewis weights of the $n\times d^k$ matrix $\bfA^{\otimes k}$, where $\bfA^{\otimes k}$ is the Khatri--Rao product of $\bfA$ with itself $k$ times, obtained by assigning the $i$th row to be the vector tensor product $\bfa_i^{\otimes k} = \bfa_i \otimes \bfa_i \dots \otimes \bfa_i$. This follows from the identity
\[
    \abs*{\angle*{\bfa^{\otimes k}, 
    \bfx^{\otimes k}}}^{p/k} = \abs*{\angle*{\bfa,\bfx}^k}^{p/k} = \abs*{\angle*{\bfa,\bfx}}^p
\]
which means that $\norm*{\bfA\bfx}_p^p = \norm*{\bfA^{\otimes k}\bfx^{\otimes k}}_{p/k}^{p/k}$. By taking $k$ to be an integer large enough so that $p/k < 4$ (i.e., $k = \floor{p/4} + 1$), we can then use the $O(\log\log n)$-pass algorithm of Theorem \ref{thm:subspace-sketch-stream-2<p<4}.

Naively, this requires $O(d^{2k}\log n)$ space, obtained by applying Theorem \ref{thm:subspace-sketch-stream-2<p<4} with $d$ set instead to $d^k$. We will now show how to improve this to $O(d^{k+1}\log n)$ space, by maintaining a coreset for the Lewis quadratic via leverage score sampling (which are equivalent to $\ell_2$ Lewis weights), using Theorem \ref{thm:1+eps-merge-reduce-subspace-sketch}. By storing a coreset of $\tilde O(d^{k})$ rows, we can implicitly maintain this by storing the original rows of $\bfA$ rather than their $k$-fold tensor products, which allows us to replace a factor of $d^k$ by $d$. 

\begin{Lemma}\label{lem:spectral-coreset}
    Let $\bfB\in\mathbb R^{n\times d}$, $k\geq 1$ an integer, $\eps>0$ an accuracy paramter, and $\delta>0$ a failure rate parameter. Then, there is a one pass streaming algorithm using at most $O(d^{k+1}\poly(\log n,\eps^{-1}))$ bits, which outputs an $s\times d$ matrix $\bfC$ with $s = O(d^k\poly(\log n,\eps^{-1}))$ such that
    \[
        \Pr\braces*{(\bfC^{\otimes k})^\top \bfC^{\otimes k} \preceq (\bfB^{\otimes k})^\top \bfB^{\otimes k} \preceq (1+\eps)(\bfC^{\otimes k})^\top \bfC^{\otimes k}} \geq 1-\delta,
    \]
    where for a matrix $\bfM\in\mathbb R^{m\times d}$, $\bfM^{\otimes k}$ denotes the $k$-fold Khatri--Rao product of $\bfM$ with itself.
\end{Lemma}
\begin{proof}
    We apply Theorem \ref{thm:1+eps-merge-reduce-subspace-sketch} with $p = 2$ to obtain our $s\times d$ matrix $\bfC$, by taking $\bfC$ to be the original rows of $\bfA$ rather than the $k$-fold tensor products. Note then that
    \[
        \forall\bfx\in\mathbb R^d, \qquad \bfx^\top\bfC^\top\bfC\bfx = \norm*{\bfC\bfx}_2^2 = (1\pm\eps)\norm*{\bfB\bfx}_2^2 = (1\pm\eps)\bfx^\top\bfB^\top\bfB\bfx,
    \]
    which is equivalent to the desired conclusion.
\end{proof}

Note that for symmetric positive definite matrices $\bfK$ and $\bfL$,
\[
    \bfL \preceq \bfK\preceq (1+\eps)\bfL \iff (1+\eps)^{-1}\bfL^{-1} \preceq \bfK^{-1}\preceq \bfL^{-1}.
\]
Thus, by applying Lemma \ref{lem:spectral-coreset} with $\bfB = \bfW^{1/2-1/p}\bfA$, we can implement the algorithm of Lemma \ref{lem:lewis-quadratic-stream}, up to a loss of $(1+\eps)$ factors. We thus obtain the following:

\begin{Theorem}[Subspace sketch in a stream, $p \geq 4$]\label{thm:subspace-sketch-stream-p>4}
    Let $\bfA\in\mathbb R^{n\times d}$, $p \geq 4$. Let $k = \floor{p/4}+1$. Then, there is a $O(\log\log n)$ pass streaming algorithm which uses $O(d^{q/2+1} + d^{k+1})\poly(\log n,\eps^{-1})$ bits of space and outputs a data structure $Q_p$ such that
    \[
        \Pr\braces*{\forall\bfx\in\mathbb R^d, \norm*{\bfA\bfx}_p \leq Q_p(\bfx) \leq O(d^{\frac12\parens*{1-\frac{q}{p}}})\norm*{\bfA\bfx}_p} \geq \frac23
    \]
\end{Theorem}

\subsection{\texorpdfstring{$O(\log n)$}{O(log n)} Pass Algorithm for \texorpdfstring{$p\geq 4$}{p >= 4}}

In this section, we implement an alternative $\ell_p$ Lewis weight computation algorithm of \cite{Lee2016} for $p\geq 2$ to obtain an algorithm with a higher number of passes with $O(\log n)$ passes, but with lower space complexity. Note that this algorithm is only useful for $p\geq 4$, given the guarantees of Algorithm \ref{alg:lewis-quadratic-stream} found in Lemma \ref{lem:lewis-quadratic-stream}. We follow the presentation of a similar algorithm by \cite{JLS2021}, which gives a simplified analysis. As with our adaptation of the \cite{CP2015}, we require some modifications in order to implement the algorithm with small bit complexity. The algorithm is very similar to Algorithm \ref{alg:lewis-quadratic-stream}, except that we need more passes, use a slightly different recursion, and need to output every iterate of the Lewis quadratic rather than the last iterate, since the final weights are the average iterate of the weights rather than the last iterate.

Recall that the algorithms of \cite{Lee2016,JLS2021} find an approximate fixed point of the recursion
\begin{equation}\label{eq:lee-lewis-weight-recursion}
    \bfw_i' \gets \bftau_i(\bfW^{1/2-1/p}\bfA).
\end{equation}
Our modified implementation of this algorithm is then found in Algorithm \ref{alg:lewis-quadratic-stream-log-n-pass}.

\begin{algorithm}
	\caption{Lewis quadratic in a stream}
	\textbf{input:} $\bfA\in\mathbb R^{n\times d}$ in a row arrival stream, $p\geq 2$, number of passes $T$. \\
	\textbf{output:} $T+1$ Lewis quadratics $\bfQ_{(t)} = \bfA^\top\bfW_{(t)}^{1-2/p}\bfA$ for $t\in[T]\cup\{0\}$.
	\begin{algorithmic}[1] 
        \State $\delta\gets \Theta(d/n)$ an integer power of $2$
	    \State $\bfQ_{(0)}\gets 0$ \Comment{$\bfQ_{(t)}$ will maintain $\bfA^\top\bfW_{(t)}^{1-2/p}\bfA$}
	    \For{$i\in[n]$}
	        \State $\bfQ_{(0)} \gets \bfQ_{(0)} + (d/n)\bfa_i\bfa_i^\top$ \Comment{Compute $\bfM = \bfA^\top\bfA$, i.e., $\bfW_{(0)} = \diag((d/n)\cdot\mathbf{1}_n)$}
	    \EndFor
	    \For{$t\in [T]$}
	        \State $\bfQ_{(t)}\gets 0$
	        \For{$i\in[n]$}
                \State $w\gets d/n$
                \For{$k\in[t]$}
                    \State $w\gets \round_{\delta}\parens*{\max\braces*{\Theta(d/n), (w\cdot\bfa_i)^\top\bfQ_{(k-1)}^{-1}(w\cdot\bfa_i)}}$ \Comment{$w$ maintains $(\bfW_{(k)})_{i,i}$} \label{line:recursion-log-n-pass}
                \EndFor
	            \State $\bfQ_{(t)}\gets\bfQ_{(t)} + w\cdot \bfa_i\bfa_i^\top$ \Comment{Compute $\bfQ_{(t)} = \bfA^\top\bfW_{(t)}^{1-2/p}\bfA$ with new $\bfW_{(t)}$}
	        \EndFor
	    \EndFor
        \Return $\bfQ_{(t)}$, $t\in[T]\cup\{0\}$
	\end{algorithmic}\label{alg:lewis-quadratic-stream-log-n-pass}
\end{algorithm}

Note that compared to the recursion of Equation \eqref{eq:lee-lewis-weight-recursion}, the recursion in Algorithm \ref{alg:lewis-quadratic-stream-log-n-pass} on line \ref{line:recursion-log-n-pass} needs to threshold by some $1/\poly(n)$, just as in Algorithm \ref{alg:lewis-quadratic-stream}.

\begin{Lemma}[Lewis quadratic in a stream]\label{lem:lewis-quadratic-stream-log-n-pass}
    Let $\bfA\in\mathbb R^{n\times d}$ and let $2\leq p <\infty$ be a constant. Then, there is an $O(\log\log n)$ pass $O(d^2\log^2 n)$ bit space algorithm (Algorithm \ref{alg:lewis-quadratic-stream-log-n-pass}) in the row arrival model, which computes $T = O(\log n)$ quadratics $\bfQ_{(t)} = \bfA^\top\bfW_{(t)}^{1-2/p}\bfA\in\mathbb R^{d\times d}$ for $t\in[T]$. Furthermore, the weights $\bfW_{(t)}$ and the quadratics $\bfQ_{(t)}$ satisfy
    \[
        \bfw_{(t)}(i) = \round_{\delta}\parens*{\max\braces*{\Theta(d/n), (\bfw_{(t-1)}(i)\cdot\bfa_i)^\top\bfQ_{(t-1)}^{-1}(\bfw_{(t-1)}(i)\cdot\bfa_i)}}
    \]
    (see line \ref{line:recursion-log-n-pass}) and
    \[
        \bfw\coloneqq\frac{3}{2T} \sum_{t=1}^T \bfw_{(t)}
    \]
    are $O(1)$-approximate $\ell_p$ Lewis weights.
\end{Lemma}
\begin{proof}
    The number of passes is immediate from the pseudocode of Algorithm \ref{alg:lewis-quadratic-stream-log-n-pass}. The space complexity follows from the fact that we maintain $O(\log n)$ matrices that are each $d\times d$ with entries bounded by $O(\log n)$ bits.

    The proof for correctness almost exactly follows the analysis in \cite{JLS2021}. We include the details for completeness. For the bound on $\norm*{\bfw}_1$, we have that
    \begin{align*}
        \norm*{\bfw}_1 &= \norm*{\frac{3}{2T} \sum_{t=1}^T \bfw_{(t)}}_1 \leq \frac{3}{2T}\sum_{t=1}^T \norm*{\bfw_{(t)}}_1 \leq \frac{3}{2T}\sum_{t=1}^T \sum_{i=1}^n \bfw_{(t)}(i) \\
        &\leq \frac{3}{2T}\sum_{t=1}^T \sum_{i=1}^n (\Theta(d/n) + \bftau_i(\bfW_{(t)}^{1/2-1/p}\bfA)) = \frac{3}{2T}\sum_{t=1}^T \Theta(d) = \Theta(d).
    \end{align*}
    
    Next, we prove that $\bfw_i \geq \bftau_i(\bfW^{1/2-1/p}\bfA)$. Define $\phi_i:\mathbb R_{\geq 0}^n \to \mathbb R$ by
    \[
        \phi_i(\bfv) = \log\parens*{\frac{\bftau_i(\diag(\bfv)^{1/2-1/p}\bfA)}{\bfv_i}}.
    \]
    Lemma A.2 of \cite{JLS2021} shows that $\phi_i$ is convex. Then,
    \begin{align*}
        \phi_i\parens*{\frac1T \sum_{t=1}^T \bfw_{(t)}} &\leq \frac1T \sum_{t=1}^T \phi_i(\bfw_{(t)}) = \frac1T \sum_{t=1}^T \log\parens*{\frac{\bftau_i(\diag(\bfw_{(t)})^{1/2-1/p}\bfA)}{\bfw_{(t)}(i)}} \\
        &\leq \frac1T \sum_{t=1}^T \parens*{\log(1.1) + \log\parens*{\frac{\bfw_{(t+1)}(i)}{\bfw_{(t)}(i)}}} \\
        &\leq 0.1 + \frac1T\log\frac{\bfw_{(T)}(i)}{\bfw_{(0)}(i)} \leq 0.1 + \frac1T\log\frac{n}{d} \leq 0.2.
    \end{align*}
    In the above, we use that for an appropriate choice of $\delta$, rounding to the nearest $\delta$ in line \ref{line:recursion-log-n-pass} gives a $1.1$-approximation of $\max\{\Theta(d/n), \bftau_i(\diag(\bfw_{(t)})^{1/2-1/p}\bfA)\}$, and that $(1.1)\bfw_{(t+1)}(i) \geq \bftau_i(\diag(\bfw_{(t)})^{1/2-1/p}\bfA)$. It follows that
    \[
        \bfw_i \geq \frac32 \exp(-0.2)\bftau_i\parens*{\bfW^{1/2-1/p}\bfA} \geq \bftau_i\parens*{\bfW^{1/2-1/p}\bfA}.
    \]
    Thus, $\bfw$ are $O(1)$-approximate $\ell_p$ Lewis weights.
\end{proof}

Then, Lemma \ref{lem:lewis-quadratic-stream-log-n-pass} gives Theorem \ref{thm:subspace-sketch-stream-p>4-log-n-pass}, just as Lemma \ref{lem:lewis-quadratic-stream} gave Theorem \ref{thm:subspace-sketch-stream-2<p<4}. We omit the proof, as it is the same as before.

\begin{Theorem}[Subspace sketch in a stream, $4\leq p < \infty$]\label{thm:subspace-sketch-stream-p>4-log-n-pass}
    Let $\bfA\in\mathbb R^{n\times d}$, $4\leq p < \infty$. Let $2 \leq q \leq p$. Then, there is a $O(\log n)$ pass streaming algorithm which uses $O(d^{q/2+1}\poly(\log n,\eps^{-1}))$ space and outputs a data structure $Q_p$ such that
    \[
        \Pr\braces*{\forall\bfx\in\mathbb R^d, \norm*{\bfA\bfx}_p \leq Q_p(\bfx) \leq O(d^{\frac12\parens*{1-\frac{q}{p}}})\norm*{\bfA\bfx}_p} \geq \frac23
    \]
\end{Theorem}

\section{Acknowledgements}

David P.\ Woodruff and Taisuke Yasuda were supported by ONR grant N00014-18-1-2562 and a Simons Investigator Award. We thank Timothy Chan, Sariel Har-Peled, Piotr Indyk, and Jeff Phillips for helpful feedback and suggestions. We thank Naren Manoj and Max Ovsiankin for pointing out an error in the proof of Theorem \ref{thm:subspace-sketch-stream-l-inf} in an earlier version of the draft. We thank anonymous reviewers for suggestions which helped improve the presentation of the draft.

\bibliographystyle{alpha}
\bibliography{citations}

\appendix

\section{Online Spectral Approximation via Leverage Score Sampling}\label{sec:online-spectral-approx}

Our result for online spectral approximation is just a tighter analysis of the algorithm from \cite{CMP2020}. This is reproduced in Algorithm \ref{alg:cmp2020-online-sample}. 

\begin{algorithm}
	\caption{Online Spectral Approximation (Figure 1, \cite{CMP2020})}
	\textbf{input:} $\bfA\in\mathbb Z^{n\times d}$ in a row arrival stream, $\eps\in(0,1)$, $\delta>0$. \\
	\textbf{output:} Online spectral approximation $\tilde\bfA$.
	\begin{algorithmic}[1] 
        \State $c\gets 8\log d / \eps^2$
        \State $\tilde\bfA_0\in\mathbb Z^{0\times d}$
        \For{$i\in[n]$}
            \If{$\bfa_i\notin\rowspan(\tilde\bfA_{i-1})$}
                \State $\tilde l_i \gets 1$
            \Else
                \State $\tilde l_i \gets \min\braces*{(1+\eps)\bfa_i^\top(\tilde\bfA_{i-1}^\top\tilde\bfA_{i-1})^-\bfa_i, 1}$
            \EndIf
            \State $p_i \gets \min\{c\tilde l_i, 1\}$
            \State $\tilde\bfA_i \gets \begin{cases}
                \begin{bmatrix}
                    \tilde \bfA_{i-1} \\ \bfa_i / \sqrt p_i
                \end{bmatrix} & \text{with probability $p_i$} \\
                \tilde\bfA_{i-1} & \text{otherwise}
            \end{cases}$
        \EndFor
        \State \Return $\tilde\bfA \coloneqq \tilde\bfA$
	\end{algorithmic}\label{alg:cmp2020-online-sample}
\end{algorithm}

We will adapt Lemma 3.3 of \cite{CMP2020}.

\begin{Lemma}\label{lem:cmp2020-lem3.3}
    Let $\tilde\bfA$ be the matrix returned by Algorithm \ref{alg:cmp2020-online-sample}. Then, with probability at least $1-1/d$,
    \[
        (1-\eps)\bfA^\top\bfA \preceq \tilde\bfA^\top\tilde\bfA \preceq (1+\eps)\bfA^\top\bfA
    \]
    and for each $i\in[n]$,
    \[
        \tilde l_i \geq \bfa_i^\top(\bfA^\top\bfA)^-\bfa_i.
    \]
\end{Lemma}
\begin{proof}
    Let $\bfu_i \coloneqq ((\bfA^\top\bfA)^-)^{1/2}\bfa_i$. We define the matrix martingale $\{\bfY_i\}_{i=0}^n\subseteq\mathbb R^{d\times d}$ exactly as in \cite{CMP2020}, that is, $\bfY_0 = 0$ and has the difference sequence $\{\bfX\}_{i=1}^n$ where if $\norm*{\bfY_{i-1}}_2 \geq \eps$, then $\bfX_i \coloneqq 0$, and otherwise,
    \[
        \bfX_i \coloneqq \begin{cases}
            (1/p_i-1)\bfu_i\bfu_i^\top & \text{if $\bfa_i$ is sampled in $\tilde\bfA$} \\
            -\bfu_i\bfu_i^\top & \text{otherwise}
        \end{cases}.
    \]
    We then have
    \[
        \bfY_{i-1} = ((\bfA^\top\bfA)^-)^{1/2}(\tilde\bfA_{i-1}^\top\tilde\bfA_{i-1} - \bfA_{i-1}^\top\bfA_{i-1})((\bfA^\top\bfA)^-)^{1/2}.
    \]

    Now note that the row space of $\tilde\bfA_i$ and $\bfA_i$ always coincide, so if $\bfa_i$ is not in $\rowspan(\bfA_{i-1})$, then $\tilde l_i = \bftau_i^{\OL}(\bfA) = 1$. Otherwise, let $\norm*{\bfY_{i-1}}_2 < \eps$. Then, $\tilde l_i \geq \bfu_i^\top \bfu_i$ exactly as calculated in \cite{CMP2020} after restricting to the row space. The predictable quadratic variation process of $\{\bfY_i\}$ can then be bounded exactly as in \cite{CMP2020}, so we conclude by the matrix freedman inequality \cite{Tro2011} that
    \[
        \Pr\braces*{\norm*{\bfY_n}_2 \geq \eps} \leq \frac1d.
    \]
    This then implies that
    \[
        (1-\eps)\bfA^\top\bfA \preceq \tilde\bfA^\top\tilde\bfA \preceq (1+\eps)\bfA^\top\bfA
    \]
    with probability at least $1-1/d$.
\end{proof}

Next, we adapt Lemma 3.4 of \cite{CMP2020}. The techniques closely follow the ideas in the proof of Theorem \ref{thm:sharp-online-leverage-scores}.

\begin{Lemma}\label{lem:cmp2020-lem3.4}
After running Algorithm \ref{alg:cmp2020-online-sample}, with probability at least $1-\exp(-d)$,
\[
    \sum_{i=1}^n \tilde l_i = O(d\log n).
\]
\end{Lemma}
\begin{proof}
Define
\[
    \Delta_i \coloneqq \log\pdet(\tilde\bfA_i^\top\tilde\bfA_i) - \log\pdet(\tilde\bfA_{i-1}^\top\tilde\bfA_{i-1}),
\]
where we take $\Delta_1 = \log\norm*{\bfa_1}_2^2$. Note that if $\bfa_i$ is not in $\rowspan(\bfA_{i-1}) = \rowspan(\tilde\bfA_{i-1})$, then with probability $1$, this quantity is at least $\log\norm*{\bfa_i^\bot}_2^2$, where $\bfa_i^\bot$ is the projection of $\bfa_i$ onto the orthogonal complement of $\rowspan(\bfA_{i-1})$ (see the proof of Theorem \ref{thm:sharp-online-leverage-scores} for more details). In turn, we have that
\[
    \E_{i-1}\exp(\tilde l_i/8 - \Delta_i) \leq \E_{i-1}\exp(1/8 - \log\norm*{\bfa_i^\bot}_2^2) \leq \frac{e^{1/8}}{\norm*{\bfa_i^\bot}_2^2}.
\]
Otherwise, we use the matrix pseudodeterminant lemma (Lemma \ref{lem:pdet-evolution-par}) in an analogous manner as in \cite{CMP2020}. That is, for $i\in[n]$ such that $\bfa_i$ is in the row span of $\tilde\bfA_{i-1}$, we show that $\bfE_{i-1}\exp(\tilde l_i/8 - \Delta_i) \leq 1$. Indeed,
\begin{align*}
    \E_{i-1}\exp(\tilde l_i/8 - \Delta_i) &= p_i \cdot e^{\tilde l_i / 8}(1 + \bfa_i^\top(\tilde\bfA_{i-1}^\top\tilde\bfA_{i-1})^-\bfa_i/p_i)^{-1} + (1-p_i)e^{\tilde l_i/8} \\
    &\leq p_i \cdot (1+\tilde l_i/4)(1 + \bfa_i^\top(\tilde\bfA_{i-1}^\top\tilde\bfA_{i-1})^-\bfa_i/p_i)^{-1} + (1-p_i)(1+\tilde l_i/4).
\end{align*}
Now if $c\tilde l_i < 1$, then $p_i = c\tilde l_i$ with $\tilde l_i = (1+\eps)\bfa_i^\top(\tilde\bfA_{i-1}^\top\tilde\bfA_{i-1})^-\bfa_i$, so
\begin{align*}
    \E_{i-1}\exp(\tilde l_i/8 - \Delta_i) &\leq c\tilde l_i\cdot (1+\tilde l_i/4)(1+1/[(1+\eps)c])^{-1} + (1-c\tilde l_i)(1+\tilde l_i/4) \\
    &= (1+\tilde l_i/4)(c\tilde l_i(1+1/[(1+\eps)c])^{-1} + 1 - c\tilde l_i) \\
    &\leq (1+\tilde l_i/4)(1 + c\tilde l_i[1-1/(4c) - 1]) \\
    &= (1+\tilde l_i/4)(1-\tilde l_i/4) \\
    &\leq 1
\end{align*}
while if $p_i = 1$, then
\begin{align*}
    \E_{i-1}\exp(\tilde l_i/8 - \Delta_i) &\leq (1+\tilde l_i/4)(1+\bfa_i^\top(\tilde\bfA_{i-1}^\top\tilde\bfA_{i-1})^-\bfa_i)^{-1} \\
    &\leq (1+\tilde l_i/4) (1+\tilde l_i)^{-1} \\
    &\leq 1.
\end{align*}
Now define $S$ as the set of $i\in[n]$ such that $\bfa_i$ is not in the row span of $\bfA_{i-1}$. Then, we inductively have that
\[
    \E\exp\parens*{\sum_{i=1}^n\tilde l_i/8 - \Delta_i} \leq \prod_{i\in S}\frac{e^{1/8}}{\norm*{\bfa_i^\bot}_2^2} \leq \frac{e^{d/8}}{\prod_{i\in S}\norm*{\bfa_i^\bot}_2^2}.
\]
As reasoned in the proof of Theorem \ref{thm:sharp-online-leverage-scores}, $\prod_{i\in S}\norm*{\bfa_i^\bot}_2^2$ is the determinant of a nonsingular integer matrix and thus at least $1$, so the above quantity is at most $\exp(d/8)$. Then by Markov's inequality, we have that
\[
    \Pr\braces*{\sum_{i=1}^n \tilde l_i > d + 8d + 8\sum_{i=1}^n \Delta_i} \leq \exp(-d).
\]
Now note that
\[
    \sum_{i=1}^n \Delta_i = \log\pdet(\tilde\bfA_n^\top\tilde\bfA_n) \leq d(1+\log\norm*{\bfA}_2^2) \leq O(d\log n)
\]
by Lemma \ref{lem:cmp2020-lem3.3}, so we conclude that with probability at least $1-\exp(-d)$,
\[
    \sum_{i=1}^n \tilde l_i = O(d\log n).
\]
\end{proof}

We may now prove Theorem \ref{thm:online-spectral-approx}:

\OnlineSpectralApproximation*
\begin{proof}
    This follows from Lemmas \ref{lem:cmp2020-lem3.3} and \ref{lem:cmp2020-lem3.4}.
\end{proof}
\section{Additional Lower Bounds for the Subspace Sketch Problem}
Note that all of our algorithms for the subspace sketch problem use at least $d^2$ space in the ``for all'' model and $d$ space in the ``for each'' model. We show that these are necessary to obtain any finite multiplicative distortion. 

\begin{Theorem}\label{thm:finite-distortion-subspace-sketch-for-all}
    Let $\bfA\in\mathbb R^{n\times d}$ and $0<p \leq \infty$. Suppose that an algorithm $\mathcal A$ solves the subspace sketch problem in the ``for all'' model with any finite distortion $\kappa<\infty$. Then, $\mathcal A$ uses at least $\Omega(d^2)$ bits of space.
\end{Theorem}
\begin{proof}
    We recall the following lemma from \cite{VWW2020}:
    \begin{Lemma}[Lemma 3.2 of \cite{VWW2020}]\label{lem:vww2020}
        For any $d>0$ and sufficiently large $t$, there exists a set of matrices $\mathcal T\subseteq\mathbb R^{d\times(d-1)}$ with integral entries in $[-t,t]$ for which $\abs*{\mathcal T} = t^{\Omega(d^2)}$ and
        \begin{enumerate}
            \item For any $T\in\mathcal T$, $\rank(T) = d-1$
            \item For any $S,T\in\mathcal T$ distinct, $\colspan([S\ T]) = \mathbb R^d$
        \end{enumerate}
    \end{Lemma}
    Suppose our input instance is $T^\top$ for a uniformly random $T\in\mathcal T$. Then, there exists $\bfx_T\in\mathbb R^d$ such that $T\bfx_T = 0$ but $S\bfx_T \neq 0$ for all $S\in\mathcal T$ not equal to $T$, since otherwise condition 2 in Lemma \ref{lem:vww2020} cannot hold. Thus, by testing $\bfx_S$ for all $S\in\mathcal T$ with the subspace sketch data structure, one can recover $T\in\mathcal T$ exactly. Then by standard information theoretic arguments, $\mathcal A$ must use at least $\Omega(d^2)$ bits of space.
\end{proof}

\begin{Theorem}\label{thm:finite-distortion-subspace-sketch-for-each}
    Let $\bfA\in\mathbb R^{n\times d}$ and $0<p<\infty$. Suppose that an algorithm $\mathcal A$ solves the subspace sketch problem in the ``for each'' model with any finite distortion $\kappa<\infty$. Then, $\mathcal A$ uses at least $\Omega(d)$ bits of space.
\end{Theorem}
\begin{proof}
    We reduce from \textsf{INDEX}. Let $A\subseteq[d]$ be Alice's subset and let $b\in[d]$ be Bob's index. Then, Alice takes the input matrix $\bfA$ to be the $1\times d$ matrix with the one row being the binary indicator vector of $A$. Then, Bob can test whether $b\in A$ or not by querying the $b$th standard basis vector $\bfe_b$. Thus, $\mathcal A$ must use at least $\Omega(d)$ bits of space.
\end{proof}

\end{document}